\documentclass[12pt]{article}
\usepackage{jheppub}
\pdfoutput=1
\usepackage{epstopdf}

\usepackage{jheppub}
\usepackage{mathrsfs}
\usepackage{psfrag}
\usepackage{color}
\usepackage{hyperref}

\usepackage{slashed}
\usepackage{feynmp-auto}
\usepackage{simplewick}
\usepackage{cancel}

\usepackage[table]{xcolor}

\usepackage{amsmath,bbm,array,amsfonts,graphicx,wrapfig,arydshln,lscape,float,multirow,longtable,rotating,makecell}
\usepackage{url}

\newcommand{\la}{\langle}
\newcommand{\ra}{\rangle}

\newcommand{\ab}[1]{\langle #1 \rangle}
\newcommand{\sqb}[1]{[ #1 ]}
\newcommand{\aMs}[3]{\langle #1|#2|#3]}  		
\newcommand{\sab}[1]{s_{#1}}

\def\dlog{d\log}

\def\N{\mathcal{N}}
\def\M{\mathcal{M}}
\def\J{\mathcal{J}}
\def\Q{\mathcal{Q}}
\def\K{\mathcal{K}}

\def\PT{\text{PT}}

\newcommand{\tw}[1]{\widetilde{#1}}
\newcommand{\lam}[1]{\lambda_{#1} }
\newcommand{\lamt}[1]{\widetilde \lambda_{#1} }
\def\twEta{\tw{\eta}}

\newcommand{\NeqFour}{\mathcal{N}=4\text{ sYM}}
\newcommand{\NeqEight}{\mathcal{N}=8\text{ SUGRA}}

\def\MHVbar{\overline{\text{MHV}}}

\definecolor{airforceblue}{rgb}{0.36, 0.54, 0.66}
\definecolor{bananayellow}{rgb}{1.0, 0.88, 0.21}
\definecolor{bittersweet}{rgb}{1.0, 0.44, 0.37}
\definecolor{blue(ncs)}{rgb}{0.0, 0.53, 0.74}
\definecolor{bole}{rgb}{0.47, 0.27, 0.23}
\definecolor{brass}{rgb}{0.71, 0.65, 0.26}
\definecolor{bronze}{rgb}{0.8, 0.5, 0.2}
\definecolor{brgreen}{rgb}{0.0, 0.26, 0.15}
\definecolor{burgundy}{rgb}{0.5, 0.0, 0.13}
\definecolor{cherry}{rgb}{1.0, 0.72, 0.77}
\definecolor{cocao}{rgb}{0.82, 0.41, 0.12}
\definecolor{citrine}{rgb}{0.99, 0.82, 0.07}

\newcommand{\tred}[1]{\textcolor{red}{#1}}
\newcommand{\tblue}[1]{\textcolor{blue}{#1}}
\newcommand{\tgreen}[1]{\textcolor{green}{#1}}
\newcommand{\tmagenta}[1]{\textcolor{magenta}{#1}}
\newcommand{\twhite}[1]{\textcolor{white}{#1}}

\newcommand{\tcyan}[1]{\textcolor{cyan}{#1}}

\newcommand{\tburgundy}[1]{\textcolor{burgundy}{#1}}

\title{Gravity On-shell Diagrams}

\author{Enrico Herrmann,$^1$}
\author{Jaroslav Trnka$^{2}$}

\affiliation{$^1$ Walter Burke Institute for Theoretical Physics,\\
California Institute of Technology, Pasadena, CA 91125, USA
}
\affiliation{$^2$ Center for Quantum Mathematics and Physics (QMAP),\\ 
Department of Physics, University of California, Davis, CA 95616, USA}

\emailAdd{eherrmann@caltech.edu, trnka@ucdavis.edu}

\preprint{April 2016}
\abstract{We study on-shell diagrams for gravity theories with any number of supersymmetries and find a compact Grassmannian formula in terms of edge variables of the graphs. Unlike in gauge theory where the analogous form involves only $\dlog$-factors, in gravity there is a non-trivial numerator as well as higher degree poles in the edge variables. Based on the structure of the Grassmannian formula for $\N=8$ supergravity we conjecture that gravity loop amplitudes also possess similar properties. In particular, we find that there are only logarithmic singularities on cuts with finite loop momentum, poles at infinity are present and loop amplitudes show special behavior on certain collinear cuts. We demonstrate on 1-loop and 2-loop examples that the behavior on collinear cuts is a highly non-trivial property which requires cancellations between all terms contributing to the amplitude.}

\preprint{
\begin{flushright}CALT-TH-2016-006\end{flushright}
}

\begin{document}
\setlength{\unitlength}{1mm}
\maketitle

\begin{fmffile}{GravityOSdiags}
%
%
\section{Introduction}
%

Within the field of scattering amplitudes, a great number of developments in the last decade or so are based on 
powerful on-shell methods \cite{BCF,BCFW,Bern:1994zx,Bern:1994cg,MaximalCuts,Britto:2004nc,Cachazo:2008vp}. Amongst others, 
BCFW-recursion \cite{BCF,BCFW} and generalized unitarity \cite{Bern:1994zx,Bern:1994cg,MaximalCuts} 
allowed to push the boundary of computable amplitudes in terms of number of loops and legs. The core idea behind 
these methods is that on-shell amplitudes break up into products of simpler amplitudes on all factorization channels. In the 
traditional picture of Quantum Field Theory, locality and unitarity dictate the form and locations of all these residues. In particular, they arise in kinematic regions where either internal particles 
or sums of external particles become on-shell. Associated with these residues are vanishing propagators and in 
this context we talk about \emph{cuts} of the amplitude. Symbolically, one can write the two types of elementary 
cuts (singularities) as \cite{ArkaniHamed:2010kv},
\begin{equation}
\raisebox{-33pt}{
\includegraphics[trim={0cm 8.5cm 0cm 8.5cm},clip,scale=.5]{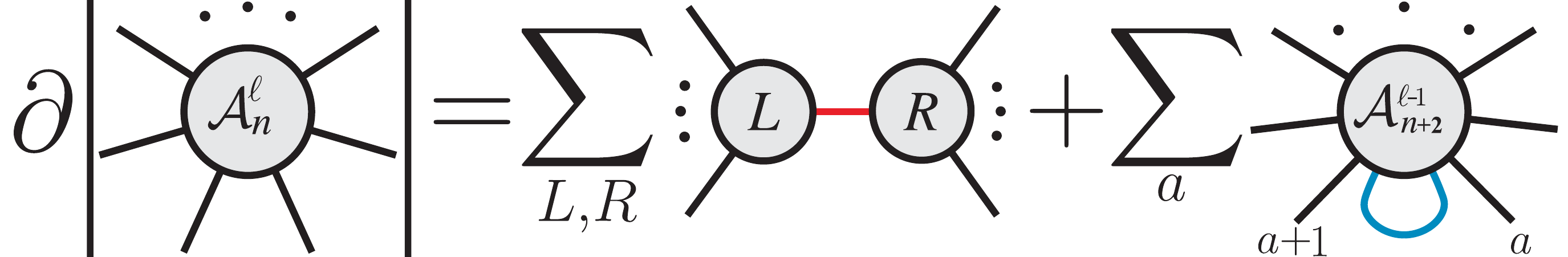}
}
\label{Sing}
\end{equation}

where the first term corresponds to the factorization into a product of lower point amplitudes (keeping the total 
loop degree fixed) while the second term is the forward limit of an $(L-1)$-loop, $(n+2)$-point amplitude. In general field theories this term suffers 
from IR-divergencies \cite{CaronHuot:2010zt} and therefore, in many cases the fundamental cut is the well-known 
\emph{unitarity cut} \cite{Cutkosky:1960sp,Mandelstam:1958xc}. Iterating 
these cuts one can calculate multi-dimensional residues by setting an increasing number of propagators to zero. 
This is known in the literature as {\it generalized unitarity} \cite{Bern:1994zx,Bern:1994cg,MaximalCuts}. 
\vskip -.4cm
$$
\includegraphics[trim={0cm 0cm 0cm 0cm},clip,scale=.2]{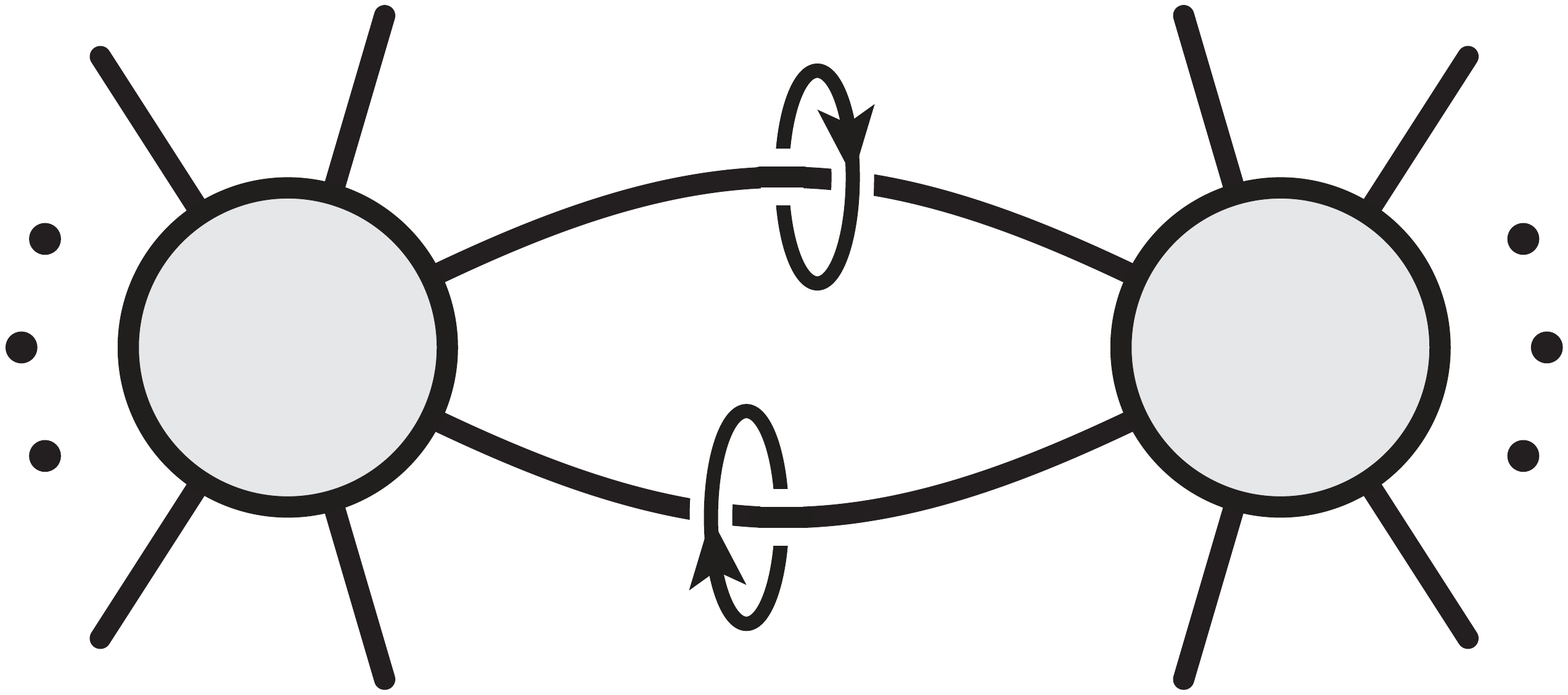}
\hspace{1cm}
\includegraphics[trim={0cm 0cm 0cm 0cm},clip,scale=.2]{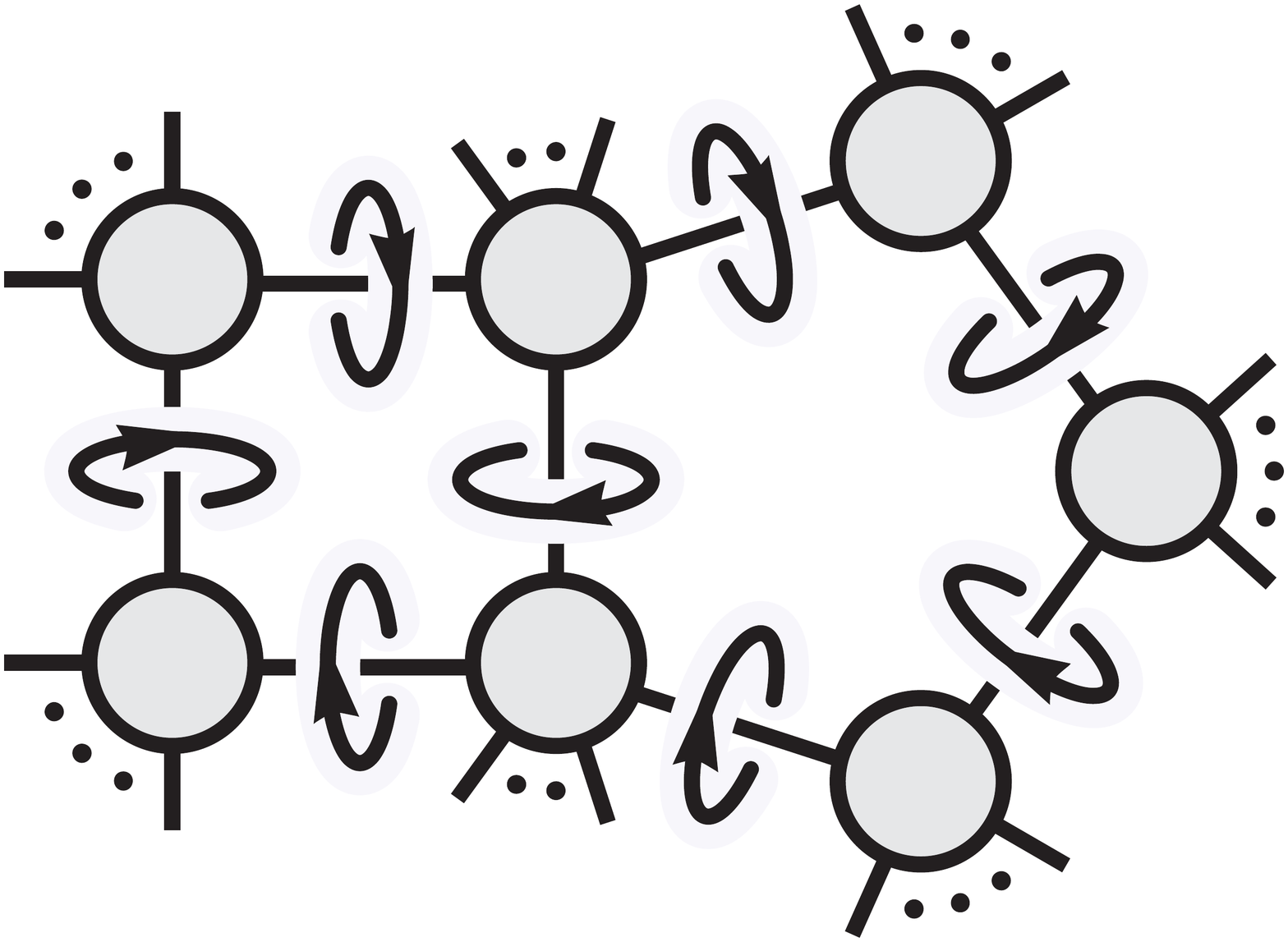}
$$
\vskip -.4cm

Generically, it is not possible to set to zero more than two propagators in a given loop while simultaneously also 
requiring real kinematics. Therefore, the loop momenta are complex when constrained by the set of 
on-shell conditions which implies that these singularities are outside the physical integration region. The main 
success of generalized unitarity then relies on the fact that the integrands are rational functions that can be 
analytically continued so that complex residues (given by a sufficient set of cuts) completely specify them.

A natural next step in this line of thought is to cut the maximum number of propagators which factorizes the amplitude 
into the simplest building blocks \cite{MaximalCuts}. The most elementary case occurs when all factors are three-point amplitudes. As we 
will describe in a moment, these are rather special due to the particular features of three-point kinematics. In this 
scenario we talk about {\it on-shell diagrams} \cite{OnshellDiagrams}. 

\subsection{On-shell diagrams}

For massless particles, the three-point amplitudes are completely fixed by Poincare symmetry to all loop orders 
in perturbation theory up to an overall constant \cite{Benincasa:2007xk}. This statement holds in any Quantum Field Theory with massless 
states and just follows from the fact that there are no kinematic invariants one can build out of three on-shell momenta. 
For real external kinematics, the on-shell conditions, $p_1^2=p_2^2=p_3^2=0$ and momentum conservation $p_1+p_2+p_3=0$ 
would force all three point amplitudes to vanish. However, for complex kinematics in $D=4$ we have two distinct solutions \cite{Witten:2003nn}
which can be conveniently written using spinor-helicity \cite{Xu:1986xb} variables $p^\mu = \sigma^\mu_{a\dot{a}} \lambda_a\widetilde{\lambda}_{\dot{a}}$. 

\begin{center}
{I.)} $\lamt{1}\sim \lamt{2}\sim \lamt{3}\ (\text{MHV})\,,$ 
\qquad
{II.)} $\lam{1}\sim\lam{2}\sim\lam{3} \ (\MHVbar)\,.$
\end{center}

Any three-point amplitude is then either of type I.) or II.). In particular, for the gluon-amplitudes in Yang-Mills 
theory we have two elementary amplitudes with MHV $(+--)$ or $\MHVbar$ $(-++)$ helicity configuration (ignoring 
higher dimensional operators that could lead to $(\pm\pm\pm)$ amplitudes, see e.g.~\cite{Broedel:2012rc}). In the maximally supersymmetric 
case of $\NeqFour$ theory these gluonic amplitudes are embedded in the MHV, resp. $\MHVbar$ superamplitudes (see e.g.~\cite{Drummond:2008vq}) 
which we denote by blobs with different colors,
\begin{equation}
	\raisebox{-38pt}{\includegraphics[scale=0.9]{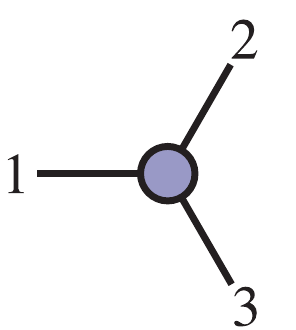}}
	\hskip -.5cm
	=\frac{\delta^4(P)\delta^8(\Q)}{\ab{12}\ab{23}\ab{31}}\,,
	\qquad\qquad
	\raisebox{-38pt}{\includegraphics[scale=0.9]{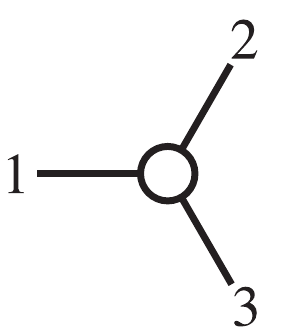}} 
	\hskip -.5cm 
	= \frac{\delta^4(P)\delta^4(\tw{\Q})}{[12][23][31]}\,,
\label{ThreeP}
\end{equation}

where $\ab{ij} = \epsilon_{\alpha\beta} \lam{i}^{\alpha}\lam{j}^{\beta}$ and $[ij] = \epsilon_{\dot{\alpha}\dot{\beta}}\lamt{i}^{\dot{\alpha}}\lamt{j}^{\dot{\beta}}$. 
Using the anti-commuting $\twEta^I,\ I=1,...,4$ variables to write the on-shell multiplet as \cite{Nair:1988bq},
$$
	\Phi(\twEta) = g^+ 
							 + \twEta^I \ \tw{g}_I
							 + \frac{1}{2!} \twEta^I\twEta^J\ \phi_{IJ}
							 + \frac{1}{3!} \epsilon_{IJKL} \twEta^I\twEta^J \twEta^K\ \tw{g}^{L}
							 + \frac{1}{4!} \epsilon_{IJKL} \twEta^I\twEta^J \twEta^K \twEta^L \ g^-
$$

the arguments of the respective delta-functions in (\ref{ThreeP}) are given by (neglecting all spinor- and $SU(4)\ R$-symmetry indices),
$$
P \equiv \lam{}\cdot\lamt{}   = \lam{1}\lamt{1} + \lam{2}\lamt{2} + \lam{3}\lamt{3},
\quad 
\Q \equiv \lam{}\cdot \tw{\eta}= \lam{1}\tw{\eta}_1 + \lam{2}\tw{\eta}_2 + \lam{3}\tw{\eta}_3, 
\quad 
\widetilde{\Q} = [12]\widetilde{\eta}_3 + [23]\widetilde{\eta}_1 + [31]\widetilde{\eta}_2\,.
$$

Here and in the following we denote $\lam{}\cdot\lamt{} \equiv \sum^n_{a=1} \lam{a}\lamt{a},\ \lam{}\cdot\twEta \equiv \sum^n_{a=1} \lam{a} \twEta_a$ as the sum over all external particles.

\medskip
Having completed the discussion of three-particle amplitudes, we are now in the position to introduce on-shell 
diagrams. For us, an on-shell diagram is any graph formed from the two types of three-point amplitudes 
(\ref{ThreeP}) connected by edges,
$$
\raisebox{4pt}{
\includegraphics[trim={0cm .1cm 0cm 0cm},clip,scale=.5]{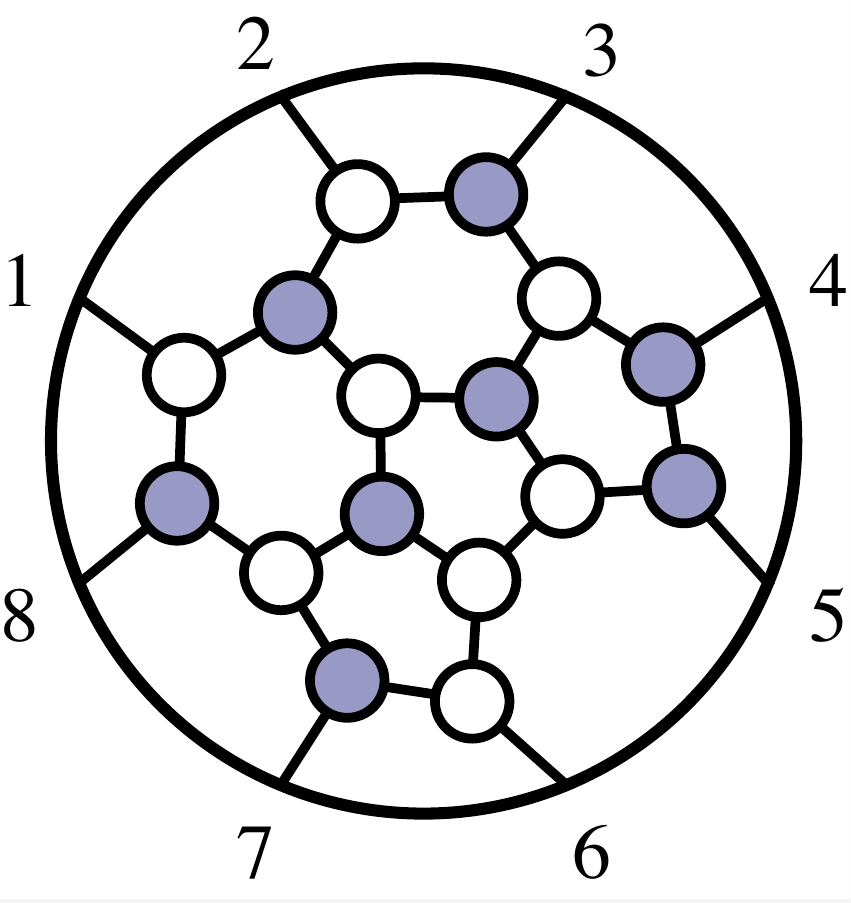}
}
\hspace{2cm}
\includegraphics[trim={0cm .1cm 0cm 0cm},clip,scale=.55]{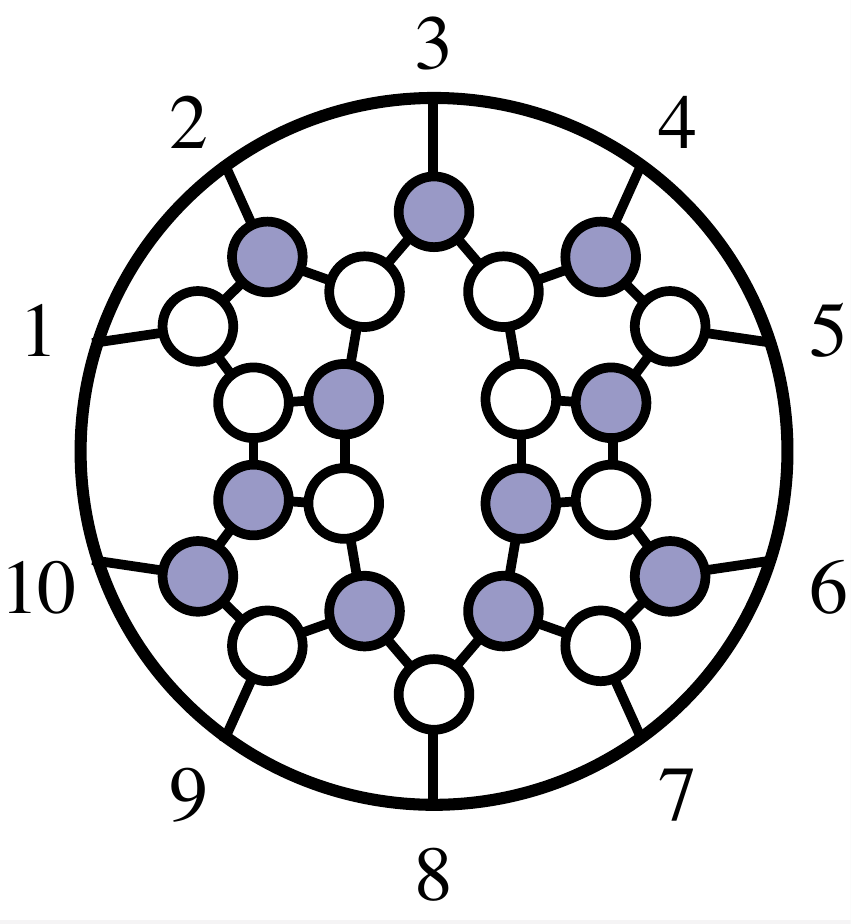}
$$

that all represent on-shell particles (both internal and external). In this section we review properties of 
on-shell diagrams in planar $\NeqFour$ and introduce all concepts relevant for our gravity discussion later. 
Further details can be found directly in \cite{OnshellDiagrams} and the review article \cite{Elvang:2013cua}. 
With this definition, the value of the diagram is given by the product of three-point amplitudes satisfying 
the on-shell conditions for all edges. In practice, the delta functions of the elementary three-point amplitudes 
can be used for solving for $\lambda_I$, $\widetilde{\lambda}_I$ and $\widetilde{\eta}_I$ of the internal particle and 
writing the overall result (including delta functions), using external data only. In this case we talk about {\it leading 
singularities} \cite{Cachazo:2008vp}. If the number of on-shell conditions exceeds the number of internal degrees of freedom, we get 
additional constraints on the external kinematics, while in the opposite case the on-shell diagram depends on some unfixed 
parameters. These cases are easily classified by a parameter $n_\delta$ counting the number of constraints on external 
kinematics $n_{\delta} = 0$, $n_\delta>0$ and $n_\delta<0$.

The simplest example of a reduced on-shell diagram ($n_\delta=0$) actually represents the color-ordered four-point tree-level amplitude 
which consists of four vertices. The simpler looking on-shell diagram with only two vertices is the residue of the amplitude 
on the $t$-channel factorization pole and imposes a constraint $(n_\delta=1)$ on the external momenta.
\begin{equation}
	\raisebox{-49pt}{\includegraphics[trim={0cm .1cm 0cm 0cm},clip,scale=.42]{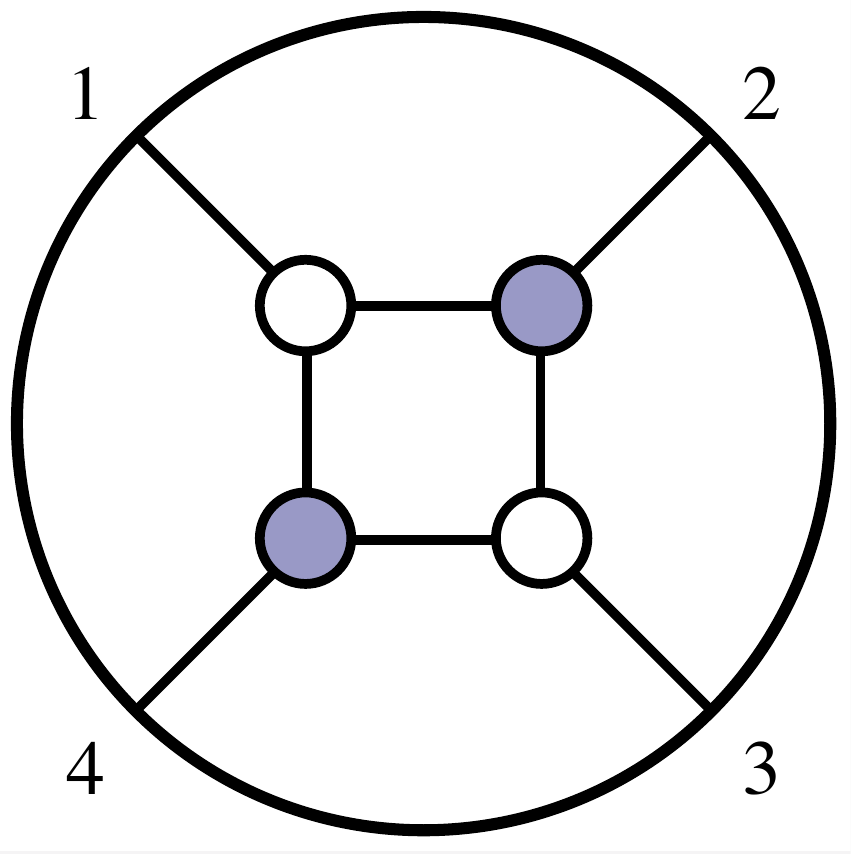}}
	\hskip .1cm 
		= \frac{\delta^4(\lam{}\cdot\lamt{})\delta^8(\lam{}\cdot\tw{\eta})}
					 {\ab{12}\ab{23}\ab{34}\ab{41}}
		\label{One4}
	\hskip 0.3cm 
	\raisebox{-49pt}{\includegraphics[trim={0cm .1cm 0cm 0cm},clip,scale=.42]{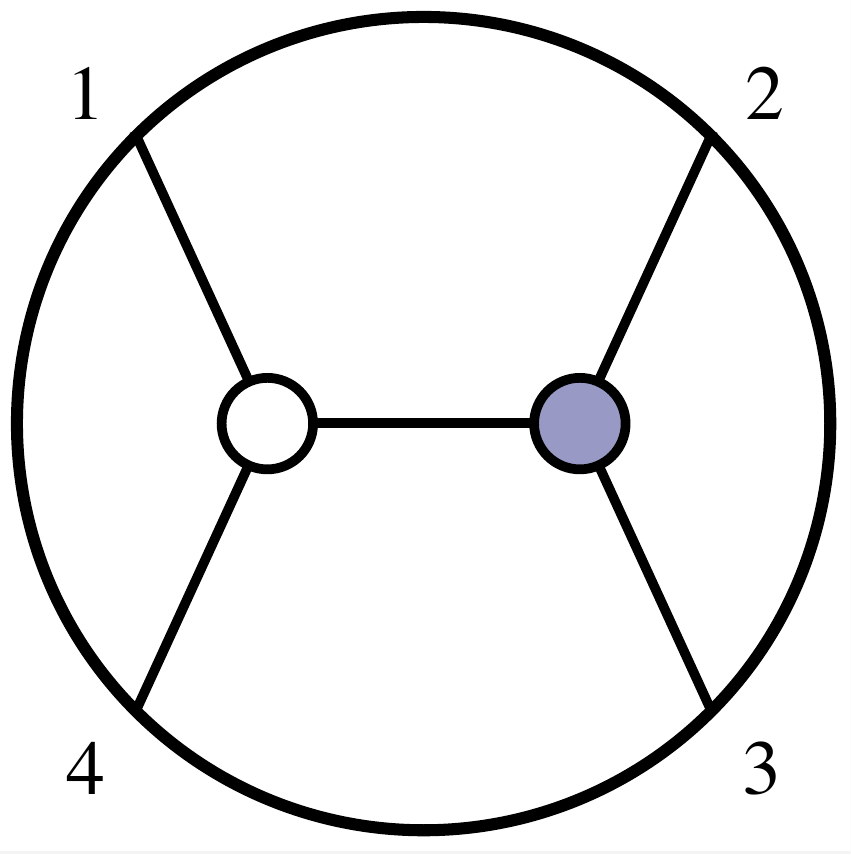}}
	\hskip .1cm 
= \frac{\delta^4(\lam{}\cdot\lamt{})\delta^8(\lam{}\cdot\widetilde{\eta})\ \delta(\ab{14})}{\ab{12}\ab{23}\ab{34}} 
\end{equation}

As an example for the third possibility $(n_\delta<0)$, we can draw a diagram which depends on one unfixed parameter $z$. 
\begin{equation}
\raisebox{-48pt}{
	\includegraphics[trim={0cm .1cm 0cm 0cm},clip,scale=.43]{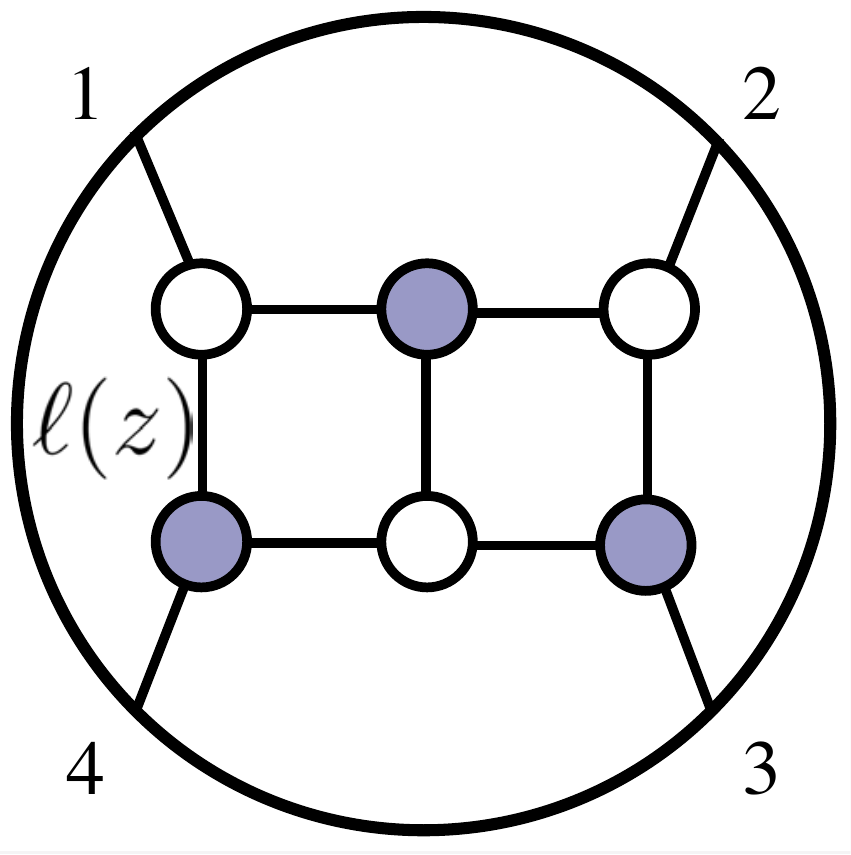}
	}
	\hskip .1cm 
	= \frac{\delta^4(\lam{}\cdot\lamt{})\delta^8(\lam{}\cdot\twEta)}{z \ab{12}\ab{23}(\ab{34}+z\ab{31})\ab{41}}
\label{2loop4ptOSplanar}
\end{equation}

The counting is easy to understand if we think about the diagram as a hepta-cut of the two loop amplitude: there are 
seven on-shell conditions imposed on two off-shell loop momenta, leaving one degree of freedom unfixed. In the 
diagram $z$ parametrizes the momentum flow along the edge between external legs $1$ and $4$, $\ell(z)=z\lambda_1\widetilde{\lambda}_4$ 
but also other internal legs will depend on $z$. In the standard approach of generalized unitarity, this diagram 
represents a {\it maximal cut} as there are no further propagators available to cut and localize the remaining degree of 
freedom. However, the amplitude does have further residues at $z=0$ and $z=\frac{\ab{34}}{\ab{13}}$. Each residue corresponds to erasing an edge of (\ref{2loop4ptOSplanar}) 
giving the one-loop on-shell diagram on the left of (\ref{One4}). This is a leading singularity of the amplitude -- all degrees of 
freedom in loop momenta are fixed by on-shell conditions. 

It turns out that not all on-shell diagrams are independent, but rather there are equivalence classes of diagrams  related by certain identity moves. 
The first is the \emph{merge and expand} move represented in (\ref{mergeExpandPlanar}). The black vertices enforce all $\lamt{}$'s to be proportional which is 
independent of the way the individual three-point amplitudes are connected,
\begin{equation}
\raisebox{-46pt}{
 \includegraphics[trim={0cm .1cm 0cm 0cm},clip,scale=.4]{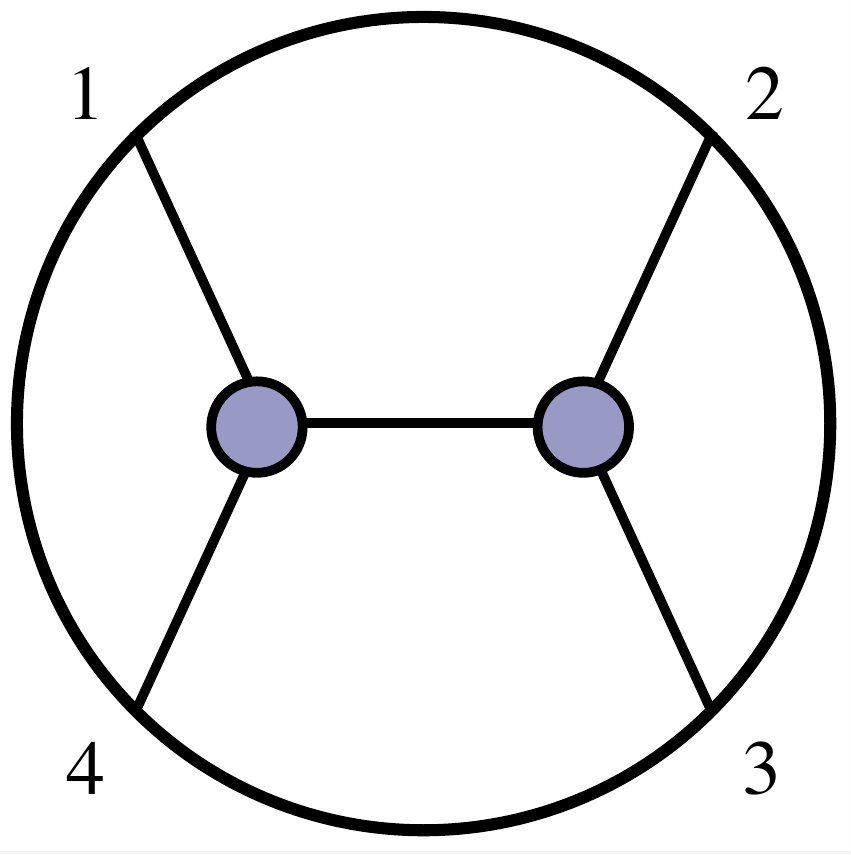}}
 \Leftrightarrow
 \raisebox{-46pt}{
 \includegraphics[trim={0cm .1cm 0cm 0cm},clip,scale=.4]{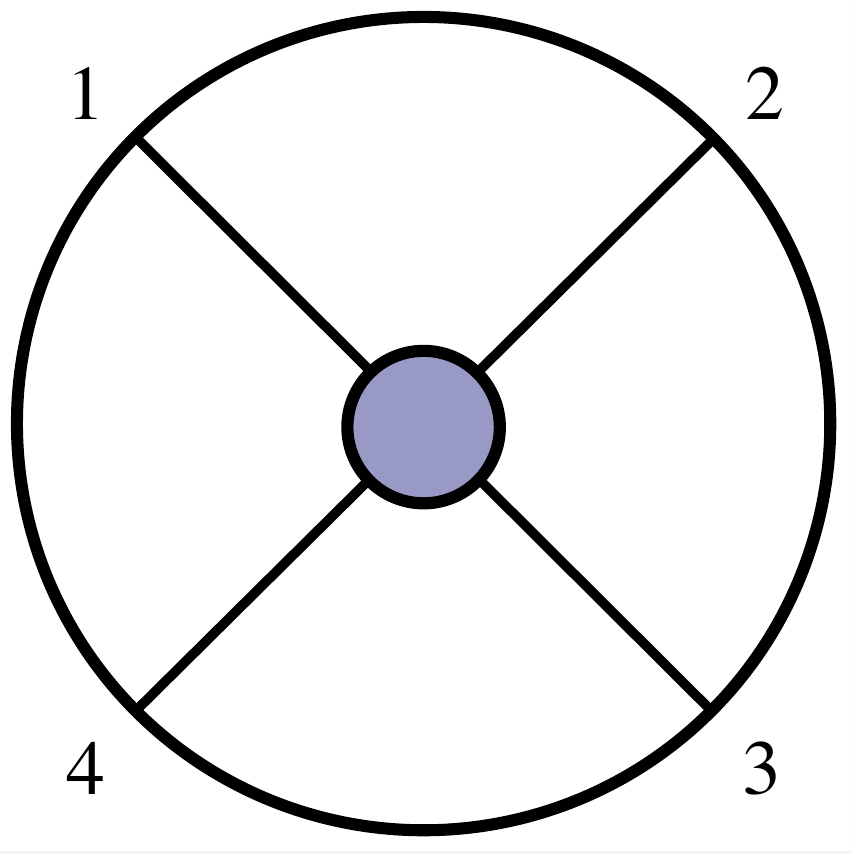}}
 \Leftrightarrow
 \raisebox{-46pt}{
 \includegraphics[trim={0cm .1cm 0cm 0cm},clip,scale=.4]{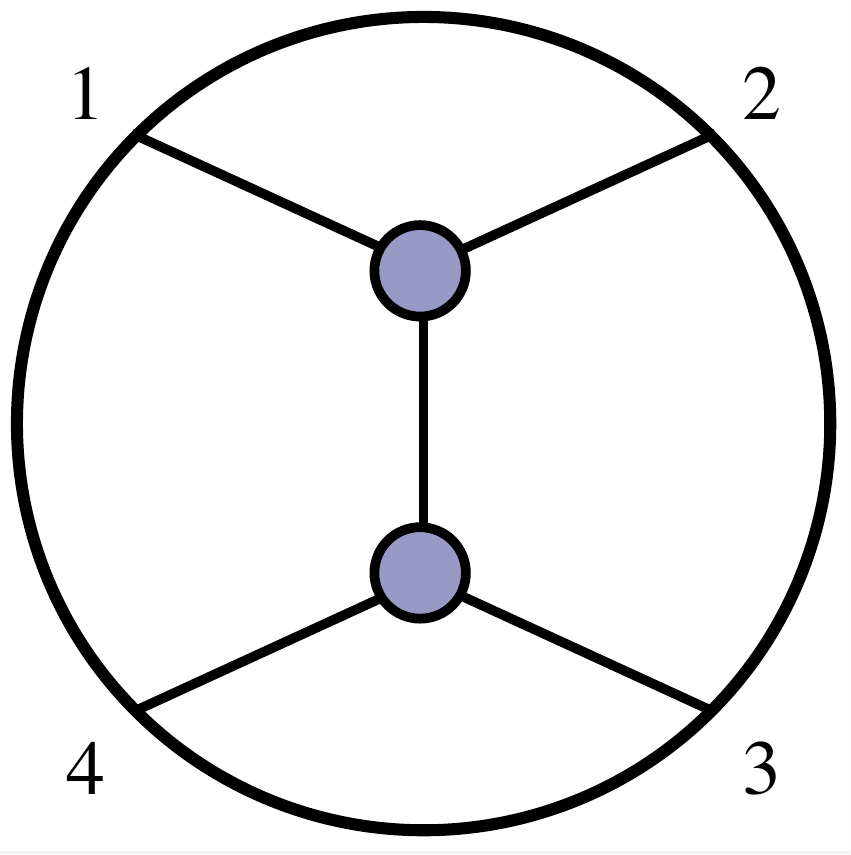}}
 \label{mergeExpandPlanar}
\end{equation}

One further nontrivial move is the \emph{square move} \cite{Hodges:2005aj} which can be motivated by the cyclic invariance of the four-particle tree level amplitude,
\begin{equation}
\raisebox{-46pt}{
 \includegraphics[trim={0cm .1cm 0cm 0cm},clip,scale=.4]{./figures/g24_graph_1}}
 \Leftrightarrow
 \raisebox{-46pt}{
 \includegraphics[trim={0cm .1cm 0cm 0cm},clip,scale=.4]{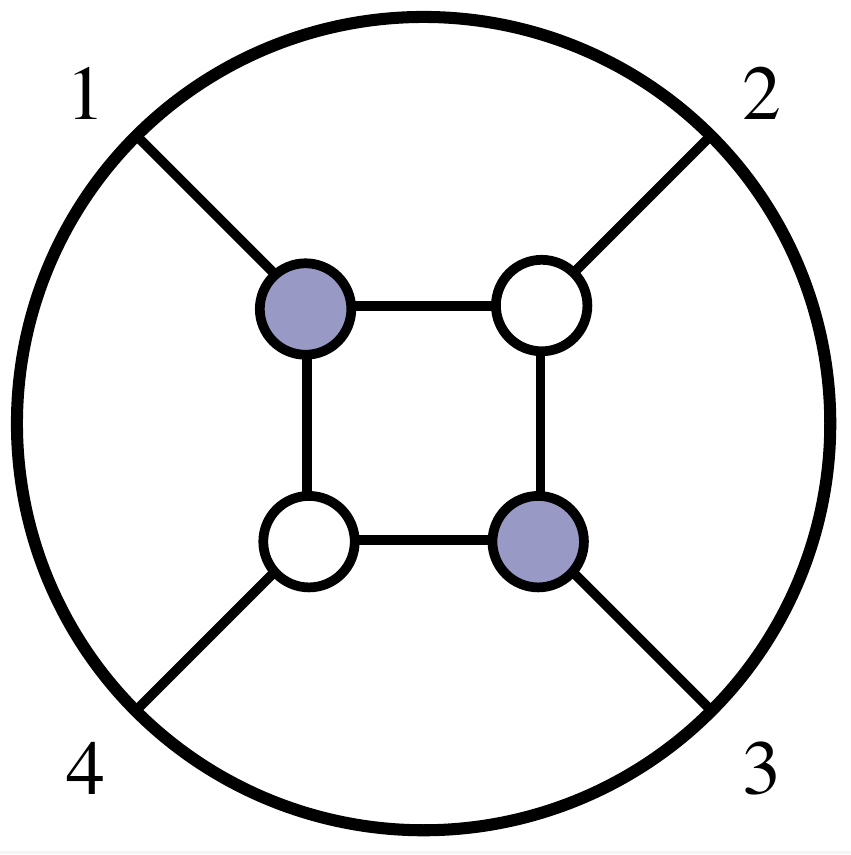}}
 \label{squareMove}
\end{equation}

Together with \emph{bubble deletion} which does not play a role in our discussion here, these are all the equivalence
moves for planar $\NeqFour$. Modulo the aforementioned moves, it is possible to give a complete classification 
of on-shell diagrams \cite{OnshellDiagrams} in this theory.

Besides representing cuts of loop amplitudes, on-shell diagrams serve directly as building blocks for constructing 
amplitudes. In particular, the BCFW recursion relation for tree-level amplitudes and loop integrands in planar $\NeqFour$ 
theory is represented by \cite{ArkaniHamed:2010kv,OnshellDiagrams} 
$$
	\includegraphics[scale=0.96]{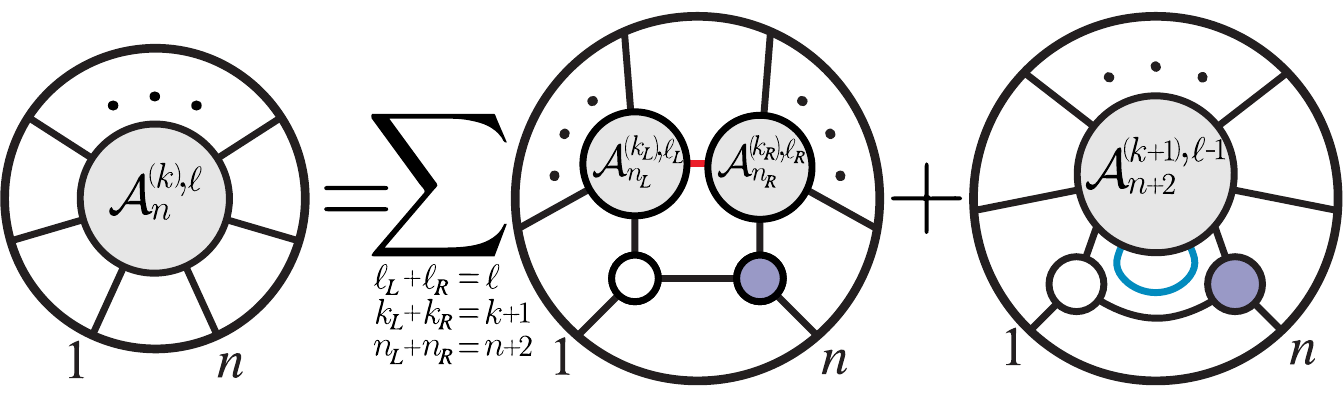}
	\label{BCFWinOSdiags}
$$

This equation is a solution to the question what on-shell function has singularities given by (\ref{Sing}). 
Here planarity was crucial because it permitted a unique definition of the {\it integrand} as a rational function with 
certain properties. It is this integrand which can be constructed recursively. The key point was the existence of global 
variables (dual variables and momentum twistors \cite{Hodges}) common to all terms in the expansion. Currently, it is the lack of global 
labels that hampers the extension of the recursion relations beyond the planar limit.

While the recursion relations are only formulated in planar $\NeqFour$ so far, the on-shell diagrams are well defined gauge 
invariant objects in any Quantum Field Theory, planar or non-planar, with or without supersymmetry. They are defined 
as products of on-shell three-point amplitudes (for theories with fundamental three point amplitudes) and at the least
represent cuts of loop amplitudes. From that point of view they encode an important amount of information about amplitudes 
in any theory and their properties are well worth studying in its own right.

%
\subsection{Grassmannian formulation} 
\label{sec:GrassmannianYM}
%

Besides viewing on-shell diagrams as gluing of three-point amplitudes integrated over the on-shell phase space (including the 
sum over all physical states that can cross the cut) there is a completely different way how to calculate on-shell 
diagrams. This \emph{dual formulation} expresses on-shell diagrams as differential forms on the (positive) Grassmannian \cite{OnshellDiagrams}. 
There are a number of ways how to motivate this picture starting from classifying configurations of points with linear dependencies to representing the permutation group in terms of 
planar bi-colored graphs \cite{postnikov}. 
Physically, the most direct way to discover the Grassmannian picture for on-shell diagrams is to think about momentum 
conservation more seriously. Starting from the innocuous equation,
\begin{equation}
\delta^4(P) \equiv\delta^4(\lam{}\cdot\lamt{})=\delta^4(\lam{1}\lamt{1} + \lam{2}\lamt{2} + \dots + \lam{n}\lamt{n})\,,
\end{equation}

one notes that this is a quadratic condition on the spinor-helicity variables. Naturally, one can ask if 
there is a way to trivialize the quadratic constraints and rewrite them as sets of linear relations between $\lam{}$s 
and $\lamt{}$s separately. The solution to this problem is to introduce an auxiliary $k$-plane in $n$-dimensions represented 
by a $(k\times n)$-matrix $C$ modulo a $GL(k)$ redundancy arising from row operations that leave the $k$-plane invariant. 
This space is known as the Grassmannian $G(k,n)$. 
Using these auxiliary variables, momentum conservation is then enforced geometrically \cite{ArkaniHamed:2009dn,ArkaniHamed:2009vw,Mason:2009qx} via the following set of delta functions (similar relations hold in twistor and momentum twistor spaces),
\begin{equation}
\delta^{(k\times 2)} (C_{\alpha a}\widetilde{\lambda}_a)\,\delta^{((n-k)\times 2)}(C^\perp_{\beta a}\lambda_a)\,,
\label{DeltaLin}
\end{equation} 

where $C^\perp$ denotes the $\big((n-k)\times n\big)$-matrix orthogonal to $C$, $C\cdot C^\perp = 0$. 
There are $2n$ delta functions in total, four of them give the overall momentum conservation while the remaining $2n-4$ constrain the parameters 
of the $C$-matrix. 

The study of Grassmannians is a vast and active topic in the mathematics community ranging, amongst others, from 
combinatorics to algebraic geometry \cite{Lusztig,postnikov,postnikov2,lauren,goncharov,knutson}. There is a close connection to on-shell diagrams which 
was simultaneously discovered both by physicists in the context of scattering amplitudes and by mathematicians (in the math 
literature these diagrams are called {\it plabic graphs}) in searching for positive parameterizations of Grassmannians. 
In particular, each on-shell diagram gives a parametrization for the $C$-matrix using a set of variables $\alpha_j$. When 
these variables are real with definite signs, the matrix $C$ has all main minors positive and then we talk about positive 
Grassmannian $G_+(k,n)$. These variables are associated with either faces or edges of the diagram. The face variables are 
more invariant but they can be used only in planar diagrams. Since in this paper we will include  
non-planar examples we use {\it edge variables} instead to parametrize the Grassmannian matrix. 

Like in the physical picture where on-shell diagrams are products of three-point amplitudes we also start our discussion with 
elementary three point vertices. We first choose a {\it perfect orientation} in which we attach arrows to all legs. 
For all black vertices two of the arrows are incoming and one outgoing while for white vertices one is incoming and 
two outgoing. Then we associate a $(2\times 3)$-matrix with the black (MHV, $k=2$) vertex and a $(1\times 3)$-matrix 
with the white ($\MHVbar,\ k=1$) vertex in the following way,

\begin{equation}
\begin{array}{ccc}
	\raisebox{-54pt}{\includegraphics[trim={0cm .1cm 0cm 0cm},clip,scale=.45]{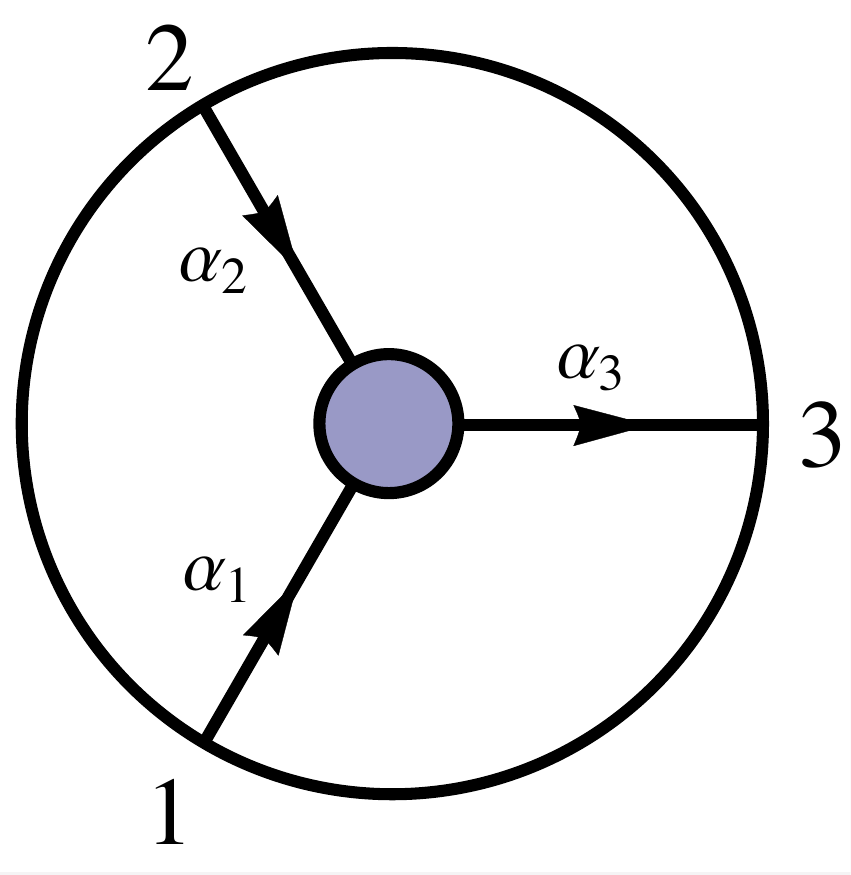}} 
	&
	\twhite{.\hspace{2cm}.}
	&
	\raisebox{-54pt}{\includegraphics[trim={0cm .1cm 0cm 0cm},clip,scale=.45]{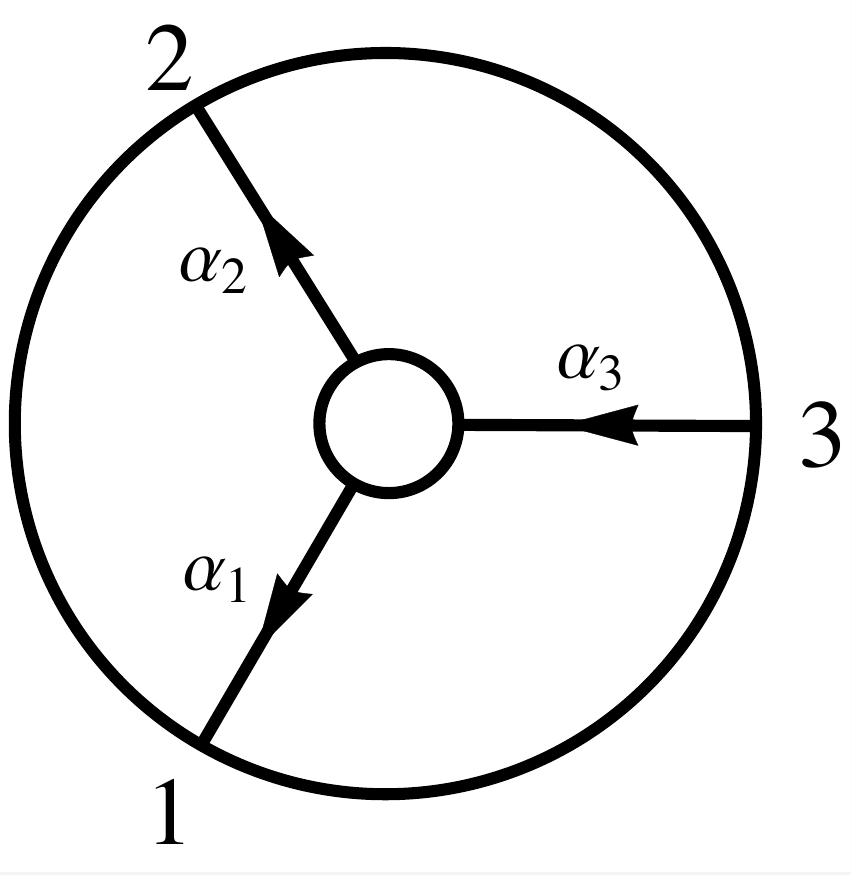}} 
	\\
	 \Updownarrow &\twhite{.}\hskip 1cm	& \Updownarrow 
	\\
	C = \begin{pmatrix}
				1 & 0 & \alpha_1 \alpha_3 \\
				0 & 1 & \alpha_2 \alpha_3
			\end{pmatrix}\,
	&
	\twhite{.}\hskip 1cm
	&
	C = \begin{pmatrix}
				\alpha_1 \alpha_3 & \alpha_2\alpha_3 & 1 
			\end{pmatrix}\,.\\
\end{array}
\end{equation}

Choosing a perfect orientation fixes a part of the general $GL(k)$-redundancy of the $C$-matrix that represents a point in 
the Grassmannian. With the remaining $GL(1)_v$-freedom we are allowed to fix any one of the 
variables associated to the half-edges of that vertex to some arbitrary value. The canonical choice would be $\alpha_3=1$, but any other 
finite, nonzero value is allowed as well. For the moment though, let us keep this freedom unfixed. 

In the next step we glue the atomic three-point vertices together into an arbitrary planar on-shell diagram to which we 
associate some bigger $(k\times n)$-matrix $C$. In the gluing process, we identify two half-edges of the vertices involved 
in the fusion to form an internal edge of the bigger on-shell diagram. Each internal edge of this big diagram is then 
parametrized by two variables $\alpha^{(1)}$ and $\alpha^{(2)}$ coming from the two vertices involved in the gluing process 
so that the $C$-matrix will only depend on their product $\alpha=\alpha^{(1)}\alpha^{(2)}$. Pictorially, this process is simple to state (the grey blob denotes the rest of the diagram),
\begin{equation}
\raisebox{-52pt}{
 \includegraphics[trim={0cm .1cm 0cm 0cm},clip,scale=.45]{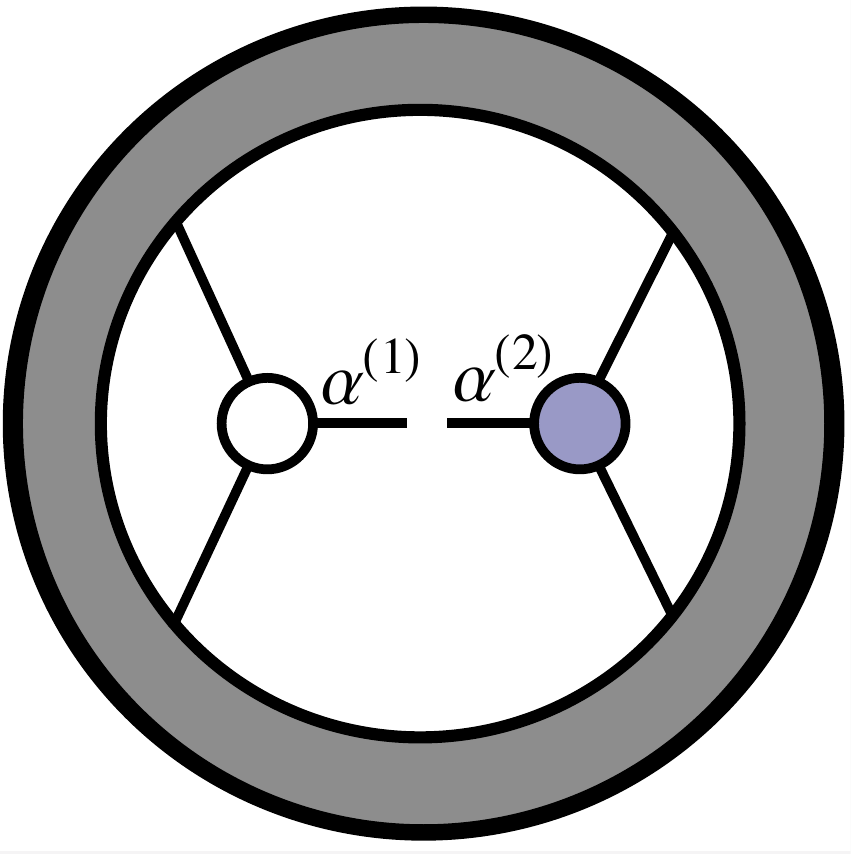}}
 \quad
 \to
 \quad
 \raisebox{-52pt}{
 \includegraphics[trim={0cm .1cm 0cm 0cm},clip,scale=.45]{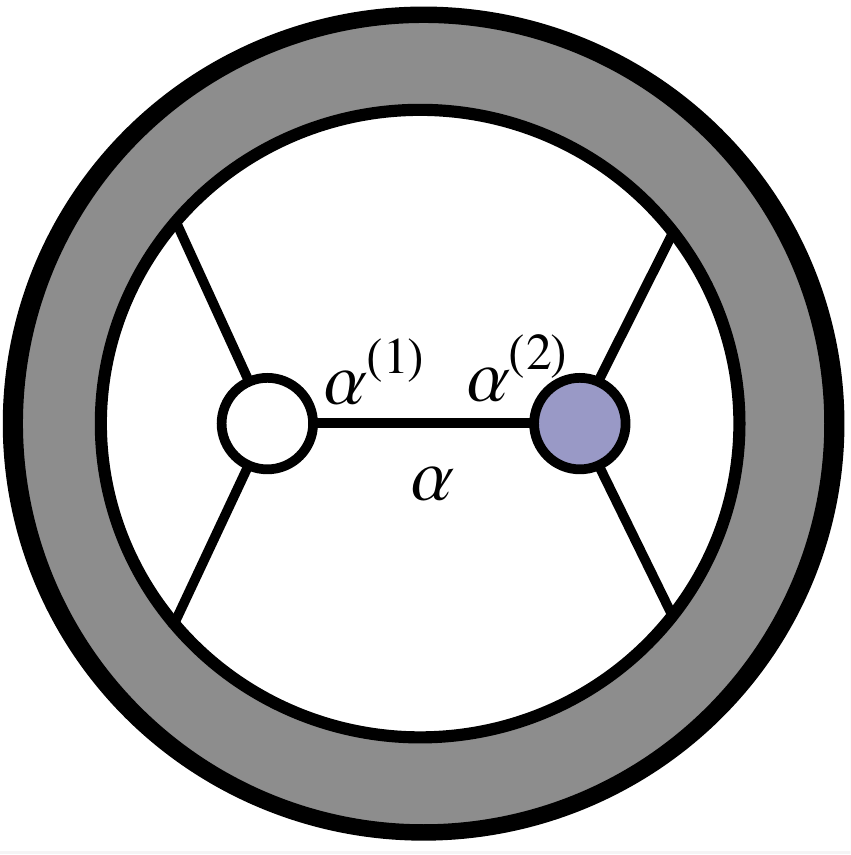}}
\end{equation}

and illustrates that it is natural to directly use edge-variables $\alpha_k$ rather than the individual vertex variables introduced by the little Grassmannians. 
The identification is as follows; in the gluing process we encounter another $GL(1)_e$ redundancy stemming from the fact 
that the internal momentum of that edge is invariant under little group rescaling $\lam{I}\to t_I \lam{I},\ \lamt{I}\to t^{-1}_I \lamt{I}$ 
which allows us to combine two of the vertex-variables into a single edge-variable. Doing this for all internal edges, we are 
only left with the $GL(1)_v$ redundancies for each vertex in the big on-shell diagram which we can use to set certain edge weights to one for instance. 

\begin{equation}
\begin{array}{ccc}
\raisebox{.2cm}{
\includegraphics[trim={0cm .1cm 0cm 0cm},clip,scale=.45]{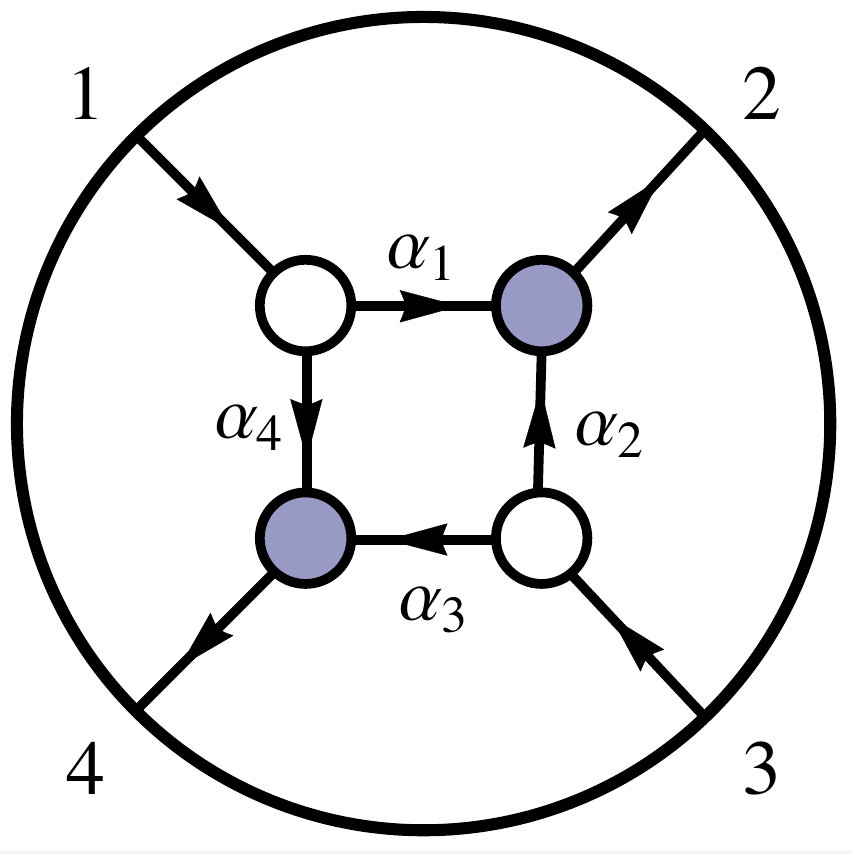}
}
&
\twhite{.\hspace{2cm}.}
&
\includegraphics[trim={0cm .1cm 0cm 0cm},clip,scale=.48]{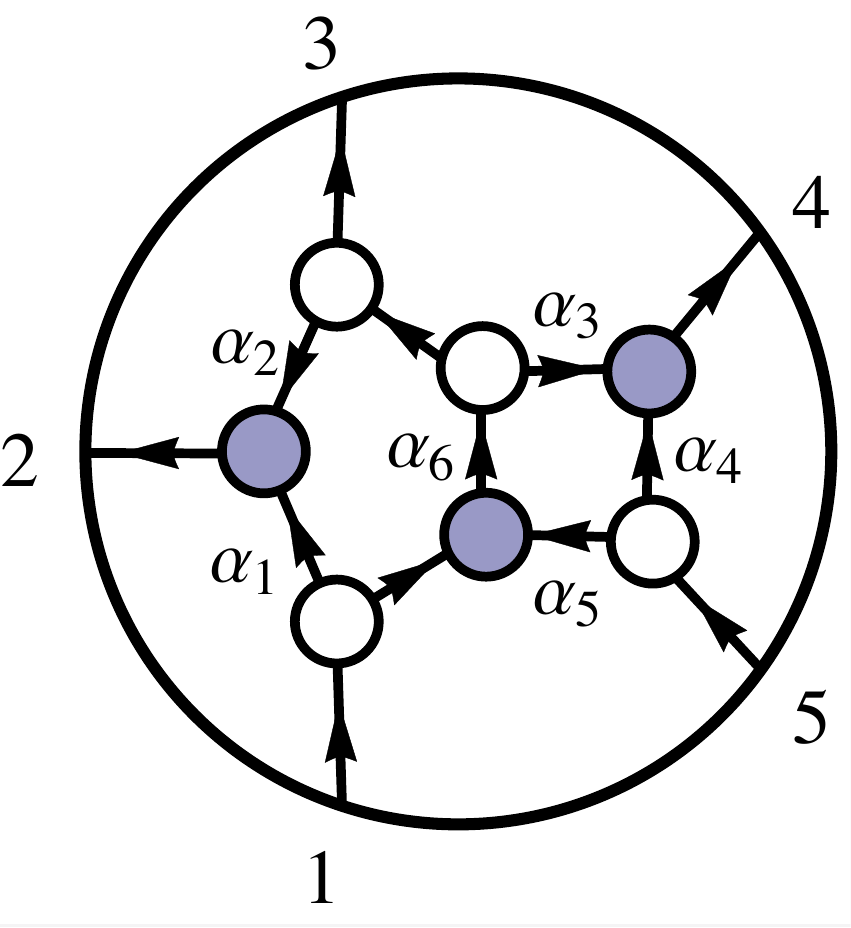}
\\
(a) & \twhite{.\hspace{2cm}.} & \twhite{.}\hspace{.2cm}(b) \\
\end{array}
\label{OSDiagExamples}
\end{equation}

In terms of edge-variables, the rule how to obtain the $C$-matrix from the graph is quite simple. First, we have to choose a 
perfect orientation for the diagram by consistently decorating all edges with arrows. The external legs with incoming arrows 
are called {\it sources}, while the external legs with outgoing arrows are called {\it sinks}. For the diagram with $k$ sources 
and $n-k$ sinks we construct a $(k\times n)$ matrix $C$. Note that these numbers do not depend on the way we choose a perfect 
orientation but are an invariant property of the on-shell diagram itself. Each row of the matrix is associated with one source 
while the columns are linked to both sources and sinks. Now each entry $C_{\alpha a}$ is calculated as 
\begin{equation}
C_{\alpha a} = \sum_{\Gamma_{\alpha \rightarrow a}} \prod_j \alpha_j\,,
\label{Entry}
\end{equation}

where we sum over all directed paths $\Gamma_{\alpha \rightarrow a}$ from the source $\alpha$ to the sink $a$ by following the arrows. Along the way we take 
the product of all edge variables. If the label $a =\alpha $ is the same source we fix the matrix entry to $1$ if $a=\alpha'$ 
is a different source the matrix entry is $0$. For the examples above the $C$-matrices are,
\begin{align}
	C^{(a)}  = \begin{pmatrix}
							1 & \alpha_1 & 0 & \alpha_4 \\
							0 & \alpha_2 & 1 & \alpha_3
						\end{pmatrix}  \,, \quad
	C^{(b)}  = \begin{pmatrix}
							1 & \alpha_1 + \alpha_2\alpha_6& \alpha_6 & \alpha_3\alpha_6 & 0 \\
							0 & \alpha_5\alpha_6\alpha_2   & \alpha_5\alpha_6 & \alpha_4+\alpha_3\alpha_5\alpha_6 & 1
						\end{pmatrix}\,.
 \label{ExampleCMatrices}
\end{align}

Different choices for the sources and sinks corresponds to different gauge fixings of the $C$-matrix that are related by 
$GL(k)$-transformations. For some gauge choices, the perfect orientation can involve closed loops. In these cases there 
are infinitely many paths from $\alpha$ to $a$ and we have to sum over all of them,
\begin{equation}
\raisebox{-52pt}{
\includegraphics[trim={0cm .1cm 0cm .0cm},clip,scale=.45]{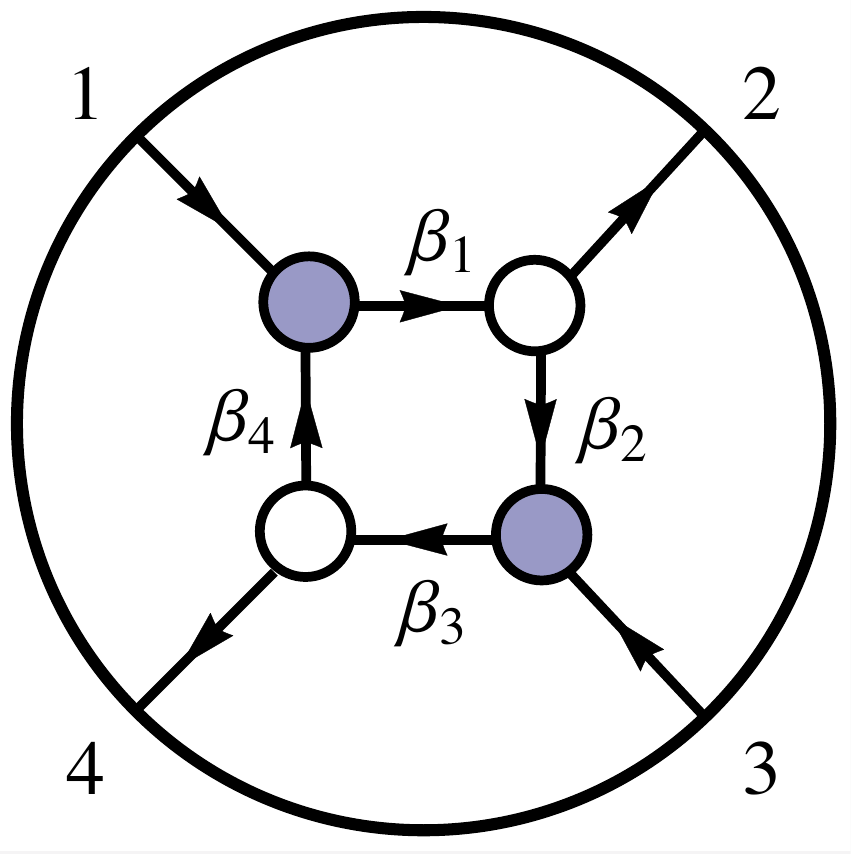}}
\Leftrightarrow
C = \begin{pmatrix}
		 1 & \beta_1 \delta & 0 & \beta_1\beta_2\beta_3 \delta \\
		 0 & \beta_3\beta_4\beta_1\delta & 1 & \beta_3 \delta
		\end{pmatrix}\,,
\label{BoxClosedLoopCMatrix}
\end{equation}

where $\delta$ is given by a geometric series,
\begin{equation}
\delta = \sum_{\sigma=0}^\infty (\beta_1\beta_2\beta_3\beta_4)^\sigma = \frac{1}{1-\beta_1\beta_2\beta_3\beta_4}\,.
\end{equation}

The important connection between the Grassmannian formulation and physics is that the same on-shell diagram that labels 
the $C$-matrix also represents a cut of a scattering amplitude in planar $\NeqFour$. The nontrivial relation is that the value 
of the on-shell diagram as calculated by multiplying three-point amplitudes is equal to the following differential form
\begin{equation}
d\Omega = \frac{d\alpha_1}{\alpha_1}\frac{d\alpha_2}{\alpha_2}\dots \frac{d\alpha_m}{\alpha_m}\,\delta(C\cdot Z)\,.
\label{Form}
\end{equation}

All the dependence on external kinematics is pushed into the delta functions,
\begin{equation}
\delta(C\cdot Z)
\equiv \delta^{(k\times 2)} (C_{ab}\widetilde{\lambda}_b)\delta^{((n-k)\times 2)}(C^\perp_{cb}\lambda_b)\,\,\delta^{(k\times \N)} (C_{ab}\widetilde{\eta}_b)
\label{DeltaExt}
\end{equation}

which linearize both momentum and super-momentum conservation $\delta^4(P)\ \delta^8(\Q)$ using the auxiliary 
Grassmannian $C$-matrix associated with the diagram. Depending on the details of the given diagram, the delta functions (\ref{DeltaExt}) 
allow us to fix a certain number of edge variables $\alpha_j$. In the case of on-shell diagrams relevant for tree-level 
amplitudes (leading singularities), all variables are fixed, while the on-shell diagrams appearing in the loop recursion 
relations have $4L$ unfixed parameters $\alpha_j$ which are related to the $4L$ degrees of freedom of $L$ off-shell loop 
momenta $\ell_i$.

So far, the $\big((n-k)\times n\big)$-matrix $C^{\perp}$ orthogonal to $C$, defined via $C\cdot C^{\perp}=0$, has played 
no significant role in our discussion but is crucial for momentum conservation in (\ref{DeltaLin}) and (\ref{DeltaExt}). 
Given a gauge fixed $C$-matrix, there is a simple rule how to obtain $C^{\perp}$. One takes the $(n-k)$ columns of the 
$C$-matrix that correspond to the $(n-k)$ sinks of the on-shell diagram. For each such column of $C$, one forms a row 
of $C^{\perp}$ by writing the negative entries of the column into the slots that correspond to the sources. The remaining 
$\big((n-k)\times(n-k)\big)$ matrix entries of $C^{\perp}$ are then filled by a $\big((n-k)\times(n-k)\big)$ identity-matrix. 
Let us apply the construction procedure to some concrete examples. For the $C$-matrices in (\ref{ExampleCMatrices}) corresponding to the on-shell 
diagrams (\ref{OSDiagExamples}), we get the following $C^{\perp}$-matrices,
\begin{align}
	C^{\perp}_{(a)} = \begin{pmatrix}
											- \alpha_1 & 1 & -\alpha_2 & 0 \\
											- \alpha_4 & 0 & -\alpha_3 & 1
										\end{pmatrix}
	\,,
	\quad
	C^{\perp}_{(b)} = \begin{pmatrix}
											- (\alpha_1+\alpha_2\alpha_6) & 1 & 0 & 0 & -\alpha_5\alpha_6\alpha_2  \\
											- \alpha_6										& 0 & 1 & 0 & -\alpha_5 \alpha_6  \\
											- \alpha_6 \alpha_3						& 0 & 0 & 1 & -(\alpha_4 + \alpha_5\alpha_6\alpha_3)
										\end{pmatrix}\,.
\end{align}  

Now that we have all ingredients together, we can go ahead and consider a simple on-shell diagram in detail. Specifically, we calculate the box on-shell 
diagram (\ref{OSDiagExamples})(a), in which case the delta functions (\ref{DeltaExt}) are equal to 
\begin{equation}
\delta (C\cdot Z) = \frac{1}{\ab{13}^4} \delta\! \left[\alpha_1 - \frac{\ab{23}}{\ab{13}}\right]\, 
																				\delta\! \left[\alpha_2 - \frac{\ab{12}}{\ab{13}}\right]\,
																				\delta\! \left[\alpha_3 - \frac{\ab{14}}{\ab{13}}\right]\,
																				\delta\! \left[\alpha_4 - \frac{\ab{43}}{\ab{13}}\right]\,
																				\delta^4(P)\delta^8(\Q)
\end{equation}

and the differential form becomes a function of external kinematics only,
\begin{equation}
d\Omega = \frac{d\alpha_1}{\alpha_1}\frac{d\alpha_2}{\alpha_2}
					\frac{d\alpha_3}{\alpha_3}\frac{d\alpha_4}{\alpha_4}\,\delta(C\cdot Z) 
			  = \frac{\delta^4(P)\delta^8(\Q)}{\ab{12}\ab{23}\ab{34}\ab{41}}\,.
\end{equation}
This is equal to formula (\ref{One4}) found by multiplying three-point amplitudes. 

The same procedure can be applied to planar on-shell diagrams in ${\cal N}<4$ sYM. The important difference is that the diagrams are necessarily oriented
unlike in the maximally supersymmetric case where the perfect orientations only played an auxiliary role for constructing the $C$-matrix. This corresponds 
to the fact that in less supersymmetric theories we need two on-shell multiplets to capture the positive and negative helicity gluons (and their respective superpartners) 
and the arrows specify which multiplet we are talking about. For the external states, we can choose the orientation of the arrows 
of a given on-shell diagram depending on the helicity structure we want to consider, but for internal legs we have to sum over all possible orientations. In addition, 
for perfect orientations with closed internal loops we have to add an extra factor, $\J$, in the measure, 
\begin{equation}
d\Omega = \frac{d\alpha_1}{\alpha_1}\frac{d\alpha_2}{\alpha_2}\dots \frac{d\alpha_m}{\alpha_m}\,{\cal J}^{{\cal N}-4}\cdot\,\delta(C\cdot Z)\,.
\end{equation}

This modification arises when passing from vertex-variables to edge-variables and ${\cal J}$ is defined as the determinant of the adjacency matrix $A_{ij}$ of the graph
\begin{equation}
{\cal J} = {\rm det}(1-A)\,.
\label{loopJacobian}
\end{equation}

The entries of $A$ are given by,
\begin{equation}
  A_{ij} = \text{weight of the \emph{directed} edge } i\to j \text{ (if any)}\,.
	\label{adjMatrix}
\end{equation}

If there is a collection of closed orbits bounding ``faces'' $f_i$, with \emph{disjoint} pairs $(f_i,f_j)$, disjoint triples $(f_i,f_j,f_k)$ etc., then the Jacobian $\J$ can
be expressed as,
\begin{equation}
 \J = 1+\sum_i f_i + \sum_{\substack{\text{disjoiont}\\ \text{pairs } i,j}}\hskip -.25cm f_i f_j + \sum_{\substack{\text{disjoiont}\\ \text{triples } i,j,k}}\hskip -.4cm f_if_jf_k + \cdots\,,
\end{equation}

where each ``face'' $f_i$ denotes the product of edge-variables along that orbit,
\begin{align}
 f_i = \prod_{r\subset \substack{\text{closed}\\ \text{ orbit}_i}} \hskip -.3cm \alpha_r\,.
\end{align}

This factor cancels in $\NeqFour$ but in the case of lower supersymmetries it is present. For further details, we refer the reader directly to
\cite{OnshellDiagrams}, Sec.~14. Here we included a brief discussion of $\J$ as it will play a role in our gravity formulas later. 

%
\subsection{Hidden properties of $\NeqFour$ amplitudes}
%

The Grassmannian formulation of on-shell diagrams make several important properties of scattering amplitudes in planar $\NeqFour$ 
completely manifest. The Yangian symmetry \cite{Drummond:2009fd} is realized as positivity preserving diffeomorphisms \cite{OnshellDiagrams}, and the recursion relations (\ref{BCFWinOSdiags}) 
make manifest that it is also present in tree-level amplitudes and the loop integrands. The loop integration breaks the Yangian symmetry 
due to the presence of IR-divergencies \cite{Drummond:2007aua} and all known regulators would break it as well. There is an ongoing search for a new regulator 
which would preserve the Yangian using integrability techniques \cite{Ferro:2012xw,Ferro:2013dga,Beisert:2014qba,Broedel:2014hca}. 

The other important property which is inherited in the formula (\ref{Form}) is the presence of logarithmic singularities only. This property 
is much stronger than just the presence of single poles since we require $\frac{dx}{x}$ behavior near any pole in the cut structure. Each on-shell 
diagram is given by a $\dlog$ form ($\dlog x \equiv \frac{dx}{x}$) in terms of edge variables multiplied by a set of delta functions (\ref{Form}). 
One can solve for $\alpha_k$ in terms of external momenta and off-shell loop momenta, and the full integrand can be written as
\begin{equation}
d{\cal I} = \sum_j \dlog f_j^{(1)} \,\dlog f_j^{(2)}\,\dlog f_j^{(3)}\dots \dlog f_j^{(4L)} \,\delta(C_j^\ast\cdot \widetilde{\eta}) \label{dlog}
\end{equation}

where the $f_j^{(k)}$ depend both on external and internal momenta and $C_j$ is the Grassmannian matrix with $\alpha_k$ substituted for. 
This representation of the loop integrand is very closely connected to the maximal transcendentality of ${\cal N}=4$ planar amplitudes 
\cite{OnshellDiagrams,LipatovTranscendentality,Eden:2006rx,BeisertEdenStaudacher}. 
For $k\leq 4$ the fermionic delta function does not depend on the loop momenta and the representation (\ref{dlog}) can in 
principle be integrated directly so that one can obtain the final result in terms of polylogarithms \cite{Goncharov:2010jf,
  Golden:2013xva, Drummond:2014ffa, Parker:2015cia, Dixon:2013eka, Dixon:2014iba, Dixon:2015iva}. For $k>4$ the loop momenta are present 
in the fermionic delta function and the result is not a $\dlog$ form in momentum space, but it still is in terms of edge variables. 
This gives rise to elliptic functions after integration which suggests that our notion of $\dlog$ forms and transcendentality should 
be generalized to include these cases. 

Finally, the on-shell diagrams make another important property completely manifest and that is the absence of poles at infinity \cite{OnshellDiagrams}. 
In other words, the loop integrand in planar $\NeqFour$ as well as individual on-shell diagrams never generate a singularity 
which would correspond to sending the loop momentum to infinity, $\ell\rightarrow\infty$. This is a consequence of dual conformal 
symmetry and the representation of on-shell diagrams (and loop integrand) makes that manifest when using momentum twistor variables.

\section{Non-planar on-shell diagrams}

On-shell diagrams are well defined for any Quantum Field Theory with fundamental three point amplitudes and do not rely on the planarity 
of graphs. We can consider an arbitrary bi-colored graph with three-point vertices, 
\vskip -.1cm
$$
\includegraphics[trim={0cm .1cm 0cm 0cm},clip,scale=.8]{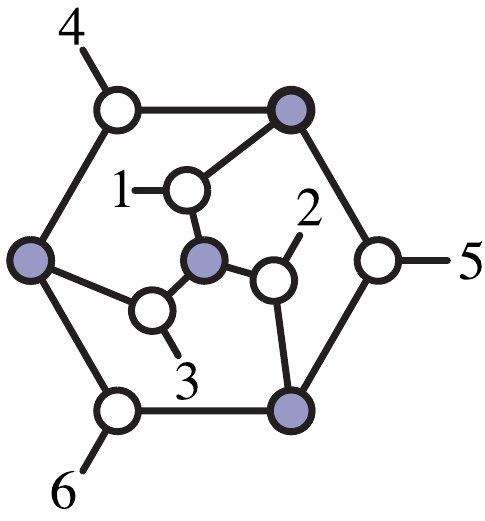}
\hspace{1cm}
\includegraphics[trim={0cm .1cm 0cm 0cm},clip,scale=.8]{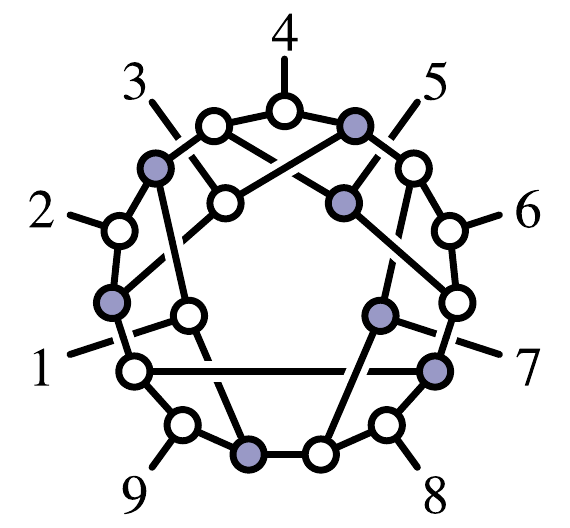}
$$

and define the on-shell function as the corresponding product of three-point amplitudes evaluated at specific on-shell kinematics 
dictated by the graph. 

To each diagram we can associate a point in the Grassmannian, represented by the matrix $C$. This identification uses the rules 
explained in the previous section: associate variables with edges $\alpha_k$, choose a perfect orientation and calculate the 
entries of the $C$-matrix using Eq.~(\ref{Entry}). If the diagram is planar and the edge variables are chosen real and with definite sign, we obtain a cell in the positive 
Grassmannian $G_+(k,n)$, in other cases we end up in some cell in a generic Grassmannian $G(k,n)$.

In general, we want to associate a form $d\Omega$ which reproduces the on-shell function given by the product of three point amplitudes,
\begin{equation}
d\Omega = df(\alpha_k)\,\delta(C\cdot Z)
\label{eq:FormSplitMeasure}
\end{equation}

where the measure $df(\alpha_k)$ depends on the theory while the delta function $\delta(C\cdot Z)$ only depends on the diagram and 
external kinematics. Therefore the problem naturally splits into two parts: finding the measure $df(\alpha_k)$ and the $C$-matrix. 
While the $C$-matrix associated to a given on-shell diagram can be found using Eq.~(\ref{Entry}), the classification of all possible 
non-planar diagrams and their associated particular subspace in $G(k,n)$ represent an important problem. For the case of MHV leading 
singularities the answer was given in \cite{Arkani-Hamed:2014bca} but understanding more general cases is a part of ongoing research \cite{Franco:2015rma}.

For a generic Quantum Field Theory the measure $df(\alpha_k)$ associated with a given diagram is not known. However, for the case of 
Yang-Mills theory the answer has been worked out in \cite{OnshellDiagrams} and turns out to be surprisingly simple,
\begin{equation}
d\Omega = \frac{d\alpha_1}{\alpha_1}\frac{d\alpha_2}{\alpha_2}\dots \frac{d\alpha_m}{\alpha_m} \,
					\J^{\N-4}\cdot\delta(C\cdot Z)\,.
\label{YMdOmega}
\end{equation}

The $\J$-factor is given by the determinant of the adjacency matrix (\ref{adjMatrix}) and the singularities coming from this part of the measure 
are closely related to the UV-sector of the theory. In $\NeqFour$ this term is absent and we get a pure $\dlog$-form. From 
the discussion so far it is clear that writing the form (\ref{YMdOmega}) did not depend on the planarity of the diagram so that the formula 
is identical to (\ref{Form}) described in the planar sector. The goal of this section is to extend the knowledge of the Grassmannian 
formulation beyond the Yang--Mills case and find the analogue of (\ref{YMdOmega}) for gravity on-shell diagrams. 

%
\subsection{First look: MHV leading singularities}
%

The leading singularities are reduced on-shell diagrams ($n_\delta=0$) associated with on-shell functions $\Omega$ (rather than forms) and they represent 
codimension $4L$ cuts of loop amplitudes. The simplest leading singularities are of MHV-type. In planar $\NeqFour$ they are all equal to the MHV 
tree-level amplitude given by the Parke-Taylor factor,
\begin{equation}
	\PT(123\dots n) = \frac{1}{\ab{12}\ab{23}\ab{34}\dots \ab{n1}}\,.
	\label{PT1}
\end{equation}

Beyond the planar limit all MHV leading singularities must be holomorphic functions $F(\lambda)$ \cite{Witten:2003nn}. Furthermore, it was shown in \cite{Arkani-Hamed:2014bca} that 
all MHV leading singularities can be decomposed into linear combinations of Parke-Taylor factors with different orderings $\sigma$,
\begin{equation}
	\Omega = \sum_\sigma c_\sigma\,\PT(\sigma_1\sigma_2\dots \sigma_n) 
	\quad \mbox{where} \quad  c_\sigma = \pm1,0\,.
	\label{PT2}
\end{equation}

This representation makes manifest the fact that all singularities are logarithmic as each Parke-Taylor factor behaves like 
$\frac{1}{x}$ near any singularity and one can infer the existence of the logarithmic form directly from the expression (\ref{PT2}). 
Following the same logic, it is very natural to look at the MHV leading singularities in $\NeqEight$ and study their 
expressions in more detail. 

Gluing together three-point amplitudes we find some suggestive expressions for a few simple on-shell diagrams
(dropping the overall (super-) momentum conserving $\delta$-functions in $\NeqEight$ $\delta^4(\lam{}\cdot\lamt{}) \delta^{16}(\lam{}\cdot\twEta)$),
\begin{large}
\vskip -.4cm
\begin{equation*}
\hspace{-.0cm}
\begin{array}{ccc}
\raisebox{-45pt}{\includegraphics[trim={0cm .1cm 0cm 0cm},clip,scale=.45]{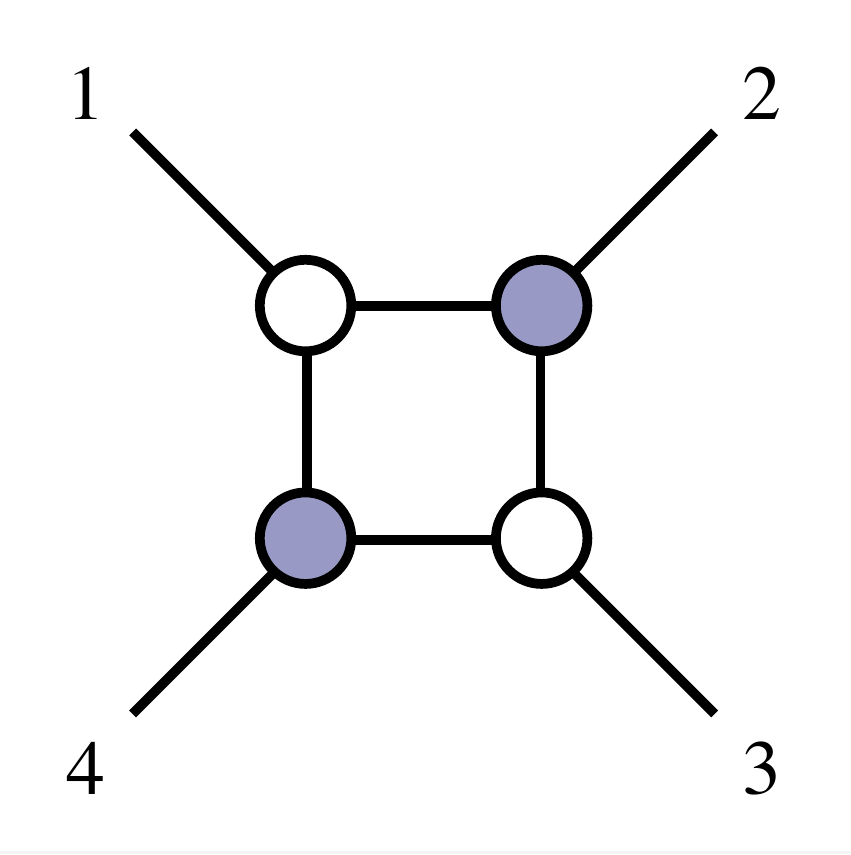}}
&
\raisebox{-45pt}{\includegraphics[trim={0cm .1cm 0cm 0cm},clip,scale=.45]{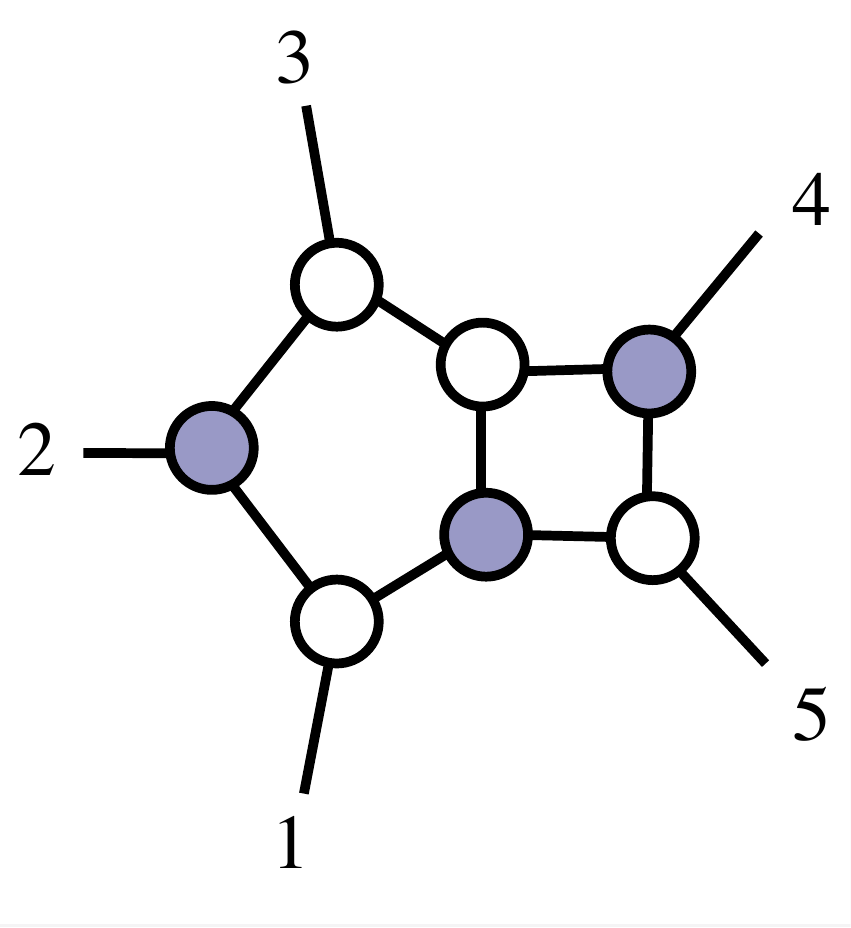}}
&
\raisebox{-45pt}{\includegraphics[trim={0cm .1cm 0cm 0cm},clip,scale=.45]{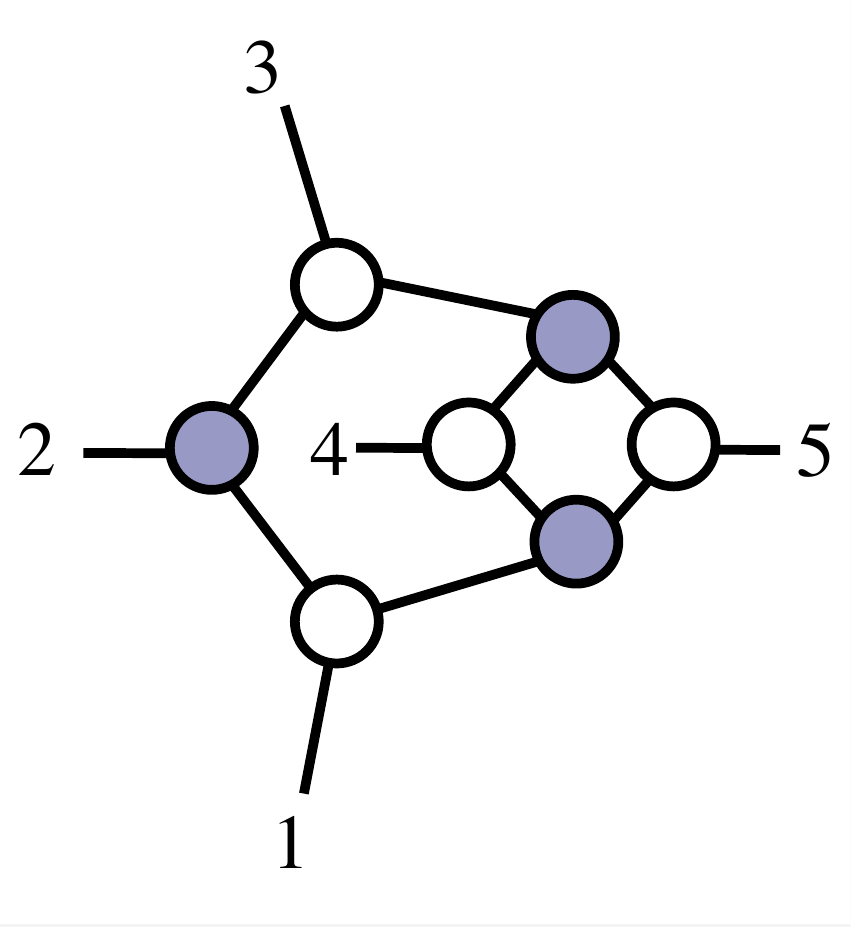}} 
\\[-14pt]
\downarrow & \downarrow & \downarrow \\
	\begin{large}
	\frac{[13][24]}
			{\ab{12}\ab{13}\ab{14}\ab{23}\ab{24}\ab{34}}\,\end{large}
&  \begin{large}\frac{[12][23][45]^2}
			{\ab{12}\ab{13}\ab{15}\ab{23}\ab{34}\ab{45}}\,\end{large}
&  \begin{large}\frac{[12][23][45]^2}
			{\ab{12}\ab{14}\ab{15}\ab{23}\ab{34}\ab{35}}\,\end{large} \\
\end{array}
\end{equation*} 
\end{large}

From these examples one could conjecture that all poles $\ab{ij}$ are linear and the numerator involves only anti-holomorphic 
brackets $[ij]$. However, looking at more complicated diagrams we learn that this is not the case and one gets both more complicated 
numerators and higher degree poles in the denominator. 
\begin{large}
\vskip -.4cm
\begin{equation*}
\hspace{-.0cm}
\begin{array}{cc}
\raisebox{-45pt}{\includegraphics[trim={0cm .1cm 0cm 0cm},clip,scale=.45]{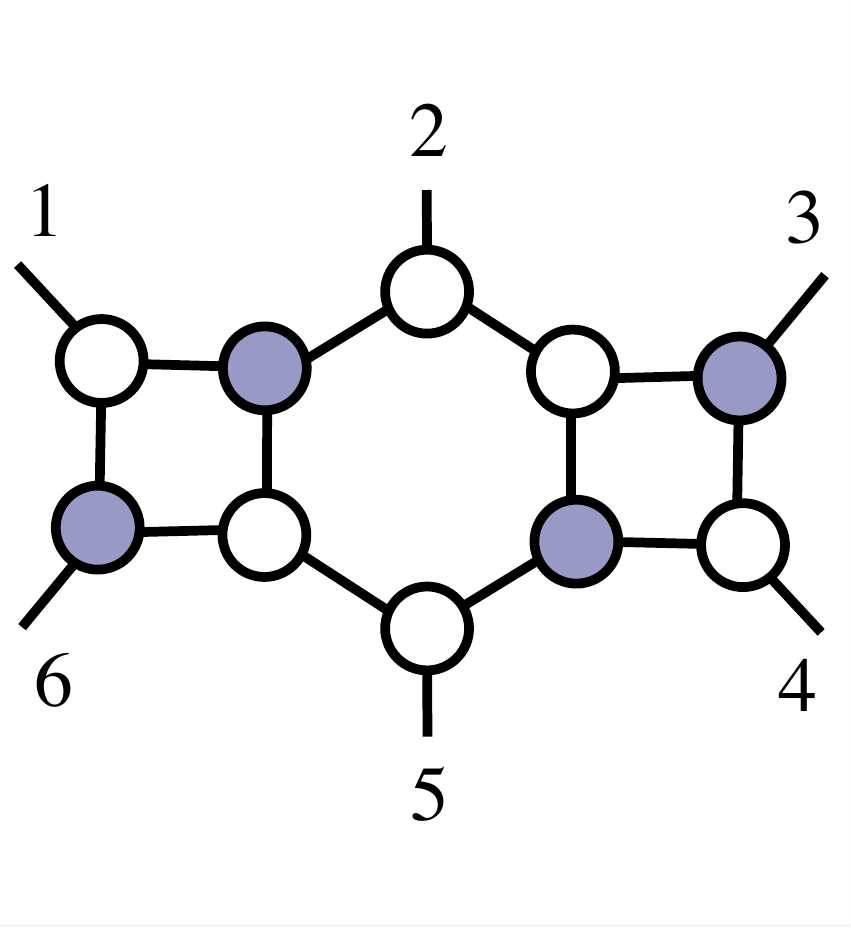}}
&
\raisebox{-45pt}{\includegraphics[trim={0cm .1cm 0cm 0cm},clip,scale=.45]{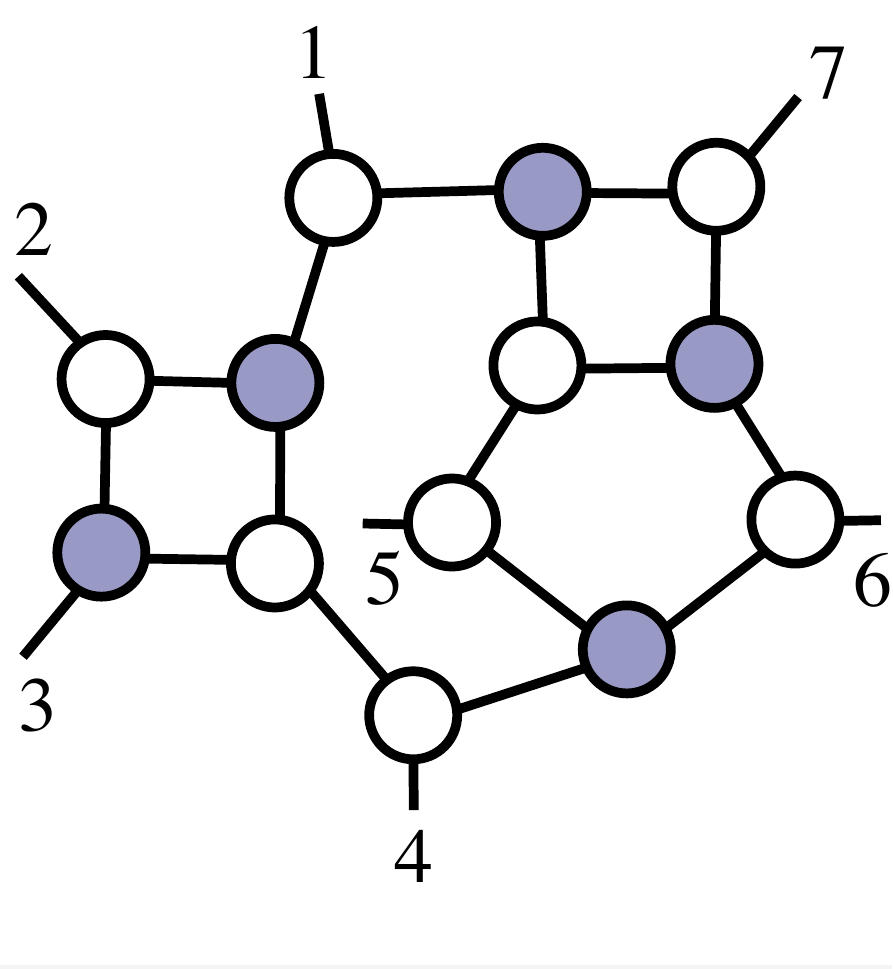}}
\\[-14pt]
\downarrow & \downarrow \\
	\begin{large}
	 \frac{\aMs{5}{Q_{16}}{2}\aMs{2}{Q_{34}}{5}[16]^2[34]^2}
	      {\ab{12}\ab{23}\ab{34}\ab{45}\ab{56}\ab{61} \tred{\ab{25}^2}}\,\end{large}
&  \begin{large}\frac{\sqb{23}^2 \aMs{1}{Q_{23}}{4} \aMs{4}{Q_{23}}{1} \aMs{1}{Q_{67}}{5} \aMs{1}{Q_{57}}{6} \aMs{4}{Q_{56}}{7}^2 }
	{\tred{\tred{\ab{14}}^3}\ab{12}\ab{15}\ab{17} \ab{23}\ab{34}\ab{45}\ab{46}\ab{56}\ab{67}}\,\end{large} \\
\end{array}
\end{equation*}
\end{large}

Analyzing the data more closely, especially looking at the on-shell solutions for the momenta of the internal edges, 
one can make the following statement: 

\begin{center}
\emph{ On-shell diagram vanishes if three momenta in a white vertex are collinear.}
\end{center}

Concretely, the white vertex already enforces the $\lam{}$'s to be proportional. If, on top of that, the $\lamt{}$'s become 
collinear as well (which implies the collinearity of momenta) the on-shell diagram vanishes. Interestingly, each factor in 
the numerator of the on-shell function exactly corresponds to such a condition which is why the number of factors in the 
numerators equals the number of white vertices in a given MHV on-shell diagram. 

Taking a closer look at the denominator of the expressions one realizes that all factors which correspond to erasing edges from 
the on-shell diagram (by sending the momentum of that edge to zero) are single poles. In contrast, all higher poles (and some single poles) 
correspond to sending the momenta of an internal loop to infinity. Such poles are completely absent in the $\NeqFour$ case 
-- this is related to the statement of no poles at infinity \cite{Log,ThreeLoopPaper,Bern:2015ple} -- but in gravity they are present. 

To clarify some of these statements, we discuss a concrete example and analyze the following on-shell diagram,
\begin{equation*}
\raisebox{-58pt}
{\includegraphics[trim={0cm .1cm 0cm 0cm},clip,scale=.45]{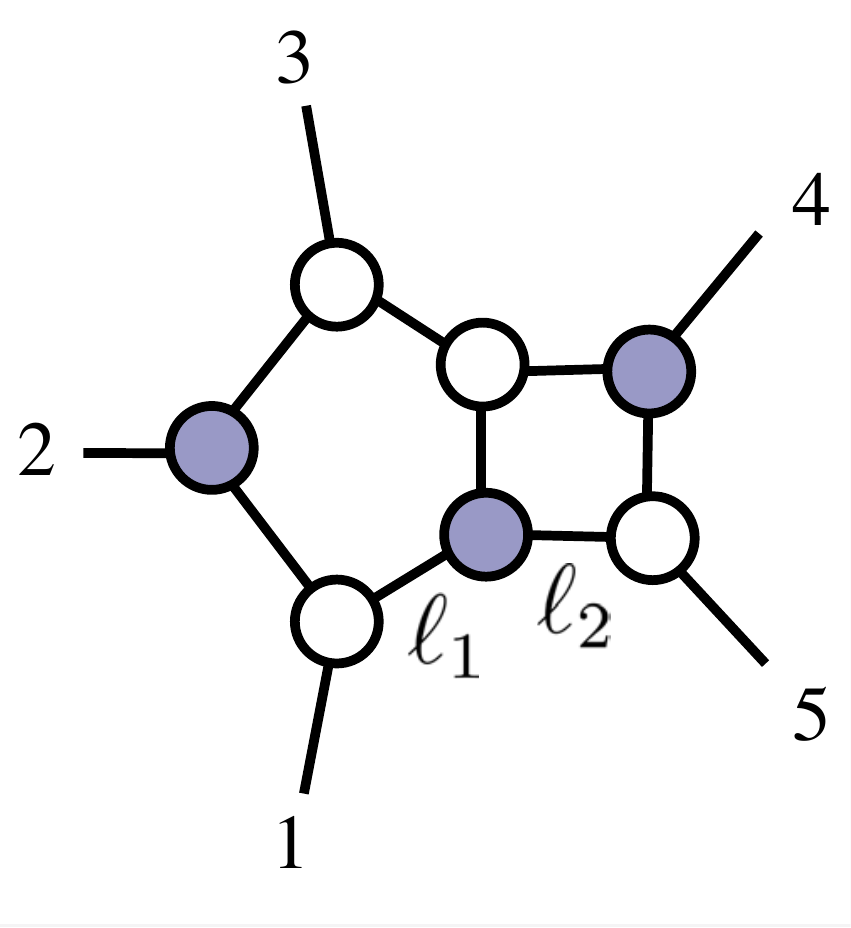}}
\hspace{.5cm}
\begin{array}{ll}
	\ell_1 = \frac{\lam{1}\ Q_{12}\cdot\lam{3}}{\tred{\ab{13}}}\,,		&	
	\ell_2 = \frac{\lam{5}\ Q_{12}\cdot\lam{3}}{\ab{35}}	\,,					\\
	\ell_1-1 = \frac{\tblue{\ab{23}}}{\tred{\ab{13}}}\lam{1}\lamt{2}\,,			  &
	\ell_2-5 = \frac{\tmagenta{\ab{34}}}{\ab{35}} \lam{5} \lamt{4}	\,,					\\
	\ell_1-Q_{12} = \frac{\tgreen{\ab{12}}}{\tred{\ab{13}}} \lam{3} \lamt{2}\,, & 
	\ell_2-Q_{45} = \frac{\tburgundy{\ab{45}}}{\ab{35}} \lam{3}\lamt{4} 			\,,		\\
	\ell_1-Q_{123}=\frac{\lam{3}\ Q_{23}\cdot\lam{1}}{\tred{\ab{13}}}\,,		& 
	\ell_1+\ell_2 = \frac{\tcyan{\ab{15}}}{\tred{\ab{13}}\ab{35}} \lam{3}\ Q_{12}\cdot\lam{3}\,. \\
	\end{array}
\end{equation*}
\vskip -.9cm
\begin{equation}
 \Omega= \frac{[12][23][45]^2} {\tgreen{\ab{12}}\tred{\ab{13}}\tcyan{\ab{15}}\tblue{\ab{23}}\tmagenta{\ab{34}}\tburgundy{\ab{45}}}\,.
 \label{5ptExampleGRForm}
\end{equation}

As explained above, most of the poles $\ab{ij}$ correspond to erasing edges in the on-shell diagram which is equivalent to setting 
the internal momentum of that edge to zero. 
In our example $\tred{\ab{13}}$ corresponds to a pole at infinity and on this pole, all momenta associated with this loop blow up. 
Finally, let's look at the structure of the numerator. Focusing on the white vertex adjacent to external leg $1$, the respective 
on-shell solutions for $\ell_1$ and $\ell_1\!-\!p_1$ as well as the external leg become collinear when $\sqb{12} = 0 \Rightarrow \lamt{2}\sim\lamt{1}$, 
$\ell_1\stackrel{[12]\to0}{\longrightarrow} \sim \lam{1}\lamt{1},\ \ell_1\!-\!p_1 \stackrel{[12]\to0}{\longrightarrow} \sim \lam{1}\lamt{1}$. As noted 
earlier, the gravity on-shell form vanishes in this limit due to the factor $[12]$ in the numerator. For the remaining white vertices, 
a similar analysis recovers all other square brackets $[ij]$ in the numerator of the gravity form (\ref{5ptExampleGRForm}). 

We can take these observations as a starting point in the search for the Grassmannian formulation of gravity on-shell diagrams. 
We learned that on-shell diagrams can have multiple poles associated with poles at infinity, and importantly the numerator factor 
must capture the curious collinear behavior observed above.

%
\subsection{Three point amplitudes with spin $s$}
%

The most natural starting point for a Grassmannian representation of gravity on-shell diagrams focuses on the atomic building blocks, the three-point amplitudes, first.
We start with a maximally supersymmetric theory of particles with spin $s$. In that case, the amount of supersymmetry is given by $\N=4s$. 
As noted before, in massless theories, the elementary three-point amplitudes are completely fixed by their little group weight to all orders 
in perturbation theory (up to an overall constant). In particular, the three-point MHV-amplitude for spin $s$ particles is given by, 
\begin{equation}
	A_3^{(2)} = \frac{\delta^4(P)\delta^{2{\cal N}}(\Q)}{\ab{12}^s\ab{23}^s\ab{31}^s}\,. 
	\label{ThreeP2}
\end{equation}

The on-shell diagram for this amplitude is just a single black vertex to which we can give a perfect orientation in exactly the same manner 
as for $\NeqFour$ discussed in section \ref{sec:GrassmannianYM}. We can use exactly the same rules from before to write the $C$-matrix,
\begin{equation}
\raisebox{-48pt}
{\includegraphics[trim={0cm .1cm 0cm 0cm},clip,scale=.4]{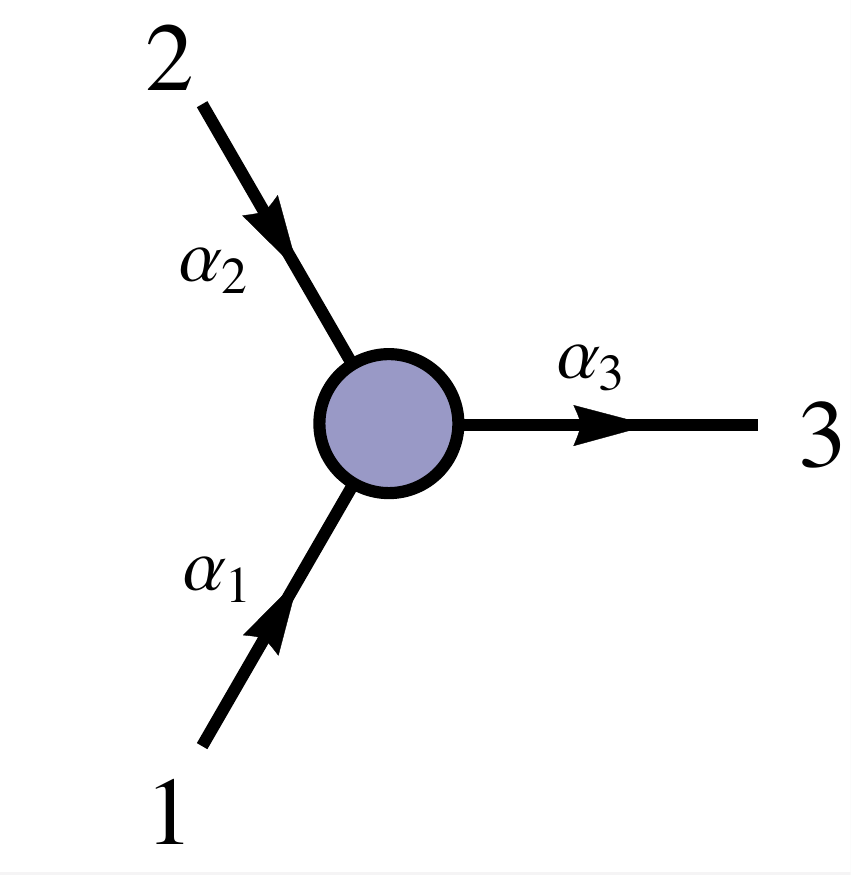}}\, \Leftrightarrow
\quad C = \left(\begin{array}{ccc} 1 & 0 & \alpha_1\alpha_3\\ 0 & 1 & \alpha_2\alpha_3\end{array}\right)\,.
\end{equation}

Here we purposely do not choose any $GL(1)_v$ gauge fixing in the vertex because gauge-independence will be one of our criteria 
for finding the correct formula. The first step towards the Grassmannian representation of (\ref{ThreeP2}) is to write the linearized delta 
functions which have a very similar form to (\ref{DeltaExt}),
\begin{equation}
\delta^{(2\times2)}(C\cdot \widetilde{\lambda})\,\delta^{(1\times2)}(C^\perp \cdot \lambda)\,\delta^{(2\times{\cal N})}(C\cdot\widetilde{\eta}) = \frac{1}{\alpha_3^2\la12\ra^{{\cal N}-1}}\delta^{(4)}(P)\delta^{(2{\cal N})}(\Q) \,.
\label{DeltaFunc1}
\end{equation}

Using the two bosonic delta-functions from $\delta^{(1\times2)}(C^\perp \cdot \lambda)$, 
we can solve for two of the auxiliary $\alpha_k$ variables,
\begin{equation}
\alpha_1 = \frac{\ab{23}}{\alpha_3\ab{12}},\quad 
\alpha_2 = \frac{\ab{13}}{\alpha_3\ab{12}}\,.
\label{Edge}
\end{equation}

The general form of the Grassmannian representation of (\ref{ThreeP2}), for which the measure depends only on the $\alpha_k$-variables 
and is permutation invariant in all three legs, is
\begin{equation}
d\Omega_\sigma = \frac{d\alpha_1}{\alpha_1^\sigma}
								 \frac{d\alpha_2}{\alpha_2^\sigma}\frac{d\alpha_3}{\alpha_3^\sigma}\,
									\delta^{(2\times2)}(C\cdot \lamt{})\,
									\delta^{(1\times2)}(C^\perp \cdot \lambda)\,
									\delta^{(2\times \N)}(C\cdot\twEta)\,,
\label{Form1}
\end{equation}

for some integer $\sigma$. We can plug (\ref{DeltaFunc1}) and (\ref{Edge}) into (\ref{Form1}) to get
\begin{equation}
d\Omega_\sigma = \frac{d\alpha_3}{\alpha_3^{2-\sigma}}\cdot 
									\frac{\delta^{(4)}(P)\delta^{(2\N)}(\Q)}
											 {\ab{12}^{\N-1-2\sigma}\ab{23}^\sigma\ab{31}^\sigma}\,.
\label{FormPluggedVersion}
\end{equation}

This expression must be permutation invariant in $\ab{12}$, $\ab{23}$, $\ab{31}$ and independent of the gauge-choice for $\alpha_3$. 
In order to ensure $GL(1)$-invariance, $\frac{d\alpha_3}{\alpha_3}$  has to factor out as the volume of $GL(1)$-transformations. These 
two requirements leave us with a unique choice: $\sigma=s=1$ which corresponds to $\NeqFour$ with the logarithmic measure. Of course, 
one can also make a special choice, $\alpha_3=\frac{1}{\ab{12}}$ so that $\alpha_1=\ab{23}$, $\alpha_2=\ab{13}$, which allows us to 
write any three point amplitude (\ref{ThreeP2}) using edge variables only. But our goal is to find a form which is independent of any such 
choices. Consequently, the form (\ref{Form1}) is not able to reproduce the gravity or any higher spin three-point amplitude.

The natural modification of the form (\ref{Form1}) involves some dimensionful, permutation invariant object $\Delta$. The $\delta(C^\perp\cdot \lambda)$ 
allows us to relate $\alpha_1\lambda_ 1 + \alpha_2\lambda_2 + \frac{1}{\alpha_3}\lambda_3 = 0$ which we use in the definition of $\Delta$ as follows, 
\begin{equation}
\Delta \equiv \la AB \ra = \la BE \ra = \la EA\ra \quad \mbox{where} \quad A=\alpha_1\lambda_1,\,\,B=\alpha_2\lambda_2,\,\, E=\frac{1}{\alpha_3}\lambda_3\,.
\label{Delta1}
\end{equation}

Note that this object has exactly the property suggested by our study of MHV leading singularities: it vanishes when all three momenta are collinear. 
Now we consider a form
\begin{equation}
d\Omega = \frac{\Delta^\rho\cdot d\alpha_1\,d\alpha_2\,d\alpha_3}{\alpha_1^{\sigma_1}\alpha_2^{\sigma_2}\alpha_3^{\sigma_3}}\,\delta^{(2\times2)}(C\cdot \widetilde{\lambda})\,\delta^{(1\times2)}(C^\perp \cdot \lambda)\,\delta^{(2\times\cal N)}(C\cdot\widetilde{\eta})\,.
\label{Form2}
\end{equation}

Repeating the same exercise that led to (\ref{FormPluggedVersion}) by solving for edge variables, converting the delta functions, 
imposing permutation invariance and the independence on $\alpha_3$ uniquely fixes $\rho = s-1$ and $\sigma_1=\sigma_2=\sigma_3 = 2s-1$. 
The modified form becomes, 
\begin{equation}
d\Omega_s =  \frac{\Delta^{s-1}\cdot d\alpha_1\,d\alpha_2\,d\alpha_3}{\alpha_1^{2s-1}\alpha_2^{2s-1}\alpha_3^{2s-1}}
	     \delta^{(2\times2)}(C\cdot \lamt{})\,\delta^{(1\times2)}(C^\perp \cdot \lambda)\,\delta^{(2\times\N)}(C\cdot\twEta)
\label{Gras1}
\end{equation}

which is a Grassmannian representation of (\ref{ThreeP2}). We would find the same unique solution even if we consider 
$\Delta=\ab{12}$ or any other function of $\alpha_1$, $\alpha_2$, $\alpha_3$ and $\ab{12}$ ($\ab{23}$ and $\ab{13}$ are 
proportional to $\ab{12}$ and $\alpha$'s). Note that this formula is well defined for all integer spins $s$ and maximal 
supersymmetry $\N=4s$. In particular, for $s=1$ it reproduces the logarithmic form of $\NeqFour$. 

There is an analogous Grassmannian representation for the $\MHVbar$ ($k=1$) three-point amplitudes, 
\begin{equation}
\raisebox{-48pt}
{\includegraphics[trim={0cm .1cm 0cm 0cm},clip,scale=.4]{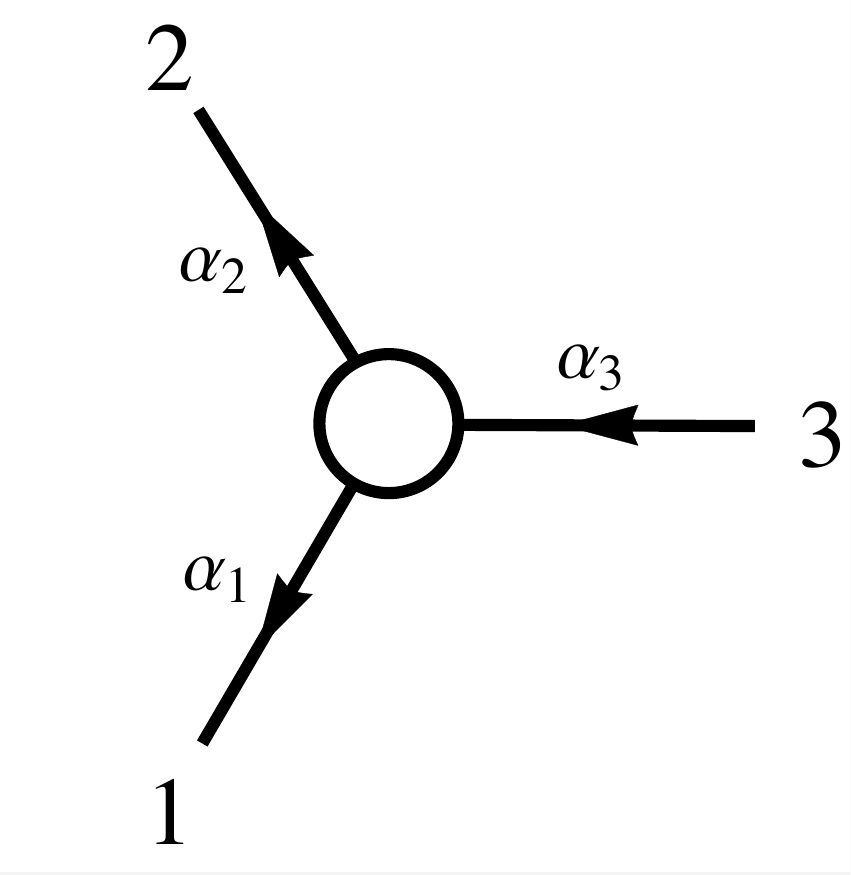}}\,,
\quad C = \left(\begin{array}{ccc} \alpha_1 \alpha_3 & \alpha_2\alpha_3 & 1\\ \end{array}\right)\,.
\end{equation}

which can be encoded by the form,
\begin{equation}
d\widetilde{\Omega}_s =  \frac{\widetilde{\Delta}^{s-1}\cdot d\alpha_1\,d\alpha_2\,d\alpha_3}{\alpha_1^{2s-1}\alpha_2^{2s-1}\alpha_3^{2s-1}}\delta^{(1\times2)}(C\cdot \widetilde{\lambda})\,\delta^{(2\times2)}(C^\perp \cdot \lambda)\,\delta^{(1\times\cal N)}(C\cdot\widetilde{\eta})\label{Gras2}
\end{equation}

where $\widetilde{\Delta} = [AB]=[BE]=[EA]$ with $A=\alpha_1\widetilde{\lambda}_1$, $B=\alpha_2\widetilde{\lambda}_2$ and $E = \frac{1}{\alpha_3}\widetilde{\lambda}_3$. 

%
\subsection{Grassmannian formula}
%

Equipped with the Grassmannian representation of the three-point amplitudes (\ref{Gras1}) and (\ref{Gras2}), 
we can write the Grassmannian representation for any spin $s$ on-shell diagram. Much like in $\NeqFour$, using the amalgamation procedure \cite{OnshellDiagrams}
to glue the three-point vertices into larger diagrams, we write the form in terms of edge variables,
\begin{align}
d\Omega_s &= \Gamma \cdot 
							\frac{d\alpha_1\,d\alpha_2\dots d\alpha_d}
									 {\alpha_1^{2s-1}\alpha_2^{2s-1}\dots \alpha_d^{2s-1}}
							\cdot \prod_{b\in B_v} \Delta_b^{s-1}\cdot 
										\prod_{w\in W_v} \widetilde{\Delta}_w^{s-1}
							\label{GenGras}\\
					&\hspace{2cm}\times 
							\J^{\N-4}\,\cdot\,\delta^{(k\times2)}(C\cdot \widetilde{\lambda})\,
																\delta^{((n-k)\times2)}(C^\perp \cdot \lambda)\,
																\delta^{(k\times \N)}(C\cdot\twEta)\nonumber
\end{align}

where $\Gamma$ denotes any color factor/coupling constant associated with the diagram. The products of $\Delta_b$ and 
$\widetilde{\Delta}_w$ are associated with the set of black ($B_v$) and white ($W_v$) vertices respectively. They can 
be easily calculated using edge variables and external spinor-helicity variables. We are going to show some explicit 
examples in section \ref{sec:GrOSdiags}.

Note that the Jacobian factor $\J$ is the same as for ${\cal N}<4$ sYM on-shell diagrams (\ref{loopJacobian}). The reason is that 
it originates from rewriting the (super-)momentum conserving delta functions in the linearized form using the $C$-matrix. In particular, it does not 
depend on the measure $df(\alpha_k)$ in (\ref{eq:FormSplitMeasure}) and therefore is the same for theories of arbitrary spin and number of supersymmetries. 
However, depending on the number of fermionic delta functions related to the amount of supersymmetry $\N$, the respective 
power $\J^{\N-4}$ changes and for $\N =4$ always cancels. While the formula has been originally derived for ${\cal N}=4s$ it is actually valid for any $s$ and any ${\cal N}$, so it also captures theories with lower supersymmetries. The reason is that ${\cal J}$ comes from solving the delta functions in the gluing three point vertices and also depends on the number supersymmetries, not the measure for a given theory.

Before proceeding further, note that the on-shell diagrams for spin $s>2$ make perfect sense as they are obtained from 
gluing elementary three point amplitudes together --which in turn are well defined. However, in Minkowski space, we know that 
there are no consistent long range forces mediated by spin $s>2$ particles \cite{Weinberg:1964ew,Weinberg:1965rz}. From the point of on-shell diagrams, we can see that 
$s=1,2$ are special if we look at the identity moves on on-shell diagrams. There are two moves satisfied by planar on-shell 
diagrams: the square move (\ref{squareMove}) and merge-expand (\ref{mergeExpandPlanar}). These moves leave invariant the cell in the positive Grassmannian $G_+(k,n)$ 
as well as the logarithmic form $d\Omega$ which calculates the value of the on-shell diagram in $\NeqFour$ theory. 

The content of the first move is the parity symmetry of a four point amplitude, and it does not really depend on planarity. 
Indeed, calculating the four point on-shell diagram (\ref{OSDiagExamples})(a) we find that for general $s$ it is equal to
\begin{equation}
\Omega_s=\left(\frac{[12]}{\ab{34}}\right)^{s-1}\cdot
				 \frac{\delta^{(4)}(P)\delta^{(2{\cal N})}(\Q)}{\ab{12}\ab{23}\ab{34}\ab{41}}
\end{equation}

which is indeed invariant under the parity flip due to the totally crossing symmetric prefactor.

The merge-expand move gets modified beyond the planar limit. In fact, it is not a two-term relation (\ref{mergeExpandPlanar}) but now involves 
a third (non-planar) contribution,
\begin{equation*}
	      0
		=
	\raisebox{-45pt}
		{\includegraphics[scale=.40]{./figures/mergeExpand_black_planar2}}
		+
		\raisebox{-45pt}
		{\includegraphics[scale=.40]{./figures/mergeExpand_black_planar1}}	
		-
		\raisebox{-45pt}
		{\includegraphics[scale=.40]{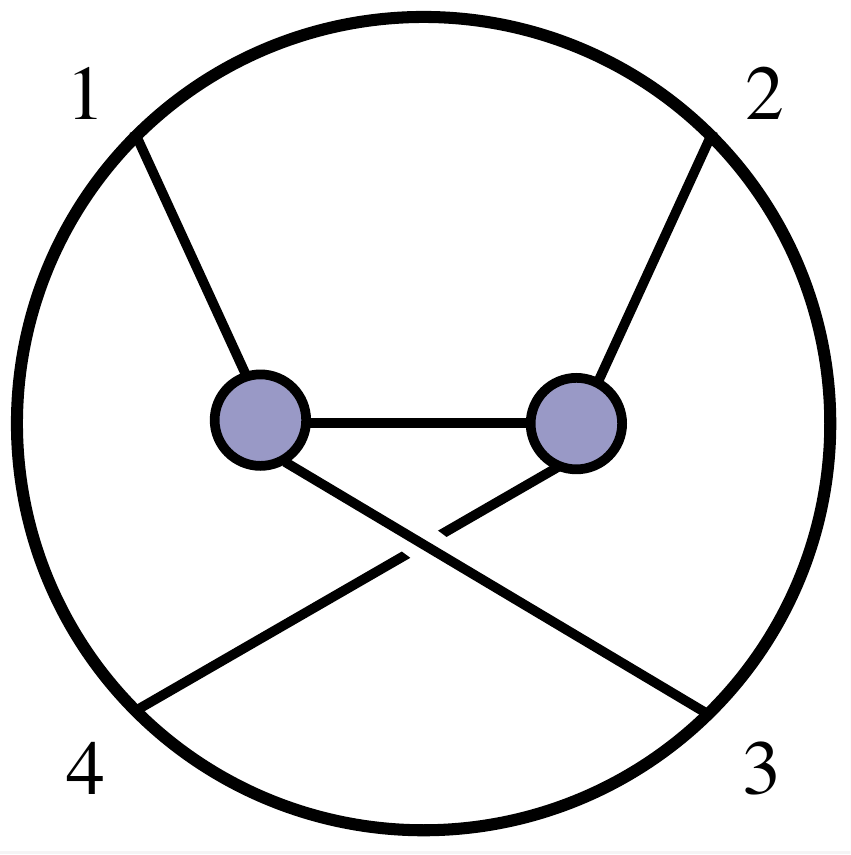}}	
\end{equation*}

Calculating all three diagrams either by gluing three point amplitudes or using the Grassmannian formula (\ref{GenGras}) 
we find that the invariance under this move requires
\begin{equation}
\Gamma_s (\ab{12}\ab{34})^{s-1} + \Gamma_t (\ab{14}\ab{23})^{s-1}=\Gamma_u (\ab{13}\ab{24})^{s-1}
\end{equation}

where $\Gamma_k$ are the group factors for $s$-, $t$- and $u$-channels. There are only two solutions to this equation: 
Either $s=1$ and $\Gamma_s+\Gamma_t=\Gamma_u$ which is nothing but the Jacobi identity for the color factors $\Gamma_s = f^{12a}f^{34a}$, 
$\Gamma_t = f^{14a}f^{23a}$, $\Gamma_u=f^{13a}f^{24a}$. Here we easily recognize $\NeqFour$. The other option for which the merge-expand 
move holds is $s=2$ and $\Gamma_s=\Gamma_t=\Gamma_u$ (and equal to some constant) due to the Shouten identity. This case corresponds to 
$\NeqEight$. All higher spin cases (as well as for $s=0$) are not consistent with the merge-expand move. 

The merge-expand move is not an essential property of on-shell diagrams, indeed the ${\cal N}<4$ SYM diagrams do not satisfy it. 
But for maximally supersymmetric theories it seems like a good guide when the theory is healthy. From now on, we will focus on the 
$s=2$ case of $\NeqEight$. For this theory, the Grassmannian representation becomes,
\begin{align}
d\Omega &=  \frac{d\alpha_1\, d\alpha_2\dots d\alpha_d}
		{\alpha_1^{3}\alpha_2^{3}\dots \alpha_d^{3}}
	\prod_{b\in B_v}\! \Delta_b 
	\prod_{w\in W_v}\! \widetilde{\Delta}_w\,		\label{GrassmannianGravity}\\
	& \hspace{2cm}\times \J^{4}\cdot \delta^{(k\times2)}(C\cdot \widetilde{\lambda})\,
	\delta^{((n-k)\times2)}(C^\perp \cdot \lambda)\,
	\delta^{(k\times 8)}(C\cdot\widetilde{\eta})\,. \nonumber
\end{align}

Note that a similar formula is valid for ${\cal N}<8$ SUGRA subject to the simple replacement ${\cal J}^4 \rightarrow {\cal J}^{{\cal N}-4}$. 
In these cases we also have to sum over all possible orientations of internal edges, in complete analogy to the Yang-Mills case.

%
\section{Properties of gravity on-shell diagrams}
\label{sec:GrOSdiags}
%
%

In this section we are going to elaborate on the Grassmannian formula for gravity (\ref{GrassmannianGravity}) obtained in the last section. 
We will show on examples how to use the formula to calculate particular on-shell diagrams and comment on their properties.

%
\subsection{Calculating on-shell diagrams}
%

After deriving the Grassmannian formulation for on-shell diagrams in $\NeqEight$ in an abstract setting, let's consider a few 
concrete examples to show that we can reproduce the correct values of the on-shell functions derived before. As a first non-trivial 
example, we consider a reduced on-shell diagram for five external particles. For the construction of the $C$-matrix, we chose a 
convenient perfect orientation. Of course, the final result will be independent of the particular choice. Since we were able to 
choose a perfect orientation without any closed loops, the Jacobian factor $\J$ in Eq.~(\ref{loopJacobian}) from converting between vertex- 
and edge-variables is trivial, $\J=1$.

In complete analogy to the Yang Mills case, we have used the $GL(1)_v$-freedom from all vertices to gauge fix several of the edge-weights 
to $1$. Starting from the gauge-fixed on-shell diagram, we can follow the same rules described in Sec.~\ref{sec:GrassmannianYM} to construct the boundary-measurement 
matrix $C$ by summing over paths from sources to sinks and multiplying the edge weights along the path.
\begin{equation}
\raisebox{-65pt}{
\includegraphics[trim={0cm .1cm 0cm .1cm},clip,scale=.53]{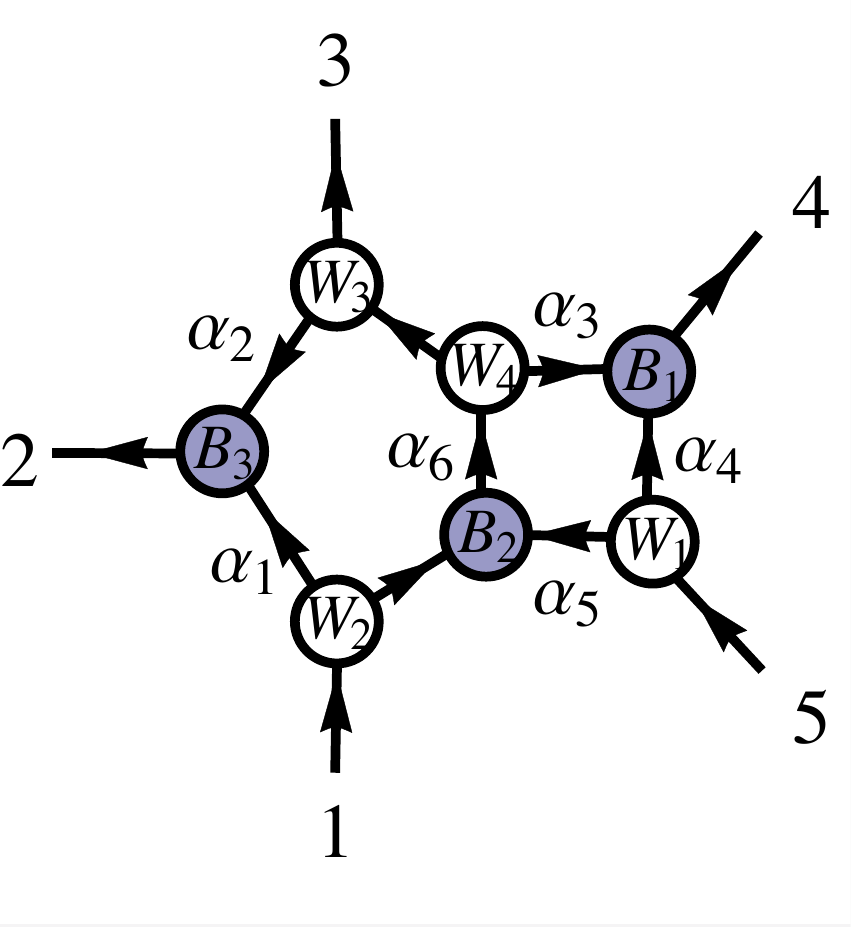}
}
C = \begin{pmatrix}
			1 & \alpha_1 + \alpha_2\alpha_6& \alpha_6 & \alpha_3\alpha_6 & 0 \\
			0 & \alpha_5\alpha_6\alpha_2   & \alpha_5\alpha_6 & \alpha_4+\alpha_3\alpha_5\alpha_6 & 1
		\end{pmatrix}
\label{GravityPentaBoxExample}
\end{equation}
The orthogonal matrix $C^{\perp}$ is then given by,
\begin{align}
	C^{\perp} & = \begin{pmatrix}
										-(\alpha_1 + \alpha_2\alpha_6) & 1 & 0 & 0 & -\alpha_5\alpha_6\alpha_2 \\
										- \alpha_6										 & 0 & 1 & 0 & -\alpha_5\alpha_6				 \\
										-\alpha_3\alpha_6							 & 0 & 0 & 1 & -(\alpha_4
																																	+\alpha_3\alpha_5\alpha_6)
								\end{pmatrix}\,.
\end{align}
We can use the $\delta^{(3\times 2)}(C^{\perp}\cdot\lam{})$ delta-functions to solve 
for all edge variables $\alpha_i$,
\begin{align}
& \alpha_1 = \frac{\ab{23}}{\ab{13}}\,, \  
	\alpha_2 = \frac{\ab{12}}{\ab{13}}\,, \
	\alpha_3 = \frac{\ab{45}}{\ab{35}}\,,	\
  \alpha_4 = \frac{\ab{34}}{\ab{35}}\,, \
	\alpha_5 = \frac{\ab{13}}{\ab{35}}\,, \
	\alpha_6 = \frac{\ab{35}}{\ab{15}}\,.
\label{5ptPentaBoxExEdgeVars}
\end{align} 
Solving for all the $\alpha_i$ induces a Jacobian $J_{C^{\perp}\cdot\lam{}} = 
\left(\ab{35}^2\ab{13}\right)^{-1}$. Plugging these solutions $\alpha_i=\alpha^*_i$ back into the remaining $\delta$-functions, we find,
\begin{align}
\delta^{(2\times 2)}(C\cdot\lamt{})  = \ab{15}^2 \delta^{4}(\lam{}\cdot\lamt{})\,, \qquad
\delta^{(2\times \N)}(C\cdot \twEta) = \frac{1}{\ab{15}^{\N}}\delta^{2\N}(\lam{}\cdot\twEta)\,.
\end{align}
As a quick sanity check, we can recover the $\NeqFour$ result,
\begin{align}
	d\Omega_{\N=4} & = \prod^{6}_{i=1}\frac{d\alpha_i}{\alpha_i} 
\delta^{(2\times 2)}(C\cdot\lamt{})\delta^{(3\times 2)}(C^{\perp}\cdot\lam{})\delta^{(2\times 4)}(C\cdot \twEta) \nonumber \\
& = \PT(12345)\, \delta^{(4)}(\lam{}\cdot\lamt{})
																					\delta^{(2\times 4)}(\lam{}\cdot \twEta)\label{EdgeSol}
\end{align}
The only missing ingredient for the gravity result are the various $\Delta_b$ and $\tw{\Delta}_w$
factors required in the definition of the measure (\ref{GrassmannianGravity}). In order 
to calculate $\Delta_b$ and $\tw{\Delta}_w$ the knowledge of the adjacent $\lam{}$ and $\lamt{}$ 
are required. Naively one could think that one has to solve for all internal momenta 
explicitly in order to construct the $\Delta$'s and $\tw{\Delta}$'s. However, the on-shell diagram knows 
about all relations between the internal $\lam{}$'s  and $\lamt{}$'s and the external kinematic data automatically. 
That is the point of constructing the $C$ matrix using the paths and there are simple rules how to read off 
$\Delta_b$ and $\tw{\Delta}_w$ directly from the diagram.

Let us first formulate the rule for the white vertices $\tw{\Delta}_w$ which is defined as a contraction of two 
incoming $\widetilde{\lambda}$ spinors in the vertex,

\begin{equation}
\tw{\Delta}_w = [ \widetilde{\lambda}_A\,\widetilde{\lambda}_B]
\end{equation}

This naively depends on the split of the internal momenta $p_I =\lam{I} \lamt{I}$ into spinors as well as the choice which 
two of the $\widetilde{\lambda}$'s to pick. However the on-shell diagram gives us the correct split automatically similar to 
how it is provided in the delta functions (\ref{GrassmannianGravity}). Furthermore, since the $\widetilde{\lambda}$-spinor is conserved in each vertex --which is exactly
the purpose of the linearized delta functions-- it does not matter which two we pick. Following the 
rules used in the construction of the $C$-matrix, we choose two of the outgoing $\widetilde{\lambda}$. Then we track 
each of them back to the external momenta following the rules:

\begin{center}
\emph{If we hit a black vertex we follow the path, if we hit a white vertex we sum over both paths. At each step we multiply by the edge variables on the way.}
\end{center}

Note that this is exactly how the $C$-matrix is constructed, just that there we start with the incoming external 
legs rather than the legs attached to an internal vertex. In case of closed internal loops, it might be necessary to sum a geometric 
series as in the construction of the $C$-matrix. 

The rule for $\Delta_b$ is similar, it is a contraction of two $\lambda$ spinors,
\begin{equation}
\Delta_b = \la \lambda_A\,\lambda_B\ra.
\end{equation}

Now we choose the two incoming arrows in the black vertex and trace them back to external legs going against the 
arrows rather than following the arrows. This can be trivially understood from the linearized delta functions, the 
$\widetilde{\lambda}$ spinors are coupled to the $C$-matrix but the $\lambda$ spinors are coupled to the $C^\perp$ which 
can be thought of as the $C$-matrix for on-shell diagrams where all black and white vertices as well as all arrows 
are flipped. 

In our example (\ref{GravityPentaBoxExample}), let us start with the white vertices. Following the arrows from the vertex $W_1$ we leave the diagram via the \emph{sinks}, and 
the spinors are,
\begin{equation}
\widetilde{\lambda}_A = \alpha_4 \widetilde{\lambda}_4,\qquad \widetilde{\lambda}_B = \alpha_5\alpha_6(\alpha_3\widetilde{\lambda}_4 + \widetilde{\lambda}_3 + \alpha_2 \widetilde{\lambda}_2)
\end{equation}
corresponding to $\tw{\Delta}_1$,
\begin{equation}
\tw{\Delta}_1 = [\widetilde{\lambda}_A\,\widetilde{\lambda}_B] = -\alpha_4\alpha_5\alpha_6 ([34] + \alpha_2[24])
\end{equation}
Similarly, for the other vertices we get,
\begin{align}
	\tw{\Delta}_2  = \alpha_1(\alpha_3 [24] + [23]), \qquad
	\tw{\Delta}_3  = \alpha_2 [23],\qquad
	\tw{\Delta}_4 =  \alpha_3 ([34]+\alpha_2 [24]).
\end{align}

For the black vertices we just go against the arrows and leave the diagram via the \emph{sources}.
\begin{align}
	\Delta_1 = \alpha_3\alpha_4 \alpha_6 \la 15\ra,\qquad
	\Delta_2 = \alpha_5 \la 15\ra,\qquad
	\Delta_3 = \alpha_1\alpha_2 \alpha_5\alpha_6 \la15\ra 
\end{align}
Collecting all terms in (\ref{GrassmannianGravity}) our formula for the on-shell diagram is (omitting $d\alpha_k$)
\begin{equation}
d\Omega\! =\! \frac{([23]\!+\!\alpha_3[24])^2([34]\!+\!\alpha_2[34])[23]\ab{15}^3}{\alpha_1\alpha_2\alpha_3\alpha_4} 
	  \delta^{(2\times 2)}(C\cdot\lamt{})\delta^{(3\times 2)}(C^{\perp}\cdot\lam{})\delta^{(2\times 8)}(C\cdot \twEta)\,. 
\label{Ex1a}
\end{equation}

Substituting the solutions for the edge variables (\ref{5ptPentaBoxExEdgeVars}), converting the $\delta$-functions and including 
the Jacobians reproduces the same gravity result (\ref{5ptExampleGRForm}) we obtained from gluing three-point amplitudes directly,
\begin{align}
	d\Omega = \frac{[12][23][45]^2}{\ab{12}\ab{23}\ab{34}\ab{45}\ab{51}\ab{13}}
										\delta^{4}(\lam{}\cdot\lamt{})\delta^{16}(\lam{}\cdot\twEta)\,.
\end{align} 

Note that the formula (\ref{Ex1a}) has only single poles in $\alpha_k$ in contrast to the cubic poles in the general 
form (\ref{GrassmannianGravity}). We will expand on this point later in this section.

%
\subsection{More examples}
%

So far we have mostly considered simple MHV examples. Here we would like to stress that our Grassmannian formulation for 
gravity on-shell diagrams is not restricted to the MHV sector but works for arbitrary $k$ as well. To illustrate this point,
let us consider a simple NMHV on-shell diagram,
$$
\raisebox{-57pt}{
\includegraphics[trim={0cm .1cm 0cm 0cm},clip,scale=.45]{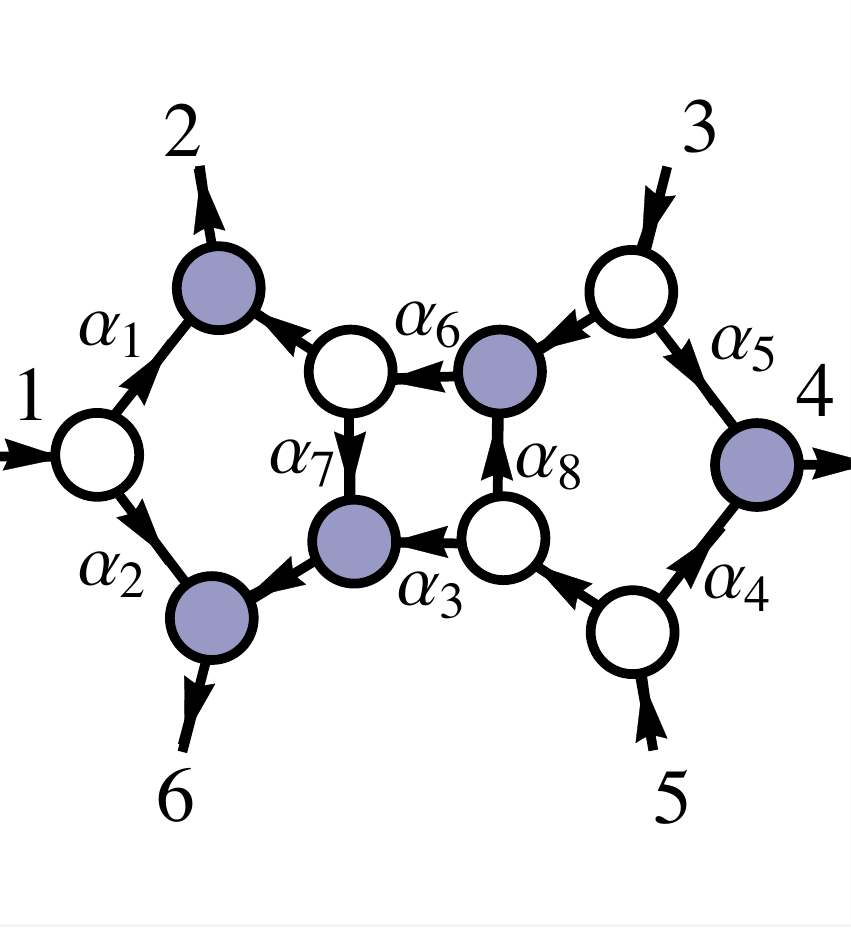}
}\Leftrightarrow
\begin{array}{c}
C = \begin{pmatrix}
      1 & \alpha_1 & 0 & 0 & 0 & \alpha_2 \\
      0 & \alpha_6 & 1 & \alpha_5 & 0 & \alpha_6\alpha_7 \\
      0 & \alpha_6\alpha_8 & 0 & \alpha_4 & 1 & \alpha_3+\alpha_6\alpha_7\alpha_8
    \end{pmatrix} \\
C^{\perp}= \begin{pmatrix}
            -\alpha_1 & 1 & -\alpha_6 & 0 & -\alpha_6\alpha_8 & 0 \\
            0 & 0 & -\alpha_5 & 1 & -\alpha_4 & 0		   \\
            -\alpha_2 & 0 & -\alpha_6 \alpha_7 & 0 & -(\alpha_3+\alpha_6\alpha_7\alpha_8) & 1
           \end{pmatrix}\,.
\end{array}
$$

Here we are going to have additional fermionic $\delta$-functions which exactly give us eight extra powers of $\twEta$ required for NMHV on-shell functions. 
Solving the bosonic $\delta$-functions for the edge variables we find,
\begin{align*}
& \alpha_1 = -\frac{[16]}{[26]}\,, \quad \alpha_2 = \frac{[12]}{[26]}\,, \quad \alpha_3 = \frac{\sab{345}}{\aMs{5}{Q_{345}}{6}}\,, \quad \alpha_4 = \frac{\ab{34}}{\ab{35}}\,, \quad
\alpha_5 = \frac{\ab{45}}{\ab{35}}\,, \\ 
& \twhite{.}\hspace{1.4cm} 
\alpha_6 = \frac{\aMs{5}{Q_{345}}{6}}{\ab{35}\sqb{26}}\,, \alpha_7 = -\frac{\aMs{5}{Q_{345}}{2}}{\aMs{5}{Q_{345}}{6}}\,,\quad
\alpha_8 = - \frac{\aMs{3}{Q_{345}}{6}}{\aMs{5}{Q_{345}}{6}}\,.
\end{align*}
Converting the $\delta$-functions,
\begin{align}
 \delta(C\cdot Z) 
  = \frac{[26]\ab{35}\ \prod^8_{i=1}\delta(\alpha_i\!-\!\alpha^*_i) }{\aMs{5}{Q_{345}}{6} \ab{35}^8[26]^8} \delta^{4}(P)
    \delta^{16}(\Q)\delta^8([26]\twEta_1+[61]\twEta_2+[12]\twEta_6)\,,
\end{align}
and writing all numerator factors $\Delta_{b_i}\,, \tw{\Delta}_{w_j}$ exactly as before, the on-shell function is,
\begin{align}
 d\Omega = \frac{\ab{12}\ab{16} [34][45]\  \delta^8([26]\twEta_1+[61]\twEta_2+[12]\twEta_6)}
		{[12][26][61]\sab{345}\ab{34}\ab{45}\ab{53} \aMs{5}{Q_{345}}{2}\aMs{3}{Q_{345}}{6}} \delta^{4}(P) \delta^{16}(\Q)\,.
\end{align}

As a further example, we can check that our Grassmannian formula for gravity on-shell diagrams also reproduces the correct result 
in cases where the graphs are non-reduced, i.e. contain additional degrees of freedom not localized by the bosonic $\delta$-functions. 
The simplest case to consider is the following,
$$
\raisebox{-51pt}{
\includegraphics[trim={0cm .1cm 0cm 0.1cm},clip,scale=.45]{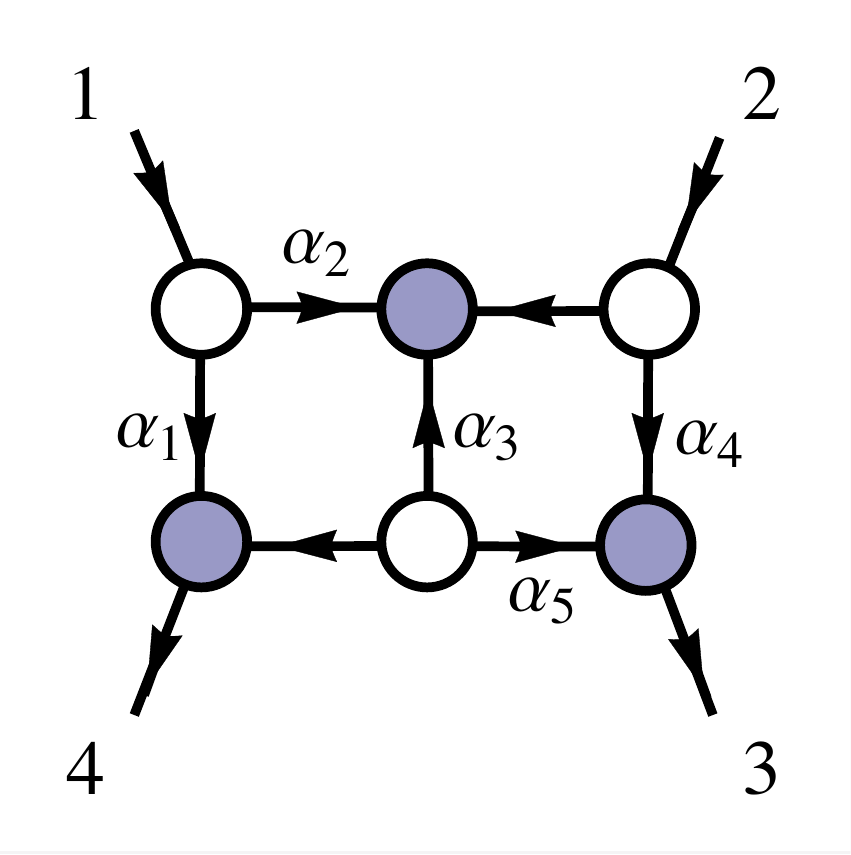}
}\Leftrightarrow
\begin{array}{c}
C = \begin{pmatrix}
      1 & 0 &  \alpha_3 \alpha_4 \alpha_5  & \alpha_1 +\alpha_2\alpha_3\\
      0 & 1 & \alpha_4 + \alpha_3\alpha_5  & \alpha_3
    \end{pmatrix} \\
C^{\perp}= \begin{pmatrix}
	    -\alpha_2\alpha_3\alpha_5 & - (\alpha_4 + \alpha_3\alpha_5) & 1 & 0\\
	    - (\alpha_1 +\alpha_2\alpha_3) & -\alpha_3 & 0 &1
           \end{pmatrix}\,.
\end{array}
$$
Choosing $\alpha_1$ to be the free parameter, we solve for the remaining edge-variables,
\begin{align*}
 \alpha_2 = \frac{\ab{42}-\alpha_1 \ab{12}}{\ab{14}}\,,\quad \alpha_3 = \frac{\ab{14}}{\ab{12}}\,, \quad
 \alpha_4 = \frac{\ab{43}-\alpha_1\ab{13}}{\ab{42}-\alpha_1\ab{12}}\,,\quad \alpha_5 =\frac{\ab{32}}{\ab{42}-\alpha_1\ab{12}}\,.
\end{align*}
As a cross check, we can again look at the Yang--Mills result $d\Omega_{\text{YM}} = \frac{1}{\alpha_1\ab{12}\ab{14}\ab{23}(\ab{43}-\alpha_1\ab{13})}$,
which agrees with the form found earlier in (\ref{2loop4ptOSplanar}) once we identify $\alpha_1\leftrightarrow-z$.

The gravity result can be obtained using our rules from the previous sections,
\begin{align}
 d\Omega = \frac{[24][23][41] }{\alpha_1 \ab{12}\ab{13}\ab{23}\ab{41}(\ab{43}-\alpha_1\ab{13})} \delta^4(P) \delta^{16}(\Q)
\end{align}

All previous examples were in the context of maximal supersymmetry. Here we will explicitly consider a non-supersymmetric case to demonstrate that our Grassmannian formula also holds there. Since the only difference to the maximally supersymmetric theory is the Jacobian $\J$ for on-shell diagrams with a perfect orientation containing closed internal cycles (c.f.~(\ref{GrassmannianGravity})), we look at the simplest diagrams, 
\begin{equation*}
\hskip -.4cm
 \raisebox{-52pt}{\includegraphics[trim={0cm .1cm .1cm .1cm},clip,scale=.45]{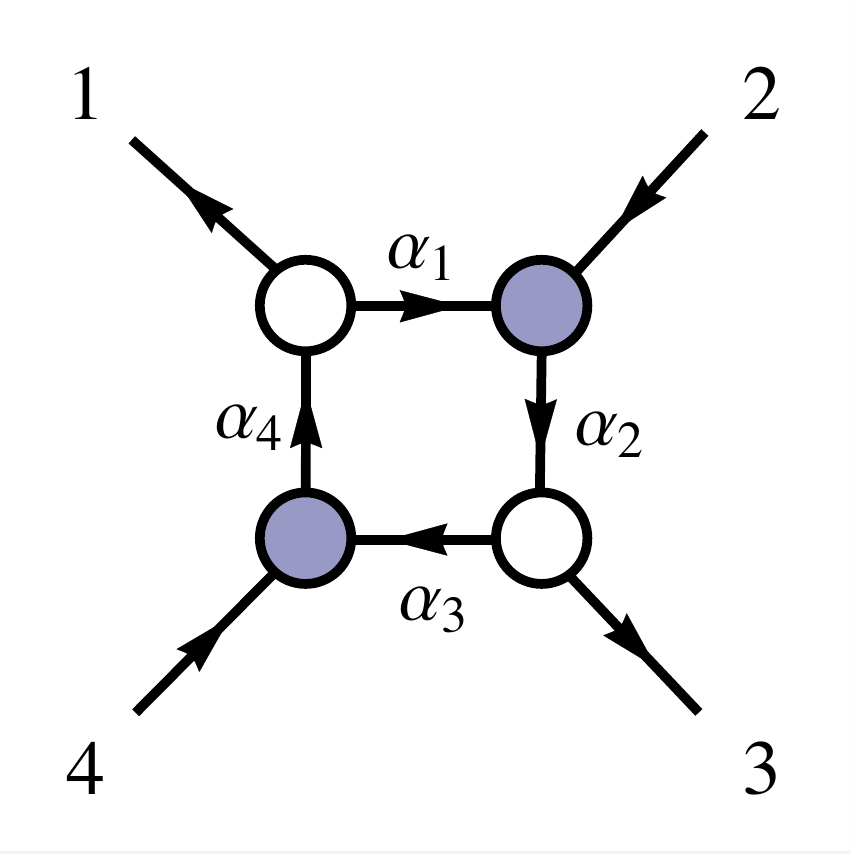}}
\hskip -.8cm 
\Leftrightarrow 
 \begin{array}{c}
 C^{(a)} = \\
 \begin{pmatrix}
		\alpha_2\alpha_3\alpha_4 \delta_a & 1 & \alpha_2 \delta_a & 0\\
		\alpha_4\delta_a & 0 & \alpha_4\alpha_1\alpha_2\delta_a & 1
 \end{pmatrix}
\end{array}
\hskip -.1cm
 \raisebox{-52pt}{\includegraphics[trim={0cm .1cm .1cm .1cm},clip,scale=.45]{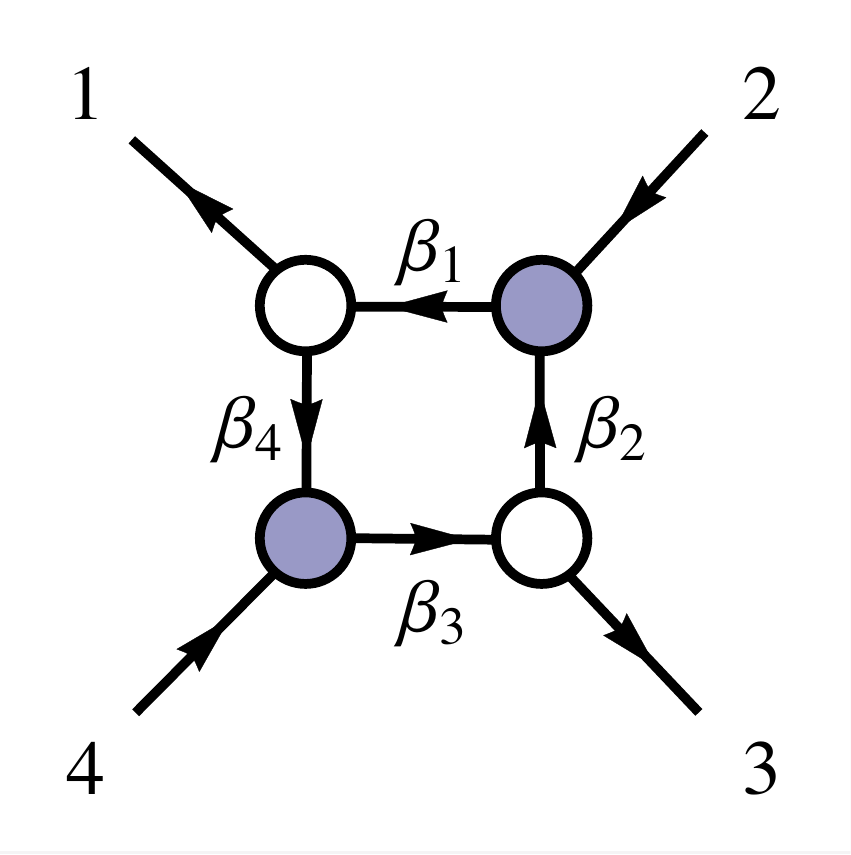}}
\hskip -.8cm 
\Leftrightarrow 
\begin{array}{c}
 C^{(b)} = \\
 \begin{pmatrix}
		\beta_1 \delta_b & 1 & \beta_1\beta_4\beta_3 \delta_b & 0\\
		\beta_3\beta_2\beta_1\delta_b & 0 & \beta_3\delta_b & 1
 \end{pmatrix}
\end{array}
\end{equation*}

As mentioned before, in order to obtain the correct result, we have to sum over all possible orientations of the internal loop which is why we include both diagrams. Introducing the usual short-hand notation for the geometric series $\delta_a = (1-\alpha_1\cdots\alpha_4)^{-1},\ \delta_b = (1-\beta_1\cdots\beta_4)^{-1}$ and solving for the edge variables we find,
\begin{align}
\delta(C^{(a)}\!\!\cdot\! Z) &= \frac{\ab{24}^4\ \delta^4(P)}{\ab{12}^2\ab{34}^2}  
												 \delta\!\left[\alpha_1\!+\!\frac{\ab{23}}{\ab{13}}\right]
												 \delta\!\left[\alpha_2\!-\!\frac{\ab{13}}{\ab{12}}\right]
												 \delta\!\left[\alpha_3\!+\!\frac{\ab{14}}{\ab{13}}\right]
												 \delta\!\left[\alpha_4\!+\!\frac{\ab{13}}{\ab{34}}\right]\,,\\
\delta(C^{(b)}\!\!\cdot\! Z) &= \frac{\ab{24}^4\ \delta^4(P)}{\ab{14}^2\ab{23}^2}  
												 \delta\!\left[\beta_1\!-\!\frac{\ab{13}}{\ab{23}}\right]
												 \delta\!\left[\beta_2\!-\!\frac{\ab{21}}{\ab{13}}\right]
												 \delta\!\left[\beta_3\!-\!\frac{\ab{13}}{\ab{14}}\right]
												 \delta\!\left[\beta_4\!-\!\frac{\ab{34}}{\ab{13}}\right]\,.
\end{align}  
We can easily find the respective numerators and Jacobians $\J$ (\ref{loopJacobian}) required for our gravity formula (\ref{GrassmannianGravity}),
\begin{align*}
	 N^{(a)} = \alpha^2_1 \alpha^2_3 \sab{12}^2\,, \quad \J^{(a)} = 1-\alpha_1\alpha_2\alpha_3\alpha_4\,, \quad
	 N^{(b)} = \beta^2_2 \beta^2_4 \sab{14}^2\,, \quad \J^{(b)} = 1-\beta_1\beta_2\beta_3\beta_4\,,
\end{align*}
to put everything together ($\N=0 \Leftrightarrow \J^{-4}$),
\begin{align}
d\Omega^{\N=0} = \frac{\ab{24}^4}{\ab{13}^4}\! \left(
								  \sab{12}^2 \frac{\ab{12}\ab{34}}{\ab{14}\ab{23}} 
								 \left[\frac{\ab{13}\ab{24}}{\ab{12}\ab{34}}\right]^{-4}\!\!\! 
								+\sab{14}^2 \frac{\ab{14}\ab{23}}{\ab{12}\ab{34}} 
								\left[\frac{\ab{13}\ab{24}}{\ab{14}\ab{23}}\right]^{-4}
								\right)\!\delta^4(P)
\end{align}
which agrees with the formula obtained by simply gluing three-point amplitudes together.  
This serves as a further verification of our Grassmannian formula for gravity on-shell diagrams (\ref{GrassmannianGravity}).
%
\subsection{Structure of singularities}
\label{sec:GravSingStruc}
%

There are two different types of singularities of on-shell diagrams. In terms of edge-variables, these are $\alpha_k \to 0$ or $\alpha_k \to \infty$ 
which correspond to either erasing edges or are associated with poles at infinity when all on-shell momenta in a given 
loop are located at $\ell\rightarrow\infty$. 

Let us show the different cases for the on-shell diagram discussed in previous subsections, and also calculated 
in section 3.1. when we first looked at gravity on-shell diagrams. 
\begin{equation*}
\raisebox{-58pt}
{\includegraphics[trim={0cm .1cm 0cm 0cm},clip,scale=.45]{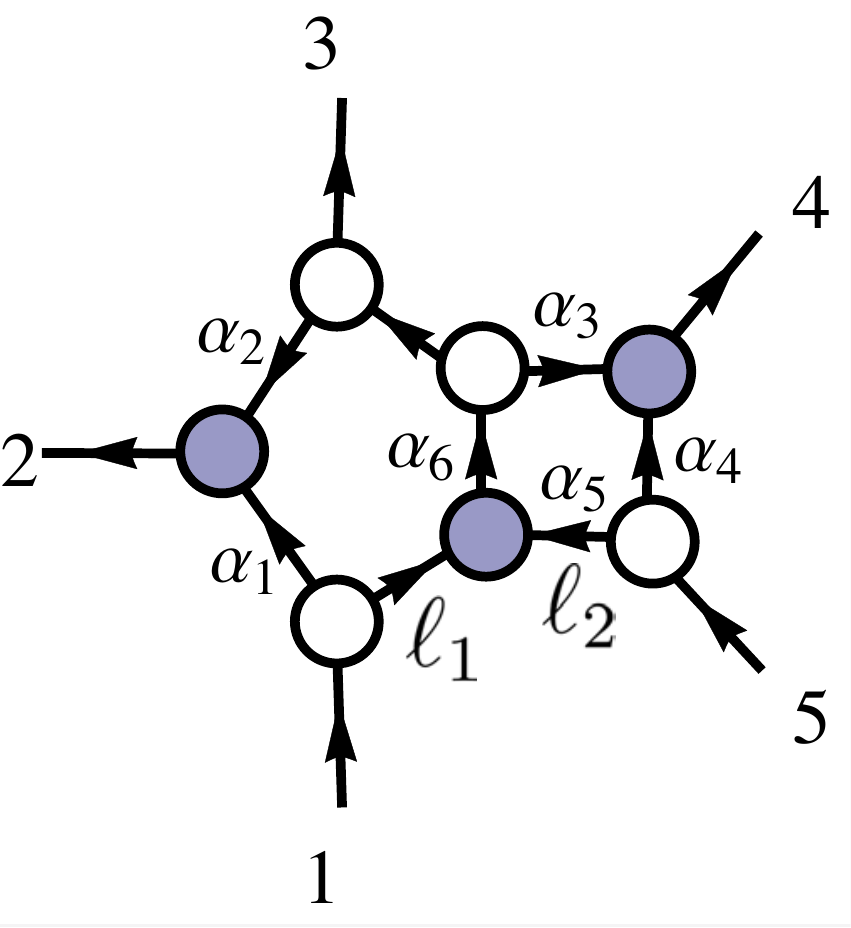}}
\hspace{.5cm}
\begin{array}{ll}
	\ell_1 = \frac{\lam{1}\ Q_{12}\cdot\lam{3}}{\tred{\ab{13}}}\,,		&	
	\ell_2 = \frac{\lam{5}\ Q_{12}\cdot\lam{3}}{\ab{35}}	\,,					\\
	\ell_1-1 = \frac{\tblue{\ab{23}}}{\tred{\ab{13}}}\lam{1}\lamt{2}\,,			  &
	\ell_2-5 = \frac{\tmagenta{\ab{34}}}{\ab{35}} \lam{5} \lamt{4}	\,,					\\
	\ell_1-Q_{12} = \frac{\tgreen{\ab{12}}}{\tred{\ab{13}}} \lam{3} \lamt{2}\,, & 
	\ell_2-Q_{45} = \frac{\tburgundy{\ab{45}}}{\ab{35}} \lam{3}\lamt{4} 			\,,		\\
	\ell_1-Q_{123}=\frac{\lam{3}\ Q_{23}\cdot\lam{1}}{\tred{\ab{13}}}\,,		& 
	\ell_1+\ell_2 = \frac{\tcyan{\ab{15}}}{\tred{\ab{13}}\ab{35}} \lam{3}\ Q_{12}\cdot\lam{3}\,. \\
	\end{array}
\end{equation*}
\vskip -.9cm
\begin{equation*}
 \alpha_1 = \frac{\tblue{\ab{23}}}{\tred{\ab{13}}}\,, \quad  
 \alpha_2 = \frac{\tgreen{\ab{12}}}{\tred{\ab{13}}}\,, \quad
 \alpha_3 = \frac{\tburgundy{\ab{45}}}{\ab{35}}\,,	\quad
 \alpha_4 = \frac{\tmagenta{\ab{34}}}{\ab{35}}\,, \quad
 \alpha_5 = \frac{\tred{\ab{13}}}{\ab{35}}\,, \quad
 \alpha_6 = \frac{\ab{35}}{\tcyan{\ab{15}}}\,.
\end{equation*}

Here we can see that four of the edge variables, $\alpha_1$, $\alpha_2$, $\alpha_3$ and $\alpha_4$, directly parametrize 
the momentum flow in a given edge. If we send one of them to zero, the zero momentum flow effectively erases that edge. 
Similarly, sending $\alpha_6\rightarrow\infty$ erases the corresponding $(\ell_1\!+\!\ell_2)$-edge. Whether the location of the pole is at $0$ or $\infty$ 
is determined by the orientation of the arrow on the edge, flipping the orientation of the arrow inverts the edge variable 
$\alpha_k\rightarrow\frac{1}{\alpha_k}$ and the location of the pole changes. Independent of the details of the orientation, the 
important statement is that all of the discussed edges are erasable by sending $\alpha_k \to 0\text{ or }\infty$. 
Note that the edge corresponding to $\alpha_5$ is not erasable. The reason 
is that if we tried to erase this edge, the remaining diagram would enforce both $[45]=\ab{13}=0$ which imposes too many constraints. 
In fact, sending $\alpha_5\to 0$ or $\infty$ blows up one of the loops with $\ell_1\to\infty$ or $\ell_2\to\infty$. 
The same happens if we set $\alpha_1$, $\alpha_2$, $\alpha_3$, $\alpha_4$ to infinity or $\alpha_6$ to zero. 
In the example above, we have already chosen a particular $GL(1)_v$ gauge-fixing, corresponding to the fact that some edge-variables are set to $1$. 
For a different gauge-fixing we could analyze these edges as well, leading to the same set of erasable edges described above.

In the case of $\NeqFour$ theory the form is logarithmic in all edge variables independent whether an edge is erasable or not. 
Furthermore, the final expression does not contain any poles that send loop-momenta to infinity so that all singularities correspond to erasing edges only. 
This is an important distinction to $\NeqEight$ where poles at infinity do appear.

Let us investigate the properties of our Grassmannian form for gravity on-shell diagrams a little more closely.  
First, it is relatively easy to see that the form (\ref{GrassmannianGravity}) has only linear poles for $\alpha_k \to 0$, when the 
corresponding edge is erasable. The denominator contains the third power of this edge variable, $\alpha_k^3$ but the numerator 
always generates two powers leaving only a single pole. We remove the erasable edge in the on-shell diagram for $\alpha_k \to 0$ 
if the arrow points from a white to a black vertex, while it is erased by $\alpha_k \to \infty$ if the arrow points from a black to a white 
vertex. The edges between same colored vertices are never removable. 

$$
\includegraphics[scale=.45]{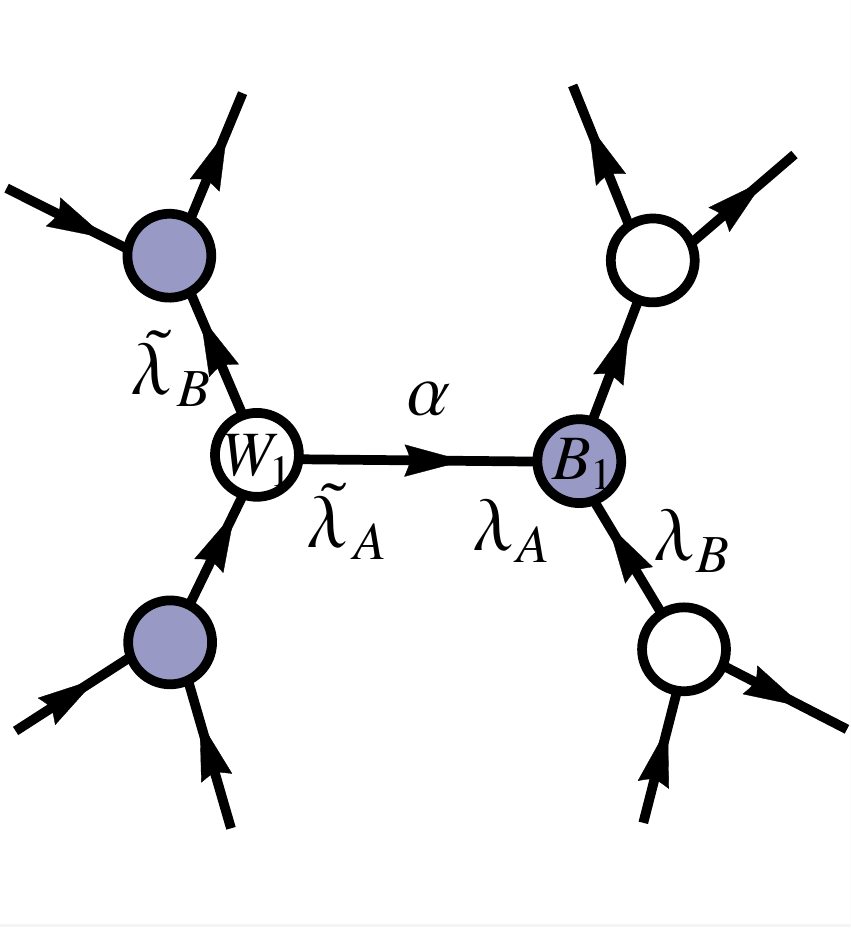}
$$

The numerator for such subgraph is given by the products of $\Delta_b$ and $\tw{\Delta}_w$. Based on our rules, we have 
$\Delta_{b_1}=\ab{\lam{A}\lam{B}}\sim \alpha$ and $\tw{\Delta}_{w_1}=[\lamt{A}\lamt{B}]\sim \alpha$, while all other $\Delta_b$ and $\tw{\Delta}_{w}$ 
do not depend on $\alpha$. Therefore the numerator generates $\sim \alpha^2$. We can also consider a modification of the subgraph 
by adding another white vertex (or in general a chain of white vertices), or consider some more distant vertex and look if they can 
possibly generate additional $\alpha$ factors in the numerator,

$$
\includegraphics[scale=.45]{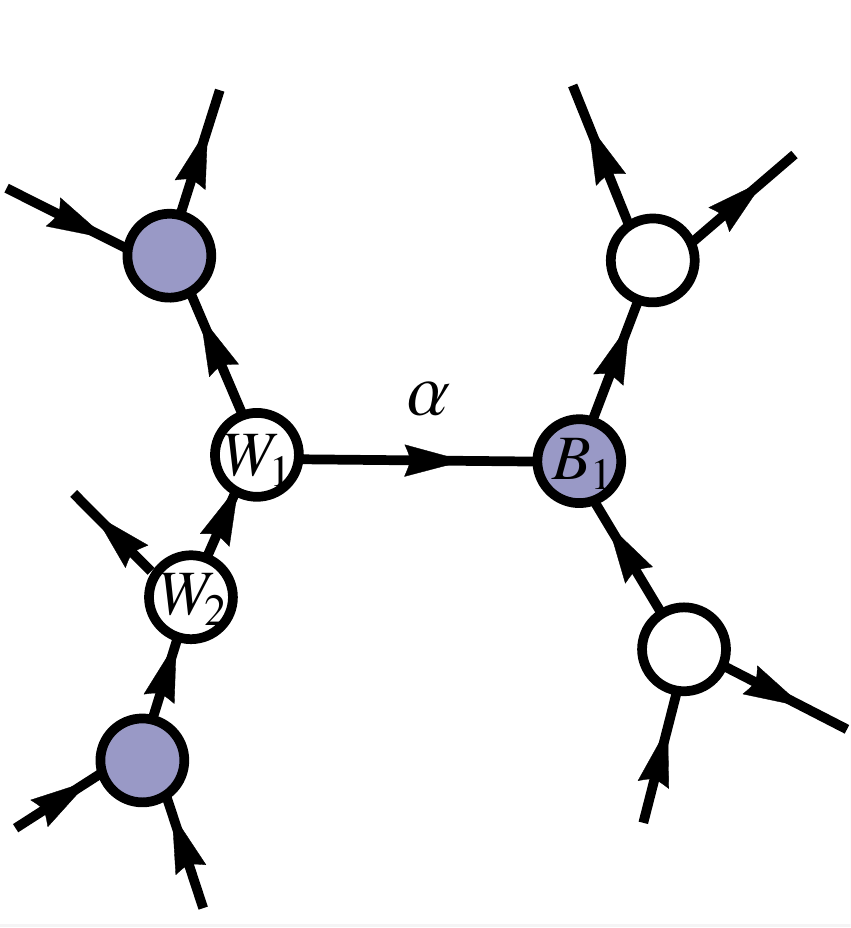}\qquad
\includegraphics[scale=.45]{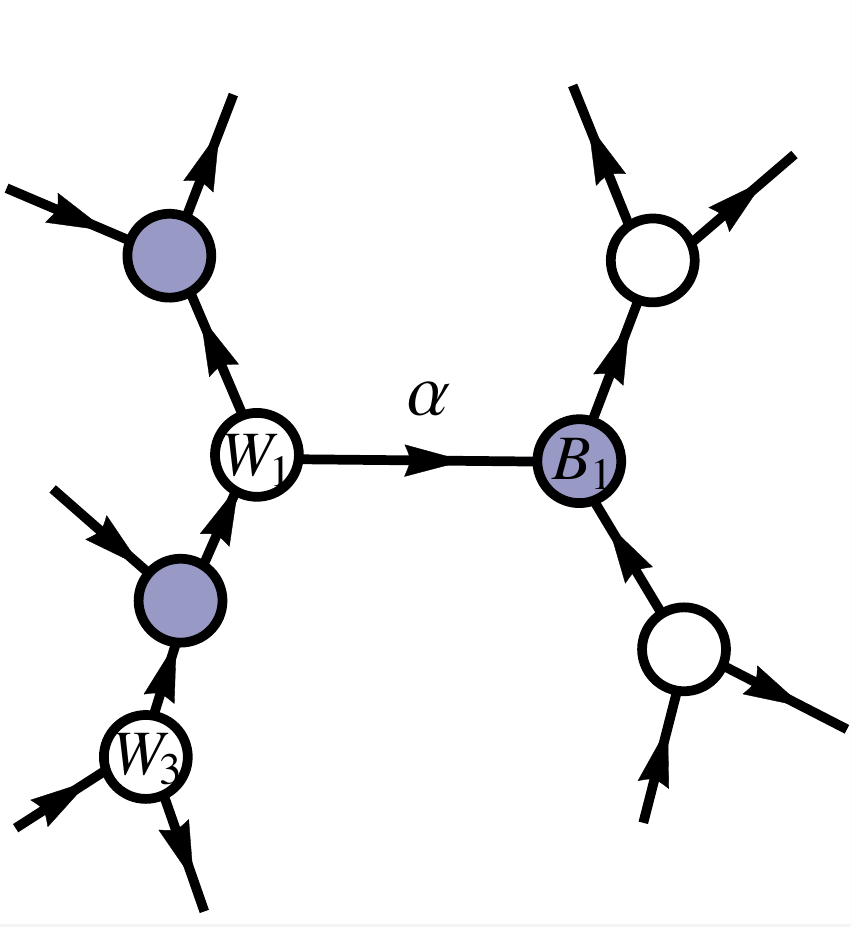}
$$

In both cases the numerator will have further $\alpha$-dependence but in either situation, it will look like $\widetilde{\Delta}_{w2},
\widetilde{\Delta}_{w3}\sim \alpha(\dots) + (\dots)$ and the linearity of the pole in $\alpha$ is not changed. The argument for 
erasable edges would be similar when the arrow points from black to white vertex. The only difference is that we have to keep track of the pole 
$\alpha\to\infty$ but we would again find a linear pole only. Alternatively, we can take the same diagram and consider 
a different perfect orientation in which the arrow again points from white to black so that the pole is localized at zero. As a result, 
all poles corresponding to erasable edges are linear. This immediately implies that all higher poles (including some simple poles) 
correspond to poles at infinity, when internal on-shell momenta in one or more loops are sent to infinity. 

\subsubsection*{Bubbles} 

Let us comment on one important property of gravity on-shell diagrams which is a trivial consequence of the formula (\ref{GrassmannianGravity}): 
any internal bubble {\bf vanishes}. Let us consider a diagram with an internal bubble.

$$
\includegraphics[scale=.45]{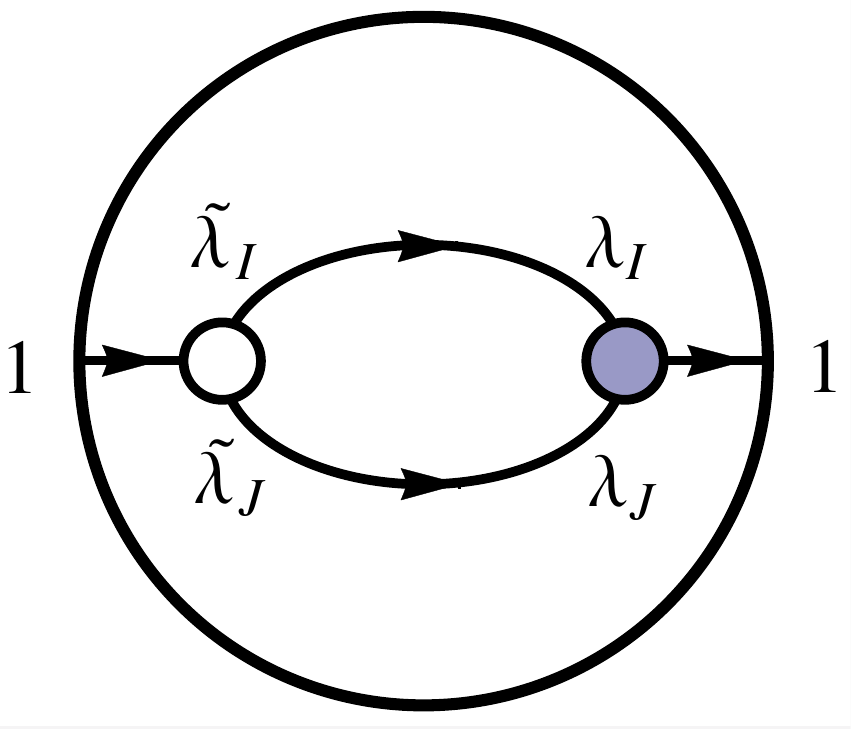}
$$

Independent of the rest of the diagram, the perfect orientation chosen and the directions of arrows, the numerator factors $\Delta_b$ 
and $\widetilde{\Delta}_w$ vanish for both vertices separately. All $\lamt{}$'s in the black vertex are proportional, so are all $\lambda$'s 
in the white vertex, which implies that $\lambda_1\sim \lambda_I\sim \lambda_J$ and $\widetilde{\lambda}_1\sim
\widetilde{\lambda}_I\sim \widetilde{\lambda}_J$ and $\Delta_b=\widetilde{\Delta}_w=0$. 

This fact will have dramatic consequences on properties of loop amplitudes. We will discuss them in greater detail in the next section. Furthermore, 
there is one interesting aspect related to vanishing bubbles: in planar $\NeqFour$, the loop integrand is expressed in terms of on-shell diagrams containing bubbles. 
In fact, via equivalence moves, one can show that four bubbles built the four degrees of freedom of the off-shell loop momentum for each loop \cite{OnshellDiagrams}. 
We do not have any recursion relations in the gravity case (as well as in $\NeqFour$ beyond the planar limit) but if such formulation exists, it must take this fact into account. In the planar case we could always use the identity moves to eliminate the bubble from the diagram in the form (see figure above). The non-planar identity moves for ${\cal N}=8$ SUGRA (and also non-planar ${\cal N}=4$ SYM) are different which might lead to a different role of  bubbles in the loop integrand.

\section{From on-shell diagrams to scattering amplitudes}

In the last sections we initiated a detailed study of gravity on-shell diagrams and gave their Grassmannian representation. This formula (\ref{GrassmannianGravity}) exhibited 
some interesting properties: (a) higher poles associated with sending internal momenta to infinity and (b) vanishing whenever three momenta in any vertex become collinear. 
As it was stressed several times, the on-shell diagrams represent cuts of loop integrands and they contain a considerable amount of information about 
the structure of loop amplitudes themselves even though we do not yet know how to express the integrand directly in terms of on-shell diagrams. There are two obvious paths beyond 
the well-understood case of planar $\NeqFour$ theory: (i) going to lower supersymmetry or (ii) going non-planar. 
The recursion relations for planar non-supersymmetric Yang-Mills theory suffers from divergencies in the forward limit term. Resolving that problem is an 
active area of research \cite{Benincasa:2015zna} and it appears to be a question of properly defining the forward limit term in these theories rather than some fundamental obstruction. 

The extension to non-planar theories, even with maximal supersymmetry, seems more difficult because it is not even clear which object should recurse 
in the first place. Beyond the planar limit we do not have global variables and loop momenta are normally associated with individual diagrams in the 
Feynman expansion, or its refined version using a set of integrals in the unitarity method. Therefore it is not clear how to associate the "loop-momentum" degrees of freedom
with those in on-shell diagrams or how to cancel spurious poles. Making progress on this problem would certainly open doors 
to many new directions of research. However, even without having the recursion relations at hand, there is an immediate question one can ask: 

\begin{center}
{\it Does the loop amplitude have the same properties as individual on-shell diagram?}
\end{center}

This analysis was done in particular examples for amplitudes in full non-planar $\NeqFour$ theory and the answer is positive \cite{Log,ThreeLoopPaper,Bern:2015ple}. Additionally, many of the 
structures present in the planar limit seem to survive in non-planar amplitudes despite the absence of good kinematic variables. We review this 
progress in the next subsection and then motivated by this success we will test the properties found for gravity on-shell diagrams on explicit 
expressions for gravity amplitudes. 

%
\subsection{Non-planar $\NeqFour$ amplitudes}
%

In the $\NeqFour$ case we are able to take the step to non-planar amplitudes. On one hand we have a detailed 
understanding of the planar sector of the theory and the properties of the amplitudes: logarithmic singularities, dual conformal \cite{DualConformalMagic,
  Alday:2007hr, Drummond:2008vq} and Yangian \cite{Drummond:2009fd}
symmetries as well as the Amplituhedron \cite{Arkani-Hamed:2013jha} construction. On the other hand, we have the non-planar on-shell diagrams which have logarithmic singularities and for MHV leading 
singularities we even know that they are expressed in terms of planar ones. 

All these ingredients lead to the following conjectures \cite{Log,ThreeLoopPaper,Bern:2015ple}:

\begin{itemize}
\item The loop amplitudes have only {\bf logarithmic singularities}, as in the planar limit. For $k>4$ (perhaps even for lower $k$) we expect 
the presence of elliptic cuts but at least for $k=2$ the logarithmic singularities must be present directly in momentum space.
\item There are {\bf no poles at infinity}. This was one of the consequence of the dual conformal symmetry of planar amplitudes, but also 
motivated by the observation about MHV leading singularities.
\end{itemize}

These conjectures were tested in \cite{Log,ThreeLoopPaper,Bern:2015ple} on the four-point amplitudes at two- and three-loops, and on the five-point amplitude at two-loops. 
These tests rely on a two-step process. First one constructs the basis of integrals ${\cal I}_k$ with the above two properties (also with unit leading singularities) and second one expands
the loop amplitudes in this basis. The correctness of the result is guaranteed by satisfying all unitarity cuts.
\begin{equation}
{\cal A} = \sum_k c_k {\cal I}_k
\label{Expan}
\end{equation}

As was argued in \cite{ThreeLoopPaper,Bern:2015ple} this is a strong evidence for a new hidden symmetry (analogue of dual conformal symmetry) in the full $\NeqFour$ theory. 

Finally, the step towards the geometric Amplituhedron-like construction was also made in \cite{Bern:2015ple}. The presence of logarithmic singularities only was one of the 
ingredients of the Amplituhedron where the $\dlog$ forms can be thought of as volumes in the Grassmannian. Moreover, motivated by the work 
\cite{Arkani-Hamed:2014dca} it was checked that all coefficients $c_k$ in (\ref{Expan}) can be fixed only from vanishing cuts. This means that the full amplitude 
is fixed entirely by homogeneous conditions providing nontrivial evidence for an Amplituhedron-type geometric formulation.

Motivated by this success we can now turn to gravity and see what structures we can carry over from on-shell diagrams directly to the amplitude. 
In particular, we want to test two statements:

\begin{itemize}
\item All singularities are logarithmic unless it is a pole at infinity.
\item The amplitude {\bf vanishes} on all collinear cuts. 
\end{itemize}

The first statement is motivated by the singularity structure of gravity on-shell diagrams described in Sec.~\ref{sec:GravSingStruc}. There, we saw that certain single poles 
correspond to erasable edges, and all higher poles are associated with sending internal momenta to infinity. The second statement is the crucial ingredient in the Grassmannian 
formula (\ref{GrassmannianGravity}) and checking it for gravity amplitudes will be a main result of this section.

%
\subsection{Gravity from Yang-Mills}
%

The relation between scattering amplitudes in Yang-Mills theory and gravity has been a long standing area of research starting by the work of Kawai, Lewellen and Tye (KLT) \cite{KLT}, 
to the recent discovery of Bern, Carrasco and Johansson (BCJ) \cite{BCJ,BCJLoop}. The BCJ-relations state that there exists a representation of the Yang-Mills amplitude (with or without 
supersymmetry) in terms of cubic graphs, 
\begin{equation}
{\cal A}_{YM} = \sum_{i\in \text{cubic}} \frac{n_ic_i}{s_i}
\label{BCJ1}
\end{equation}

where $n_i$ are kinematic numerators, $c_i$ are color factors and $s_i$ is the denominator of the cubic graph given by Feynman propagators BCJ \cite{BCJ,BCJLoop} states that 
whenever the color factors $c_i$ satisfy the Jacobi identity $c_i+c_j = c_k$ then the numerators satisfy the same relation $n_i+n_j=n_k$. Once we have 
(\ref{BCJ1}) the gravity amplitude can be then obtained by the simple formula\footnote{There is a natural identification of coupling constants 
which does not play a role in our discussion and we suppress them altogether.},
\begin{equation}
{\cal M}_{GR} = \sum_{i\in \text{cubic}} \frac{n_i\widetilde{n}_i}{s_i}
\label{BCJ2}
\end{equation}

where the set of numerators $\widetilde{n_i}$ do not necessarily have to satisfy the Jacobi relation, i.e. they can belong to a non-BCJ 
representation of the Yang-Mills amplitude. If we start with two copies of $\NeqFour$ then we obtain an $\NeqEight$ amplitude.  
There is a dictionary for the squaring relations between amplitudes in lower supersymmetric theories with different matter content (see e.g. \cite{Johansson:2015oia}) 
and even for some effective field theories \cite{Chen:2013fya}. The BCJ-relations are a conjecture which was proven for tree-level amplitudes and tested up 
to high loop order for loop amplitudes, there it is a statement about integrands. 

In order to prove that the amplitudes in $\NeqEight$ have only logarithmic singularities (except poles at infinity) we first assume the loop 
BCJ-relations (\ref{BCJ2}) and also the statement that the $\NeqFour$ amplitudes can always be expressed in (\ref{Expan}) where all basis 
integrals ${\cal I}_k$ have only logarithmic singularities. This is certainly true up to high loop order \cite{Log,ThreeLoopPaper,Bern:2015ple} and it is reasonable to assume it 
holds to all loops. Then we can use one copy of the Yang-Mills amplitude written in this manifest $\dlog$ form, and the other copy written in the BCJ-form 
(\ref{BCJ1}). The gravity amplitude is then given by (\ref{BCJ2}). While the numerator in the $\dlog$ form $\widetilde{n_i}$ already guarantees that 
term-by-term all singularities are logarithmic in the Yang-Mills amplitude, then the expression (\ref{BCJ2}) will also have only logarithmic singularities 
term-by-term. This is not true for poles at infinity as adding the extra numerator $n_i$ introduces further loop momentum dependence in the numerator, but for finite $\ell$ all singularities 
stay logarithmic. This argument was already used in \cite{ThreeLoopPaper} but we repeat it here because it is in perfect agreement with the results we get from the gravity 
on-shell diagrams.

Let us comment on the poles at infinity explicitly. The on-shell diagrams have higher poles at infinite momentum and this is what we also expect from the BCJ-form (\ref{BCJ2}) 
as adding two copies of $n_i$ increases the power counting in the numerator. Indeed, looking at the explicit results we can see that the loop amplitudes in $\NeqEight$ 
do have poles at infinity. The simplest example is the 3-loop four-point amplitude. The cut represented by the following (non-reduced) on-shell diagram,
\begin{equation}
\raisebox{-52pt}{
\includegraphics[scale=.45]{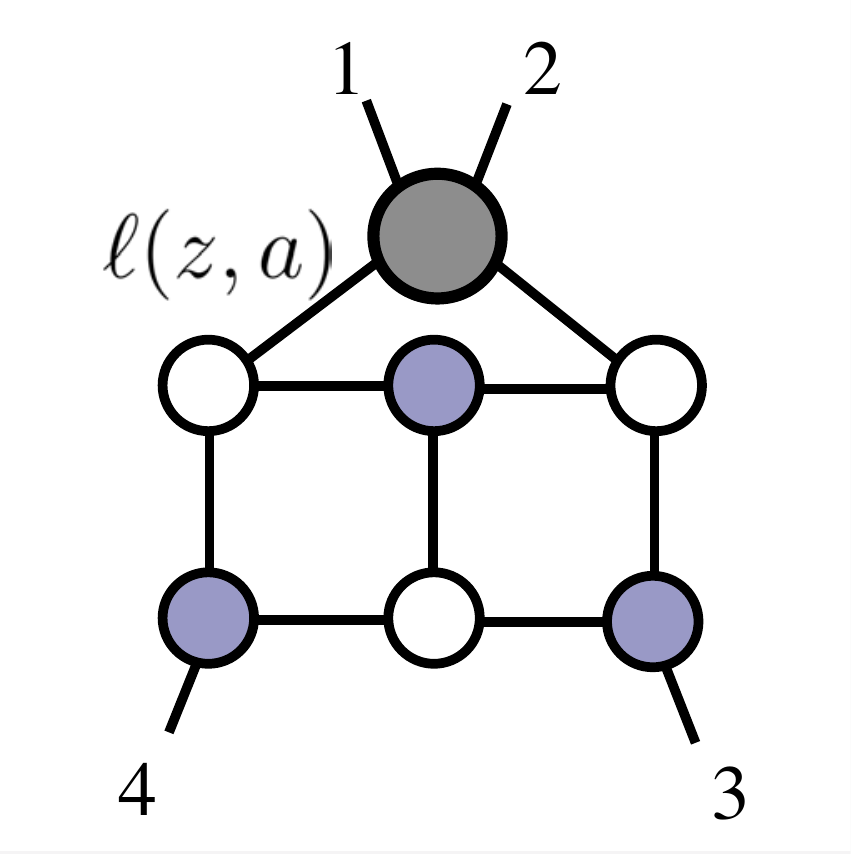}}
\stackrel{a\to0}{\longrightarrow} 
\raisebox{-52pt}{
\includegraphics[scale=.45]{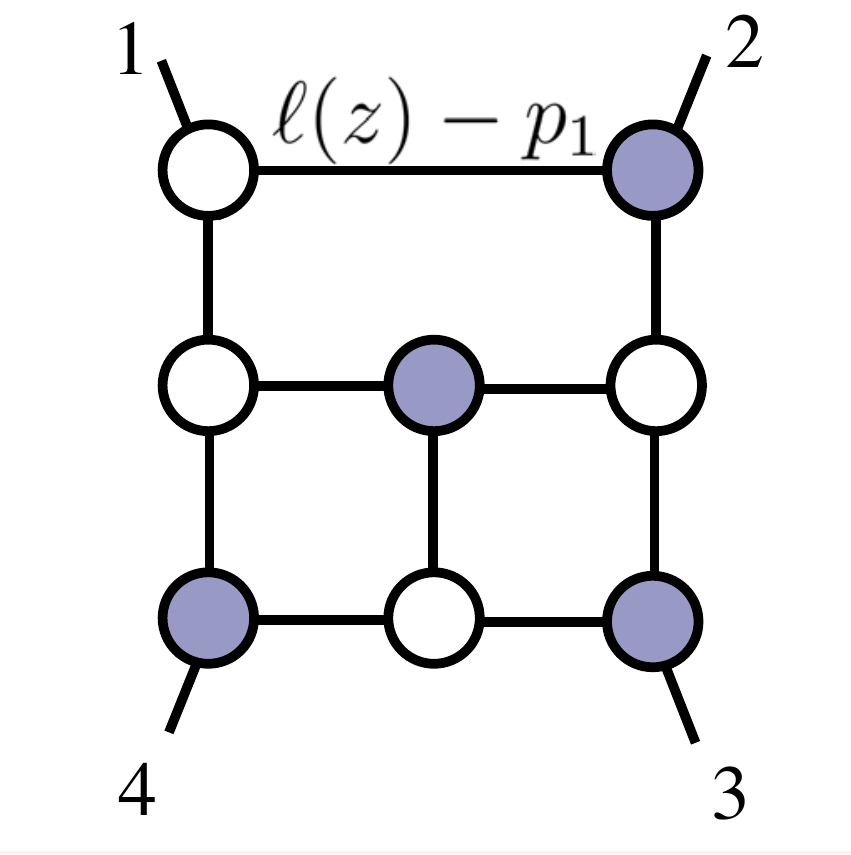}}
\sim \int \frac{dz}{z} \times F(\xcancel{z})\,,
\label{tennisCourtPoleInfty}
\end{equation}

has a pole at $z\to\infty$, corresponding to $\ell\to\infty$. The detailed expression for the $z$-independent function $F(\xcancel{z})$ 
is not particularly illuminating but can be obtained by either gluing together 
tree-amplitudes or by evaluating the known representation of the gravity amplitude \cite{Manifest3} on the cut.
Starting with the cut on the left hand side of (\ref{tennisCourtPoleInfty}), the relevant loop momentum $\ell$ is parameterized by two degrees of freedom, $a$ and $z$, 
$$
  \ell(a,z) = (1-a)\lam{1}\lamt{1} + a \lam{2}\lamt{2} + \frac{a(1-a)}{z} \lam{2}\lamt{1} + z \lam{1}\lamt{2}\,.
$$

By localizing $a\to0$, we go to the maximal cut and select a unique contribution where no further cancellations are possible. Since we are on the maximal cut, 
the gravity numerator in the diagrammatic expansion of the amplitude can be obtained by squaring the respective $\NeqFour$ numerator of any representation and we take \cite{ThreeLoopPaper},
$$
 N^{\text{GR}}\Big|_{\text{cut}} \sim  stu \M^{\text{tree}}_4 \cdot \Big[s(\ell+p_4)^2\Big]^2\Big|_{\text{cut}}\,,
$$

where $stu \M^{\text{tree}}_4=\left(\frac{\sqb{34}\sqb{41}}{\ab{12}\ab{23}}\right)^2$ is the totally crossing symmetric prefactor depending on external kinematics only.
The important observation is that the integrand in (\ref{tennisCourtPoleInfty}) behaves like $\frac{dz}{z}$ leading to the pole at infinity in $\ell(z)\to \infty$. 
At higher loops we even get multiple poles at infinity \cite{ThreeLoopPaper}. In general, poles at infinity 
can indicate potential UV-divergencies after integration as is the case for the bubble integral. However, a direct association of poles at infinity with a UV-divergence is not 
possible. The triangle integral for example also has a pole at infinity but it is UV-finite. Finding a precise rule between the interplay of poles at infinity and the UV-behavior 
of gravity amplitudes is an active area of research and would have a direct bearing on the UV-finiteness question of $\NeqEight$ \cite{Bern:2006kd}.

\subsection{Collinear behavior}
\label{subsec:CollBehavior}

Based on the numerator factors in the Grassmannian formula for gravity on-shell diagrams (\ref{GrassmannianGravity}) it is natural to conjecture 
that the residue of loop amplitudes on cuts that involve a three-point vertex (where the grey blob is any tree or loop amplitude),
$$
\includegraphics[trim={0cm .1cm 0cm 0cm},clip,scale=.45]{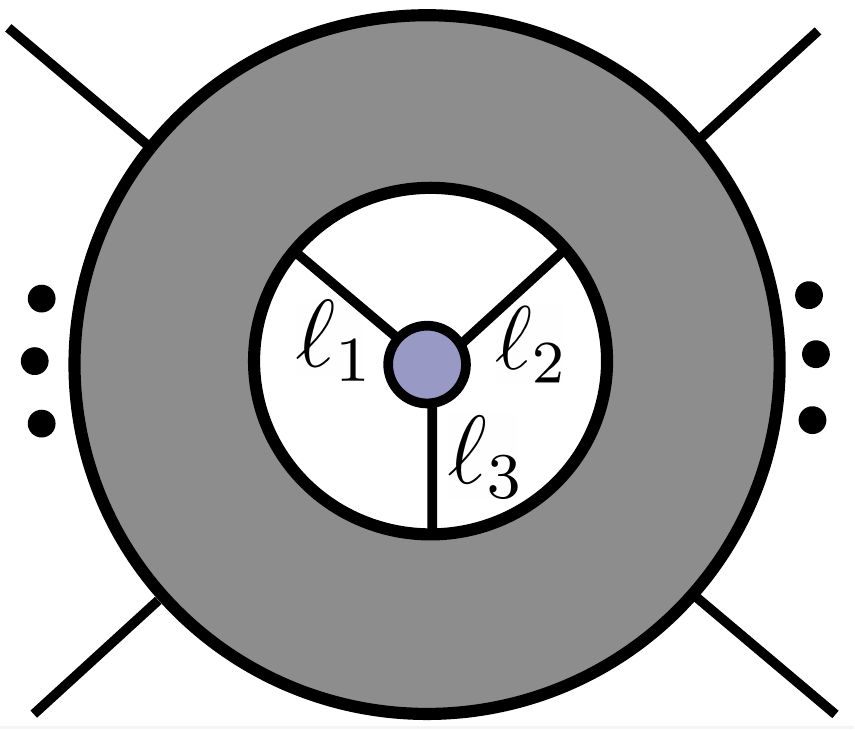} \hspace{2cm}
\includegraphics[trim={0cm .1cm 0cm 0cm},clip,scale=.45]{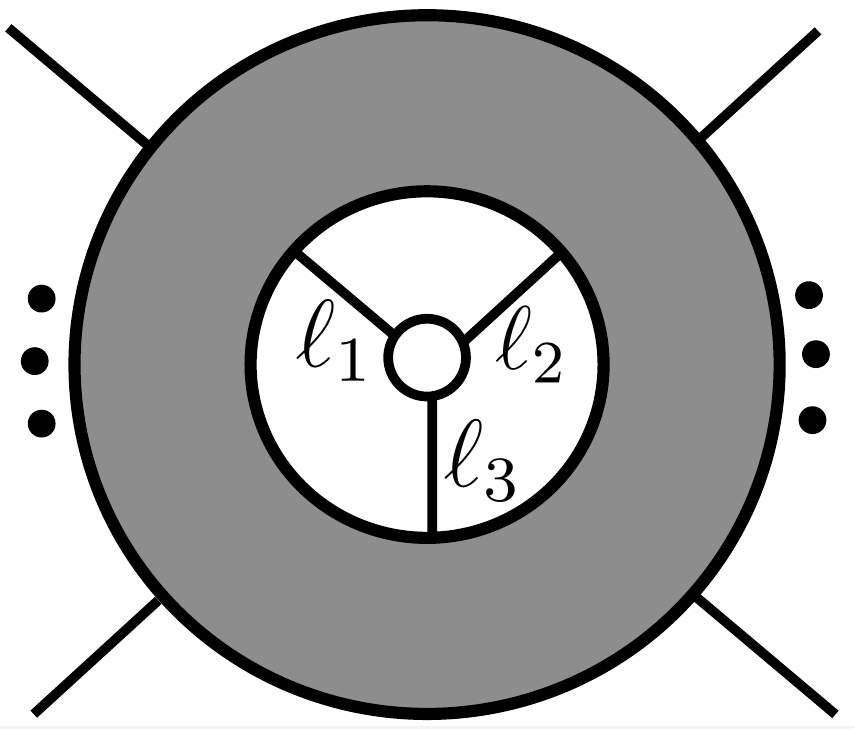}
$$
factorize in a particular way, 
\begin{align}
{\cal M} = \la \ell_1\,\ell_2\ra \cdot {\cal R} \quad&\mbox{for MHV vertex, i.e. $\widetilde{\lambda_{\ell_1}}\sim\widetilde{\lambda_{\ell_2}}\sim\widetilde{\lambda_{\ell_3}}$}\,,\\ 
{\cal M} = [\ell_1\,\ell_2]\cdot \overline{{\cal R}}\quad&\mbox{for $\MHVbar$ vertex, i.e. $\lambda_{\ell_1}\sim\lambda_{\ell_2}\sim\lambda_{\ell_3}$}\,,
\end{align}

where ${\cal R}$ and $\overline{\cal R}$ are functions regular in $\ab{\ell_1\ell_2}$ and $\sqb{\ell_1\ell_2}$ respectively. 
If both $\ell_1$ and $\ell_2$ are external particles this reduces to the well known behavior of gravity amplitudes in the collinear limit \cite{Bern:1998xc,Bern:1998sv}, 
\begin{equation}
{\cal M} \sim \frac{[12]}{\la12\ra}\cdot \widetilde{{\cal M}} \quad\mbox{for $\ab{12}\to 0$},\qquad {\cal M} \sim \frac{\la12\ra}{[12]}\cdot \widetilde{{\cal M}} \quad\mbox{for $[12]\to 0$}\,.
\end{equation}

Let us stress that our claim is more general as one or both of the $\ell_k$ can be loop momenta and there is no such statement available in the literature. 
It is fair to say that this statement does not follow from formula (\ref{GrassmannianGravity}) for on-shell diagrams but it is rather motivated 
by it. The reason is that the lower cuts can not be directly written as the sums of on-shell diagrams. There are some extra $1/s_{ij}$ factors 
one has to add when going from on-shell diagram to generalized cuts, and therefore our statement does not immediately apply to the other cuts. 
If we calculate the residue of the amplitude on the cut when the three point amplitude (say MHV) factorizes then this piece factorizes 
$\la \ell_1\ell_2\ra$ but it is not guaranteed that the rest of the diagram does not give additional $\frac{1}{\la \ell_1\ell_2\ra}$ and 
cancel this factor. This does not happen in the case of on-shell diagrams but it could for generalized cuts. Our conjecture is that indeed 
it does not happen and any cut of the amplitude of this type would be proportional to $\la \ell_1\ell_2\ra$. 
We will test this conjecture explicitly on several examples.

%
\subsubsection*{Four point one-loop}
%

The four-point one-loop $\NeqEight$ amplitude was first given by Green, Schwarz and Brink \cite{Green:1982sw} as a sum of three box integrals\footnote{The gravitational 
coupling constant $(\kappa/2)^{n-2}$ for n-pt tree level amplitudes and $(\kappa/2)^n$ for n-pt one-loop amplitudes will be suppressed ($\kappa=\sqrt{32\pi G_N}$).},
\begin{align}
\label{eqn:M41loop}
 \M^{1}_4(1234) = i stu \M^{\text{tree}}_4(1234) \Big[I^1_4(s,t)+I^1_4(t,u)+I^1_4(u,s)\Big]\,,
\end{align}

where the corresponding tree amplitude $ \M^{\text{tree}}_4(1234)$ carries the helicity information. Multiplying by $stu$ one finds the totally permutation invariant 
four-point gravity prefactor, see e.g.~\cite{BjerrumBohr:2005xx},
\begin{align}
 stu \M^{\text{tree}}_4(1234) = \underbrace{\left(\frac{\sqb{34}\sqb{41}}{\ab{12}\ab{23}}\right)^2}_{\equiv \K_8}\,.
\end{align}

The one-loop box integrals $I^1_4(\_,\_)$ are defined without the usual $st$-type normalization which was put into the permutation invariant prefactor $\K_8$. All integrals 
have numerator $N=1$ and therefore do not have unit leading singularity $\pm1,0$ on all residues,
\newline \twhite{.}
\begin{equation}
\twhite{.}\hspace{1.4cm}
\raisebox{-25pt}{
  \begin{fmfchar*}(22,20)
    \fmfforce{(sw)}{one}
    \fmfforce{(nw)}{two}
    \fmfforce{(ne)}{three}
    \fmfforce{(se)}{four}
    \fmfforce{(0w,.5h)}{leftV}
    \fmfforce{(.1w,.1h)}{dl}
    \fmfforce{(.1w,.9h)}{ul}
    \fmfforce{(.9w,.9h)}{ur}
    \fmfforce{(.9w,.1h)}{dr}
    \fmf{plain}{dl,one}
    \fmf{plain}{ul,two}
    \fmf{plain}{ur,three}
    \fmf{plain}{dr,four}
    \fmf{fermion,label=$\ell$,label.side=left}{dr,dl}
    \fmf{plain}{dl,ul}
    \fmf{plain}{ul,ur}
    \fmf{plain}{ur,dr}
    \fmfv{label=$I^1_4(s\,,;t)=$}{leftV}
    \fmfv{label=$1$}{one}
    \fmfv{label=$2$}{two}
    \fmfv{label=$3$}{three}
    \fmfv{label=$4$}{four}
    %
   \fmffreeze
 \end{fmfchar*} 
 \hspace{2cm}
 \begin{fmfchar*}(22,20)
    \fmfforce{(sw)}{one}
    \fmfforce{(nw)}{two}
    \fmfforce{(ne)}{three}
    \fmfforce{(se)}{four}
    \fmfforce{(0w,.5h)}{leftV}
    \fmfforce{(.1w,.1h)}{dl}
    \fmfforce{(.1w,.9h)}{ul}
    \fmfforce{(.9w,.9h)}{ur}
    \fmfforce{(.9w,.1h)}{dr}
    \fmf{plain}{dl,one}
    \fmf{plain}{ul,two}
    \fmf{plain}{ur,three}
    \fmf{plain}{dr,four}
    \fmf{fermion,label=$\ell$,label.side=left}{dr,dl}
    \fmf{plain,rubout=3mm}{dl,ur}
    \fmf{plain}{ur,ul}
    \fmf{plain}{ul,dr}
    \fmfv{label=$I^1_4(t\,,;u)=$}{leftV}
    \fmfv{label=$1$}{one}
    \fmfv{label=$2$}{two}
    \fmfv{label=$3$}{three}
    \fmfv{label=$4$}{four}
    %
   \fmffreeze
 \end{fmfchar*} 
 \hspace{2cm}
 \begin{fmfchar*}(22,20)
    \fmfforce{(sw)}{one}
    \fmfforce{(nw)}{two}
    \fmfforce{(ne)}{three}
    \fmfforce{(se)}{four}
    \fmfforce{(0w,.5h)}{leftV}
    \fmfforce{(.1w,.1h)}{dl}
    \fmfforce{(.1w,.9h)}{ul}
    \fmfforce{(.9w,.9h)}{ur}
    \fmfforce{(.9w,.1h)}{dr}
    \fmf{plain}{dl,one}
    \fmf{plain}{ul,two}
    \fmf{plain}{ur,three}
    \fmf{plain}{dr,four}
    \fmf{fermion,rubout=3mm}{ur,dl}
    \fmf{plain}{dl,ul}
    \fmf{plain}{ul,dr}
    \fmf{plain}{dr,ur}
    \fmfv{label=$I^1_4(u\,,;s)=$}{leftV}
    \fmfv{label=$\ell$,label.angle=12,label.dist=5mm}{dl}
    \fmfv{label=$1$}{one}
    \fmfv{label=$2$}{two}
    \fmfv{label=$3$}{three}
    \fmfv{label=$4$}{four}
    %
   \fmffreeze
 \end{fmfchar*} }
 \label{Boxes1}
\end{equation}
\twhite{.}

As there is no unique origin in loop momentum space, there is a general problem how to label the loop momentum $\ell$ in individual diagrams; 
we will come back to this point shortly. In the definition (\ref{Boxes1}), we chose an arbitrary origin for the loop momentum routing in each of the three boxes.

Let us consider a double cut of the amplitude where $\ell^2=(\ell-p_1)^2=0$ which chooses natural labels on the cut. 
For complex momenta, there are two solutions to the on-shell conditions. Here we choose the one with $\ell=\lambda_1\widetilde{\lambda}_\ell$ 
for some $\tw{\lambda}_{\ell}$, which corresponds to the cut diagram. The grey blob corresponds to five point $(L-1)$ loop amplitude, but in our case $L=1$ and it is just tree,
\begin{equation}
\raisebox{-45pt}{
\includegraphics[trim={0cm .1cm 0cm 0cm},clip,scale=.45]{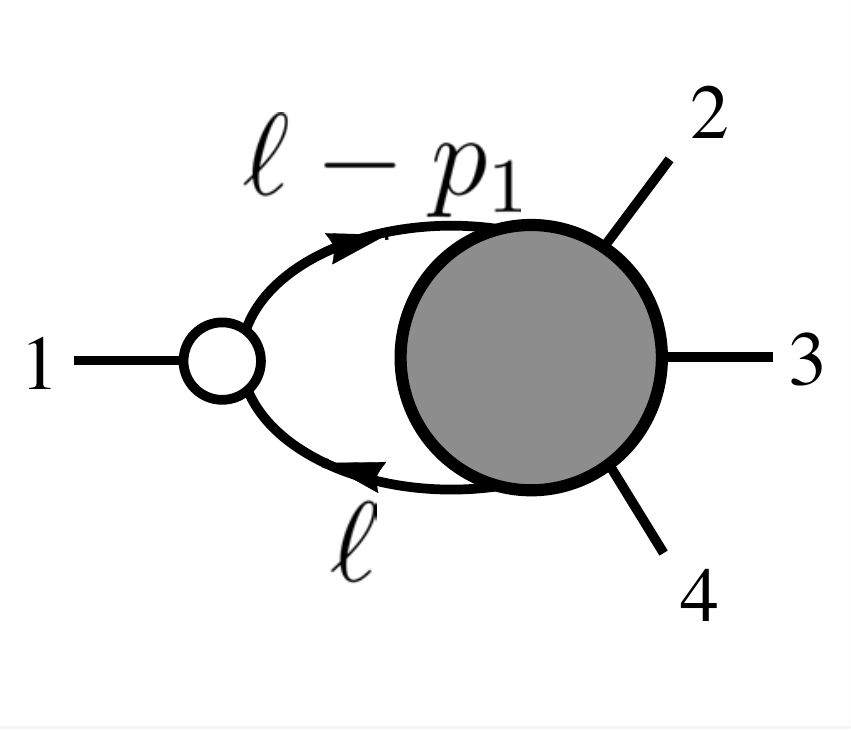}}
\label{OneLoop4ptChiralDoubleCutPic}
\end{equation}

Note that for $\ell^2=0$ the loop momentum $\ell$ becomes null and can be written as, $\ell=\lambda_\ell\widetilde{\lambda}_\ell$ so that the other 
propagator factorizes, $(\ell-p_1)^2=\ab{\ell 1}[\ell 1]$. The solution we chose sets $\ab{\ell 1}=0$ and the Jacobian of 
this double cut is,
\begin{equation}
\J = \frac{1}{[\ell 1]}\,. 
\label{JacDoubleCut}
\end{equation}

Using the box-expansion of the one-loop amplitude (\ref{eqn:M41loop}) we can calculate the residue on this cut for all three boxes (\ref{Boxes1}) individually and get,
\begin{align}
& \Big[I^1_4(s,t)+I^1_4(t,u)+I^1_4(u,s)\Big] \Bigg|_{\ell = \lambda_1\widetilde{\lambda}_\ell} = \nonumber\\
&=\frac{1}{[\ell 1]} \left[
			      \frac{1}{(\ell-p_1-p_2)^2(\ell+p_4)^2} 
			    + \frac{1}{(\ell-p_1-p_3)^2(\ell+p_4)^2}
			    + \frac{1}{(\ell-p_1-p_2)^2(\ell+p_3)^2} 
			  \right] \Bigg|_{\ell = \lambda_1\widetilde{\lambda}_\ell}\nonumber\\ 
&= \frac{1}{[\ell 1]} \left[\frac{1}{\ab{12}([12]-[\ell 2])\ab{14}[\ell 4]}  
			       + \frac{1}{\ab{13}([13]-[\ell 3])\ab{14}[\ell 4]}
			       + \frac{1}{\ab{12}([12]-[\ell 2])\ab{13}[\ell 3]}
			   \right]\nonumber\\
&= \frac{[\ell 1]\cdot [34]\ab{14}}{[\ell1]\cdot [\ell 3][\ell 4]([12]-[\ell 2])([13]-[\ell 3])\ab{12}\ab{13}\ab{14}}\label{BoxSum1}
\end{align}

From the Jacobian (\ref{JacDoubleCut}), each term contains a factor $\frac{1}{[\ell 1]}$ but combining all three boxes we generate an expression with $[\ell 1]$ 
in the numerator which cancels $\J$. However, this is not enough. Our conjecture was that on this cut the amplitude behaves like 
$\sim [\ell 1]$. The computation above seems to immediately contradict the conjecture but due to labeling issues mentioned earlier, the calculation is incomplete. In labeling the box 
diagrams in (\ref{Boxes1}), we made a particular choice. We could have labeled the three boxes in a different way,
\newline \twhite{.}
\begin{equation}
\twhite{.}\hspace{1.4cm}
\raisebox{-25pt}{
  \begin{fmfchar*}(22,20)
    \fmfforce{(sw)}{one}
    \fmfforce{(nw)}{two}
    \fmfforce{(ne)}{three}
    \fmfforce{(se)}{four}
    \fmfforce{(0w,.5h)}{leftV}
    \fmfforce{(.1w,.1h)}{dl}
    \fmfforce{(.1w,.9h)}{ul}
    \fmfforce{(.9w,.9h)}{ur}
    \fmfforce{(.9w,.1h)}{dr}
    \fmf{plain}{dl,one}
    \fmf{plain}{ul,two}
    \fmf{plain}{ur,three}
    \fmf{plain}{dr,four}
    \fmf{plain}{dr,dl}
    \fmf{fermion,label=$\ell$,label.side=left}{ul,dl}
    \fmf{plain}{ul,ur}
    \fmf{plain}{ur,dr}
    \fmfv{label=$\tw{I}^1_4(s\,,;t)=$}{leftV}
    \fmfv{label=$1$}{one}
    \fmfv{label=$2$}{two}
    \fmfv{label=$3$}{three}
    \fmfv{label=$4$}{four}
    %
   \fmffreeze
 \end{fmfchar*} 
 \hspace{2cm}
 \begin{fmfchar*}(22,20)
    \fmfforce{(sw)}{one}
    \fmfforce{(nw)}{two}
    \fmfforce{(ne)}{three}
    \fmfforce{(se)}{four}
    \fmfforce{(0w,.5h)}{leftV}
    \fmfforce{(.1w,.1h)}{dl}
    \fmfforce{(.1w,.9h)}{ul}
    \fmfforce{(.9w,.9h)}{ur}
    \fmfforce{(.9w,.1h)}{dr}
    \fmf{plain}{dl,one}
    \fmf{plain}{ul,two}
    \fmf{plain}{ur,three}
    \fmf{plain}{dr,four}
    \fmf{plain}{dr,dl}
    \fmf{fermion,rubout=3mm}{ur,dl}
    \fmf{plain}{ur,ul}
    \fmf{plain}{ul,dr}
    \fmfv{label=$\tw{I}^1_4(t\,,;u)=$}{leftV}
    \fmfv{label=$1$}{one}
    \fmfv{label=$2$}{two}
    \fmfv{label=$3$}{three}
    \fmfv{label=$4$}{four}
    \fmfv{label=$\ell$,label.dist=.5cm,label.angle=64}{dl}
    %
   \fmffreeze
 \end{fmfchar*} 
 \hspace{2cm}
 \begin{fmfchar*}(22,20)
    \fmfforce{(sw)}{one}
    \fmfforce{(nw)}{two}
    \fmfforce{(ne)}{three}
    \fmfforce{(se)}{four}
    \fmfforce{(0w,.5h)}{leftV}
    \fmfforce{(.1w,.1h)}{dl}
    \fmfforce{(.1w,.9h)}{ul}
    \fmfforce{(.9w,.9h)}{ur}
    \fmfforce{(.9w,.1h)}{dr}
    \fmf{plain}{dl,one}
    \fmf{plain}{ul,two}
    \fmf{plain}{ur,three}
    \fmf{plain}{dr,four}
    \fmf{plain,rubout=3mm}{ur,dl}
    \fmf{fermion,label=$\ell$,label.side=left}{ul,dl}
    \fmf{plain}{ul,dr}
    \fmf{plain}{dr,ur}
    \fmfv{label=$\tw{I}^1_4(u\,,;s)=$}{leftV}
    \fmfv{label=$1$}{one}
    \fmfv{label=$2$}{two}
    \fmfv{label=$3$}{three}
    \fmfv{label=$4$}{four}
    %
   \fmffreeze
 \end{fmfchar*} }
 \label{Boxes2}
\end{equation}
\twhite{.}

\noindent
which gives a different residue on the cut (\ref{OneLoop4ptChiralDoubleCutPic}), 
\begin{align}
& \Big[\tw{I}^1_4(s,t)+\tw{I}^1_4(t,u)+\tw{I}^1_4(u,s)\Big] \Bigg|_{\ell = \lambda_1\tw{\lambda}_{\ell}} = \nonumber\\
&\frac{1}{[\ell 1]} \left[\frac{1}{(\ell-p_1-p_4)^2(\ell+p_2)^2}
			      + \frac{1}{(\ell-p_1-p_4)^2(\ell+p_3)^2}
			      + \frac{1}{(\ell-p_1-p_3)^2(\ell+p_2)^2} 
			 \right] \Bigg|_{\ell = \lambda_1\tw{\lambda}_{\ell}}\nonumber\\ 
&= \frac{[\ell 1]\cdot [23]\ab{12}}{[\ell1]\cdot [\ell 2][\ell 3]([13]-[\ell 3])([14]-[\ell 4])\ab{12}\ab{13}\ab{14}}
\label{BoxSum2}
\end{align}

Summing over both expression (\ref{BoxSum1}) and (\ref{BoxSum2}) (we should include a factor $\frac12$ but that is irrelevant here) 
and using $[23]\la12\ra=[34]\la14\ra$ we get
\begin{equation}
\M^{1}_4(1234)\Bigg|_{\ell = \lambda_1\widetilde{\lambda}_\ell} \hskip -.4cm \sim
    \frac{[23]\ab{12}[24]\cdot [\ell 1]^2}{[\ell1]\cdot [\ell 2][\ell 3][\ell 4]([12]-[\ell 2])([13]-[\ell 3])([14]-[\ell 4])\ab{12}\ab{13}\ab{14}},
\end{equation}

so that our conjecture indeed passes this check as the amplitude vanishes for $[\ell 1]=0$, i.e. $\ell\sim p_1$. This example clearly demonstrates 
that the symmetrization over labels is important in getting the correct result. Note that the sum over six terms naturally arises when one starts directly
from the cut-picture (\ref{OneLoop4ptChiralDoubleCutPic}). To get all contributions, one is instructed to expand the five-point tree in all possible ways and find 
the contributions of all basis integrals. This procedure automatically takes into account all labellings of loop-momenta.

%
\subsubsection*{Four point two-loop}
%
%

We will now test the same property for the four-point two-loop amplitude which is given as a sum of planar- and 
non-planar double-box integrals including a numerator factor \cite{Bern:1998ug},
\newline\twhite{.}
\begin{equation}
\twhite{.}\hspace{1.6cm}
 \raisebox{-25pt}{
 \begin{fmfchar*}(40,20)
    \fmfset{arrow_len}{3mm}
    \fmfforce{sw}{one}
    \fmfforce{nw}{two}
    \fmfforce{ne}{three}
    \fmfforce{se}{four}
    \fmfforce{0.15w,.5h}{phantomvertex}
    \fmfforce{(.15w,.85h)}{lu}
    \fmfforce{(.15w,.15h)}{ld}
    \fmfforce{(.85w,.85h)}{ru}
    \fmfforce{(.85w,.15h)}{rd}
    \fmfforce{(.5w,.85h)}{mu}
    \fmfforce{(.5w,.15h)}{md}
    \fmf{plain}{ld,one}
    \fmf{plain}{lu,two}
    \fmf{plain}{ru,three}
    \fmf{plain}{rd,four}
    \fmf{plain}{md,ld}
    \fmf{plain}{ld,lu}
		%
    \fmf{plain}{lu,mu,ru,rd,md,mu}
    %
    \fmfv{label=$1$,label.dist=0.05mm}{one}
    \fmfv{label=$2$,label.dist=0.05mm}{two}
    \fmfv{label=$3$,label.dist=0.05mm}{three}
    \fmfv{label=$4$,label.dist=0.05mm}{four}
    \fmfv{label=$I^{(P)}_{(1234)}=s^2\times$}{phantomvertex}
    \fmffreeze
 \end{fmfchar*}
 }
 \hspace{3cm}
 \raisebox{-25pt}{
  \begin{fmfchar*}(40,20)
    \fmfset{arrow_len}{4mm}
    \fmfforce{0.1w,0h}{one}
    \fmfforce{0.1w,1h}{two}
    \fmfforce{(.4w,.5h)}{three}
    \fmfforce{(w,.5h)}{four}
    \fmfforce{0.15w,.5h}{phantomvertex}
    \fmfforce{(.2w,.9h)}{lu}
    \fmfforce{(.2w,.1h)}{ld}
    \fmfforce{(.5w,.5h)}{boxl}
    \fmfforce{(.9w,.5h)}{boxr}
    \fmfforce{(.7w,.9h)}{boxu}
    \fmfforce{(.7w,.1h)}{boxd}
    \fmf{plain}{ld,one}
    \fmf{plain}{lu,two}
    \fmf{plain}{boxl,three}
    \fmf{plain}{boxr,four}
    \fmf{plain}{lu,boxu}
    \fmf{plain}{ld,lu}
    \fmf{plain}{boxd,ld}
    \fmf{plain}{boxl,boxu}
    \fmf{plain}{boxd,boxl}
    \fmf{plain}{boxr,boxu} 
    \fmf{plain}{boxd,boxr}
		%
    \fmfv{label=$4$,label.dist=0.05mm}{one}
    \fmfv{label=$3$,label.dist=0.05mm}{two}
    \fmfv{label=$2$,label.dist=0.05mm}{three}
    \fmfv{label=$1$,label.dist=0.05mm}{four}
    \fmfv{label=$I^{(NP)}_{(1234)}=s^2\times$}{phantomvertex}
    \fmffreeze
 \end{fmfchar*}
 }
 \label{2loop4ptBasisInts}
\end{equation}
\twhite{.}
\begin{align}
\M_{4}^{2} &= \frac{\K_{8}}{4}\sum_{\sigma\in\mathfrak{S}_4}
						  \Big[ I_\sigma^{(P)} + I_\sigma^{(NP)} \Big] \,,
\label{eqn:two_loop_amplitude}
\end{align}
where the sum over $\sigma$ runs over all 24 permutations of $\mathfrak{S}_4$.

The full calculation can be performed numerically, but here we present a simplified version in which we calculate the 
residue on $\ell^2=\ab{\ell 1} = [\ell 1]=0$ which sets $\ell=\alpha p_1$ directly. When combining all pieces, the numerator again 
generates $[\ell 1]^2$ so that the residue on the $\frac{1}{[\ell 1]}$ pole vanishes quadratically.
Going directly to the kinematic region where $\ell = \alpha p_1$ we are only able to see a pure vanishing $\M^2_4(1234)\Big|_{\ell = \alpha p_1} = 0$, 
but even this weaker statement requires an intricate cancellation between a large number of different terms. 

\begin{figure}[ht!]
\centering
\includegraphics[trim = {0cm 4cm 0cm 4cm},clip,scale=.6]{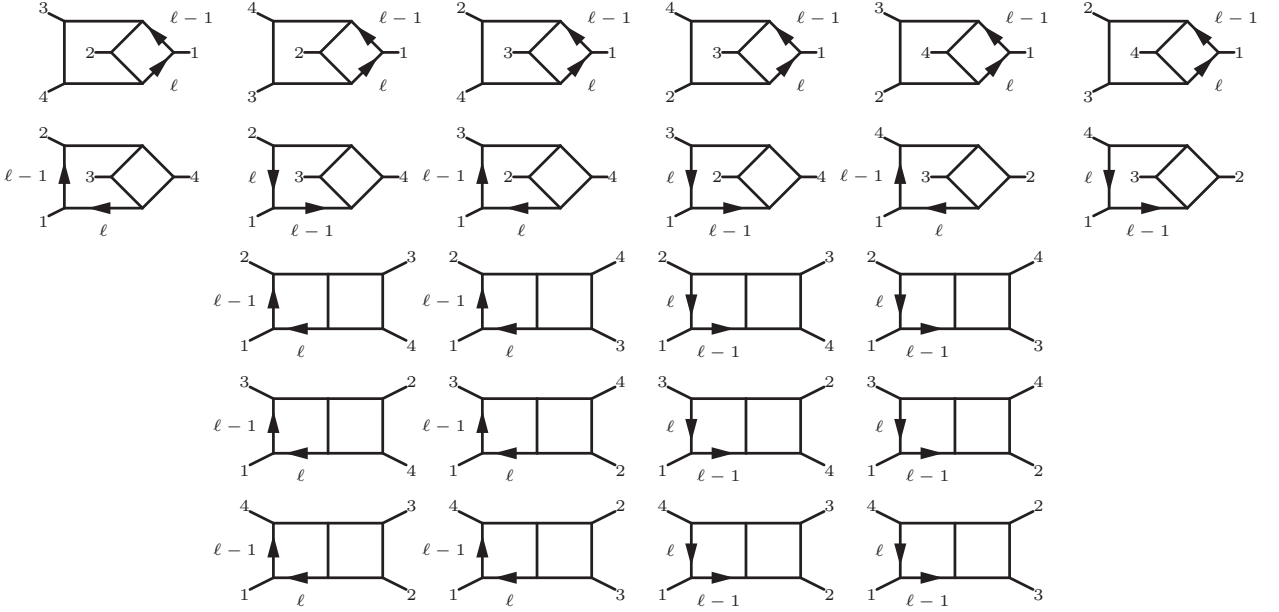}
\vskip -.4cm
\caption{\label{fig:2loopCollallInts}Contributing integrals on the collinear cut.}
\end{figure}

Starting with the collinear cut $\ell^2= \ab{\ell 1} = \sqb{\ell 1} = 0$, there are 24 terms contributing. If we look at the nonplanar integrals, for collinear kinematics $\ell = \alpha p_1$, we can use one factor of $s$ of the numerator (\ref{2loop4ptBasisInts}) 
to decompose the pentagon as a sum of boxes. This is only possible for this special kinematics.
\begin{equation}
\raisebox{-75pt}{
 \includegraphics[trim={0cm 5cm 0cm 5cm},clip,scale=.49]{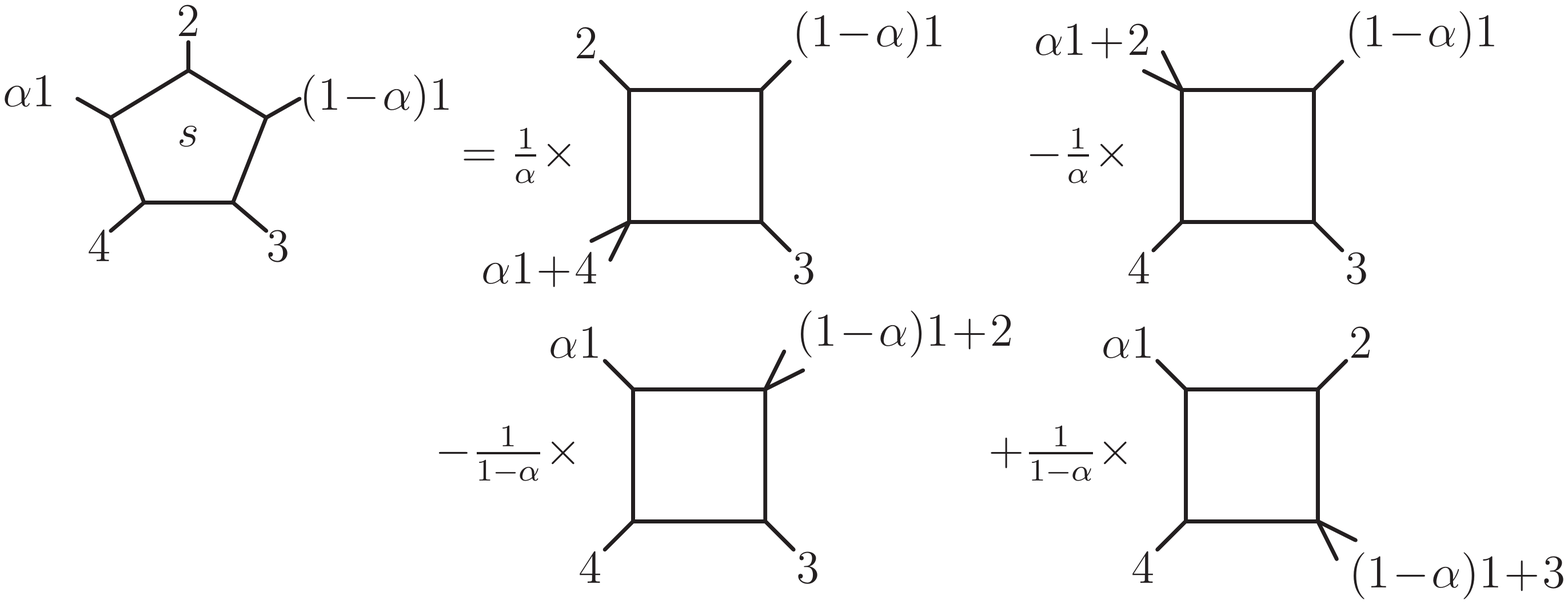}
 }
 \label{pentDecomp}
\end{equation}

If one uses the pentagon decomposition (\ref{pentDecomp}) on all nonplanar integrals in the first line of Fig.~\ref{fig:2loopCollallInts} 
and rewrites the $\frac{1}{\alpha}$ and $\frac{1}{1-\alpha}$ coefficients of the boxes in terms of propagators by multiplying and dividing by appropriate Mandelstam variables, 
one can see that all the planar double-boxes cancel. Each nonplanar integral in the first line cancels exactly two planar 
double boxes, so that the counting works perfectly. 
The remaining two terms of the decomposition that come with a plus sign are almost as straight-forward. One has to collect 
all these terms and re-express them as non-planar integrals. Combined with the non-planar integrals of the second line in Fig.~\ref{fig:2loopCollallInts}, 
one can show that they always come in the combination $(s+t+u) = 0$ so that they also cancel. This concludes our calculation and indeed we find our conjecture to hold.
All signs work out such that the two-loop four-point amplitude in fact vanishes on the collinear cut $\ell = \alpha p_1$. 
%
%
\subsubsection*{Internal collinear region}
%

Finally we can show one more example when the collinear region is between internal loops only corresponding to the cases described in the beginning of Sec.~\ref{subsec:CollBehavior}.
The simplest example where we can study this kinematic region is for the two-loop four-point amplitude discussed above. Instead of going to the triple cut $\ell^2_1=\ell^2_2=(\ell_1+\ell_2)^2 =0$ 
we can cut one more propagator to simplify the analysis by limiting the number of contributing terms,
\begin{equation}
\raisebox{-42pt}{
  \includegraphics[scale=.45]{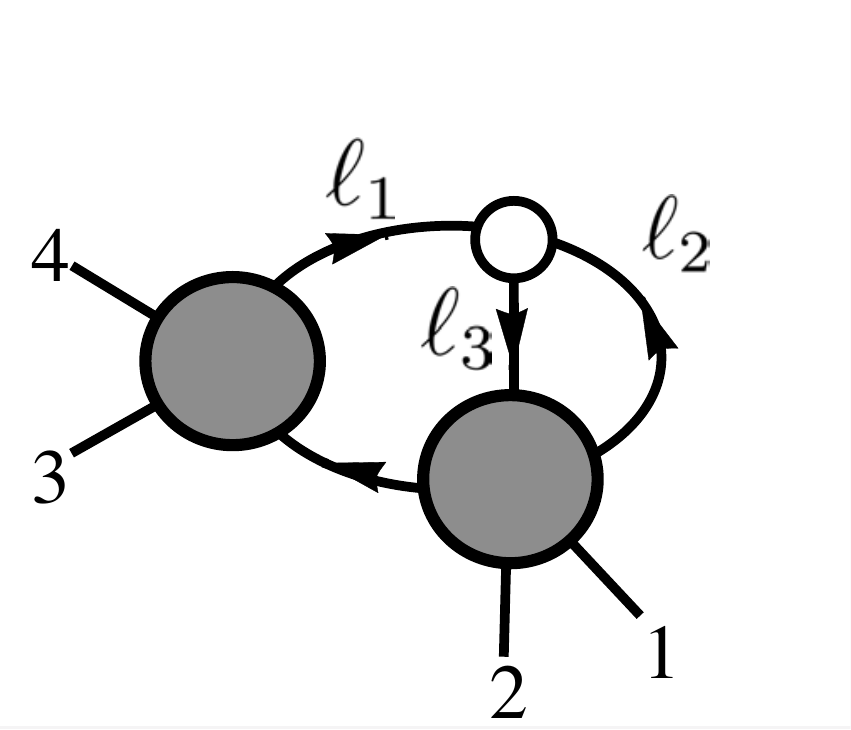}}
 \label{twoLoopIntCollRegion}
\end{equation}

Parameterizing the cut solution on $\ell^2_1=\ell^2_2=0$ as
$$
\ell_1 = \Big[\lam{1}+\alpha_1\lam{2}\Big]\Big[\alpha_2\lamt{2}+\alpha_3\lamt{1}\Big]\,, \qquad
\ell_2 = \Big[\lam{1}+\beta_1\lam{2}\Big]\Big[\beta_2\lamt{2}+\beta_3\lamt{1}\Big]\,,
$$

the third propagator $\ell^2_3 \equiv (\ell_1+\ell_2)^2$ factorizes and we cut $\ab{\ell_1\ell_2} = 0$ by setting $\beta_1 = \alpha_1$. The remaining part of the facotrized propagator becomes,
$[\ell_1\ell_2] = [21](\alpha_2\beta_3-\alpha_3\beta_2)$. As mentioned before, we simplify our life by further cutting $(\ell_1+p_3+p_4)^2 = 0$ which sets $\alpha_3 = 1-\alpha_1\alpha_2$.

Blowing up the blobs in (\ref{twoLoopIntCollRegion}) into planar and non-planar double-boxes (\ref{2loop4ptBasisInts}) of different labels and combining all $(8+4)$ terms, 
we checked numerically that the two-loop amplitude behaves as,
\begin{align}
 M^2_4 \Big|_{\ab{\ell_1\ell_2}=0} \sim \frac{[\ell_1\ell_2]^2}{[\ell_1\ell_2]^{\twhite{1}}} \cdot \overline{\mathcal{R}}\,,
\end{align}
where the numerator generates the $[\ell_1\ell_2]^2$-factor consistent with our conjecture.

%
\section{Conclusion}
%

In this paper we studied on-shell diagrams in gravity theories. We wrote a Grassmannian representation using edge variables and our formulation includes a non-trivial numerator factor in the measure as well as higher degree poles in the denominator. We showed that all higher poles correspond to cases where internal momenta in the loop are sent to infinity while all erasable edges are represented by single poles only. The numerator factor can be interpreted as a set of collinearity conditions on the on-shell momenta. We provide several examples in the paper for both leading singularities as well as diagrams with unfixed parameters. Because on-shell diagrams are also cuts of gravity loop amplitudes it is natural to conjecture that loop amplitudes share the same properties. We tested this conjecture on the cases of 1-loop and 2-loop amplitudes in ${\cal N}=8$ SUGRA and found a perfect agreement. Unlike in the Yang-Mills case these properties of on-shell diagrams can not be
implemented term-by-term and require non-trivial cancellations between diagrams (even at four-point one-loop). 

There was one aspect of gravity on-shell diagrams we did not discuss in more detail: {\it poles at infinity}. While absent in gauge theory they are present in gravity on-shell diagrams as poles of arbitrary degree. Poles at finite locations in momentum space correspond to erasing edges in on-shell diagrams but there is no such interpretation for poles at infinity. It is not clear how to embed them in the Grassmannian and what is the on-shell diagrammatic interpretation for them. This also prevents us from writing homological identities between different on-shell diagrams which was an important ingredient in the Yang-Mills case. Finally, the poles at infinity are closely related to the UV-behavior of gravity loop amplitudes and further study of their role in on-shell diagrams could lead to new insights there.

\medskip

{\bf Note:} While this work was completed, \cite{HeslopLipsteinGravOS} appeared which has some overlap with our results.

\subsection*{Acknowledgments}
We thank Nima Arkani-Hamed, Zvi Bern, Jacob Bourjaily, Sean Litsey and James Stankowicz for interesting discussions. Most figures are drawn with the Mathematica package \cite{Bourjaily:2012gy}. E.~H. is supported in part by a Dominic Orr Graduate Fellowship and by DOE Grant $\#$ DE-SC0011632.

\bibliographystyle{JHEP}
\phantomsection         
            \bibliography{References}
            \clearpage
            
\end{fmffile}

\end{document}